\documentclass[a4paper]{article}

\usepackage[english]{babel}
\usepackage[utf8x]{inputenc}
\usepackage[T1]{fontenc}

\usepackage[a4paper,top=3cm,bottom=2cm,left=3cm,right=3cm,marginparwidth=1.75cm]{geometry}

\usepackage{dsfont}
\usepackage{amsthm}
\usepackage{amsmath}
\usepackage{amssymb}
\usepackage{graphicx}    
\usepackage[colorinlistoftodos]{todonotes}
\usepackage[colorlinks=true, allcolors=blue]{hyperref}
\usepackage{calrsfs} 
\usepackage{cancel} 
\usepackage[normalem]{ulem}
\usepackage{bbold}						
\DeclareMathAlphabet{\mathcal}{OMS}{cmsy}{m}{n} 
\usepackage{stmaryrd}					
\usepackage{bm}
\usepackage{fancyvrb}
\usepackage[all]{xy}					 
\usepackage{authblk}			

\newcommand{\ba}{\textbf{a}}
\newcommand{\bA}{\textbf{A}}
\newcommand{\bb}{\textbf{b}}
\newcommand{\bB}{\textbf{B}}

\newcommand{\bh}{\textbf{h}}
\newcommand{\bx}{\textbf{x}}
\newcommand{\by}{\textbf{y}}

\newcommand{\br}{\textbf{r}}
\newcommand{\bF}{\textbf{F}}
\newcommand{\bJ}{\textbf{J}}
\newcommand{\bI}{\textbf{I}}

\newcommand{\bD}{\textbf{D}}
\newcommand{\bC}{\textbf{C}}
\newcommand{\be}{\textbf{e}}
\newcommand{\bE}{\textbf{E}}
\newcommand{\bu}{\textbf{u}}
\newcommand{\bv}{\textbf{v}}

\newcommand{\bT}{\textbf{T}}
\newcommand{\bp}{\textbf{p}}
\newcommand{\bP}{\textbf{P}}

\newcommand{\bS}{\textbf{S}}
\newcommand{\bs}{\textbf{s}}
\newcommand{\bw}{\textbf{w}}

\newcommand{\bK}{\textbf{K}}
\newcommand{\bQ}{\textbf{Q}}

\newcommand{\boldeta}{\boldsymbol{\eta}}

\newcommand{\bmu}{\boldsymbol{\mu}}

\newcommand{\bOmega}{\boldsymbol{\Omega}}

\newcommand{\balpha}{\boldsymbol{\alpha}}
\newcommand{\bbeta}{\boldsymbol{\beta}}
\newcommand{\blambda}{\boldsymbol{\lambda}}
\newcommand{\bnu}{\boldsymbol{\nu}}
\newcommand{\bxi}{\boldsymbol{\xi}}
\newcommand{\bgamma}{\boldsymbol{\gamma}}
\newcommand{\bGamma}{\boldsymbol{\Gamma}}
\newcommand{\btheta}{\boldsymbol{\theta}}

\newcommand{\rmd}{\mathrm{d}}

\newcommand{\mbR}{\mathbb{R}}

\newcommand{\mbL}{\mathbb{L}}
\newcommand{\mbZ}{\mathbb{Z}}

\newcommand{\mbS}{\mathbb{S}}

\newcommand{\mcM}{\mathcal{M}}

\newcommand{\mcD}{\mathcal{D}}
\newcommand{\mcP}{\mathcal{P}}

\newcommand{\mcH}{\mathcal{H}}
\newcommand{\mcE}{\mathcal{E}}
\newcommand{\mcA}{\mathcal{A}}
\newcommand{\mcO}{\mathcal{O}}
\newcommand{\mcS}{\mathcal{S}}
\newcommand{\mcC}{\mathcal{C}}
\newcommand{\mcL}{\mathcal{L}}
\newcommand{\mcI}{\mathcal{I}}

\newcommand{\tpss}{\tilde{p}_{\rm ss}}

\newcommand{\reversed}[1]{\bar{#1}}
\newcommand{\bvss}{\bv_{\rm ss}}    
\newcommand{\bJss}{\bJ_{\rm ss}}
\newcommand{\Jss}{J_{\rm ss}}
\newcommand{\pss}{p_{\rm ss}}
\newcommand{\Pss}{P_{\rm ss}}
\newcommand{\bPss}{\bP_{\rm ss}}

\newcommand{\bbpssl}{\reversed{p}_{\rm ss}^{\blambda}}

\newcommand{\ie}{\textit{i.e. }}
\newcommand{\eg}{\textit{e.g. }}
\newcommand{\dime}{\mathrm{dim}}

\newcommand{\Div}{\mathrm{div}}
\newcommand{\bnabla}{\boldsymbol{\nabla}}

\newcommand{\halpha}{\hat{\alpha}}

\newcommand{\wnabla}{\widetilde{\nabla}}
\newcommand{\wbE}{\widetilde{\bE}}
\newcommand{\wbOmega}{\widetilde{\bOmega}}
\newcommand{\tblambda}{\tilde{\blambda}}
\newcommand{\tlambda}{\tilde{\lambda}}

\newcommand{\llangle}{\left\langle}
\newcommand{\rrangle}{\right\rangle}

\definecolor{monVert}{RGB}{0,200,0}

\definecolor{monOrange}{RGB}{250,150,0}

\definecolor{lavande}{RGB}{170,150,255}

\definecolor{monBleu}{RGB}{25,58,239}
\definecolor{monVert}{RGB}{126,211,33}
\definecolor{monRouge}{RGB}{208,2,27}
\definecolor{bleuClair}{RGB}{153,204,255}
\definecolor{bleuClair2}{RGB}{102,178,255}
\definecolor{monRose}{RGB}{255,102,178}

\newcommand{\via}{\textit{via }}

\newcommand{\mkt}{\mathfrak{t}}
\newcommand{\mkm}{\mathfrak{m}}
\newcommand{\mke}{\mathfrak{e}}
\newcommand{\mbT}{\mathbb{T}}
\newcommand{\mcT}{\mathcal{T}}

\newcommand{\mcW}{\mathcal{W}}

\newcommand{\mkX}{\mathfrak{X}}
\newcommand{\mcY}{\mathcal{Y}}
\newcommand{\mkT}{\mathfrak{T}}

\title{Geometric theory of (extended) time-reversal symmetries in stochastic processes -- Part I: finite dimension}

\author[1,2]{J. O'Byrne}
\author[1]{M.E. Cates}

\affil[1]{DAMTP, Centre for Mathematical Sciences, University of Cambridge, Wilberforce Road, Cambridge CB3 0WA, United Kingdom}
\affil[2]{Laboratoire Jean Perrin, Sorbonne Universit\'e, 4 Place Jussieu, Paris 75005, France}

\date{}

\begin{document}
\maketitle

\begin{abstract}
In this article, we analyze three classes of time-reversal of a Markov process with Gaussian noise on a manifold. 
We first unveil a commutativity constraint for the most general of these time-reversals to be well defined. Then we give a triad of necessary and sufficient conditions for the stochastic process to be time-reversible. While most reversibility conditions in the literature require knowledge of the stationary probability, our conditions do not, and therefore can be analytically checked in a systematic way. 
We then show that the mathematical objects whose cancellation is required by our reversibility conditions play the role of independent sources of entropy production.
Furthermore, we give a geometric interpretation of the so-called irreversible cycle-affinity as the vorticity of a certain vector field for a Riemannian geometry given by the diffusion tensor. We also discuss the relation between the time-reversability of the stochastic process and that of an associated deterministic dynamics: its Stratonovitch average. 
Finally, we show that a suitable choice of a reference measure -- that can be considered as a prior or a gauge, depending on the context -- allows to study a stochastic process in a way that is both coordinate-free and independent of the prescription used to define stochastic integrals. 
When this reference measure plays the role of a gauge choice, we interpret our previous results through the lens of gauge theory and prove them to be gauge-invariant.
\end{abstract}

\newpage 

\tableofcontents

\newpage

\newpage

\section{Introduction}
\label{sec:intro}

Statistical physics aims at describing, with a reduced set of variables, the large-scale, long-time behavior of assemblies of elementary agents, possibly in contact with reservoirs. For systems at equilibrium, this program can be explicitly carried out and 
time-average observables can be computed as ensemble averages with respect to the appropriate Boltzmann-Gibbs measure. 
The success of the theory of equilibrium statistical physics allowed to improve our understanding and control of many systems, with a wide variety of applications over the last century.

Naturally, physicists have then tried to extend the theory out of equilibrium. One approach to tackle this problem consists in modeling the degrees of freedom we want to integrate out by random variables. In the context of Markov processes, it has been understood that the systems of equilibrium statistical mechanics and thermodynamics correspond to stationary states of dynamics satisfying a dynamical property called detailed balance~\cite{maxwell1867iv,boltzmann2022lectures,onsager1931reciprocal}. This detailed balance is a time-reversal symmetry (TRS) -- or reversibility -- property of the distribution of trajectories in the steady state.

Perhaps contrary to what one might think at first sight, the meaning of ``reversibility'', or ``time-reversal symmetry'', turns out to be subtle in stochastic processes and can take various forms depending on the physical context.
For instance, the overdamped motion of a particle in a homogeneous thermal bath and in a potential is described by the overdamped Langevin equation that reads
\begin{equation}
\dot{\br} =-\mu\nabla V(\br) + \sqrt{2\mu kT}\boldeta \ , 
\label{overdamped_Langevin}
\end{equation}
where $\mu$ is the mobility, $T$ the temperature, $k$ the Boltzmann constant, and $\boldeta$ a Gaussian white noise of zero mean and unit variance. Dynamics~\eqref{overdamped_Langevin} is well know to have a path probability $\mcP$ which, in steady state, is invariant under the time-reversal that just consists in the inversion of the time variable. In other words, in steady state, any trajectory has the same probability to be traveled in one way or the other: $\mcP[(\br_t)_{t\in[0,\mcT]}]=\mcP[(\br_{\mcT-t})_{t\in[0,\mcT]}]$. 
But in a context where the inertia of the particle of mass $m$ is not negligible, this particle follows the Kramers dynamics:
\begin{eqnarray}
\dot{\br} &=&\bv \label{underdamped_Langevin_01}\\ 
m\dot{\bv} &=& -\nabla V(\br) - \mu^{-1}\bv + \sqrt{2\mu^{-1}kT} \boldeta \ , \label{underdamped_Langevin_02}
\end{eqnarray}
which is again reversible, but one has to reverse the velocity $\bv$ alongside the time variable: $\mcP[(\br_t,\bv_t)_{t\in[0,\mcT]}] = \mcP[(\br_{\mcT-t},-\bv_{\mcT-t})_{t\in[0,\mcT]}]$. 
Finally, if the particle has now a charged $q$ and is embedded in a magnetic field $\bB$, the dynamics reads:
 \begin{eqnarray}
\dot{\br} &=& \bv \label{Langevin_Magnetic_field_01} \\ 
m\dot{\bv} &=& q\bv\times\bB -\nabla V(\br) - \mu^{-1}\bv + \sqrt{2\mu^{-1}kT} \boldeta \ . \label{Langevin_Magnetic_field_02}
\end{eqnarray}
This dynamics is still thermodynamically reversible, as the Lorentz force does no work, but the corresponding time-symmetry now also requires the reversal of $\bB$, alongside the time and velocity variables: $\mcP_\bB[(\br_t,\bv_t)_{t\in[0,\mcT]}] = \mcP_{-\bB}[(\br_{\mcT-t},-\bv_{\mcT-t})_{t\in[0,\mcT]}]$. To account for the thermodynamic reversibility of this system, the necessary flipping of the magnetic field as we reverse time is due to the fact that $\bB$ results from the motion of charged particles, not explicitly described in the model, that would run backward when time is reversed, hence generating a reversed magnetic field.

In this article, we will call \textit{T-reversal} the time-reversal under which dynamics~\eqref{overdamped_Langevin} is invariant, \ie the one that corresponds to just inverting the time variable. A time-reversal similar to that under which dynamics~\eqref{underdamped_Langevin_01}-\eqref{underdamped_Langevin_02} is invariant, \ie that consists in inverting some degrees of freedom together with the time-variable, will be called a \textit{MT-reversal}, the ``M'' standing for the ``mirror'' map applied to those degrees of freedom. Finally, we will call \textit{EMT-reversal}, the ``E'' standing for ``extended'', the type of time-reversal that reverses the time variable, some degrees of freedom, and directly modify the drift of the dynamics (and also possibly its diffusion tensor). This is the type of time-reversal under which dynamics~\eqref{Langevin_Magnetic_field_01}-\eqref{Langevin_Magnetic_field_02} is invariant.
In these definitions, we allow the map that directly transform drift and diffusion tensor and the mirror map that reverse degrees of freedom to be trivial, \ie to do nothing. With this convention, these three notions of time-reversal are not mutually exclusive but satisfy the inclusion relations: T-reversal $\subset$ MT-reversal $\subset$ EMT-reversal.

Note that, although these three types of time-reversal allow a good description of the effect of running time backward in many stochastic models of physical systems, it is not the case for all such stochastic models.
Indeed, a stochastic model can represent a physical system in too coarse-grained a manner for this effect of running time backward in the real physical system to be correctly captured by the model~\cite{o2022time}. For instance, consider the following basic run-and-tumble model of the motile bacterium \textit{Escherichia coli}: its position $\br$ satisfies the dynamics $\dot{\br}=v_0\bu$, where the amplitude $v_0$ of its self-propulsion velocity is constant, while its orientation $\bu$ follows a Poisson process to get reoriented uniformly on the unit sphere.
\textit{Escherichia coli} self-propels thanks to flagella that are attached to the rear part of its body~\cite{berg2004coli}. The flagella are connected to a motor that can turn in two directions. In one direction, the bacterium is self-propelled in a rather straight line, while as the motor rotates in the opposite direction (which it does briefly and at random between extended ``runs'' of forward motion), the flagella perform a complicated motion that makes the body tumble.  
Now, as we run time backward, it is hard to tell what would happen to the internal machinery of the bacterium that allows it to self-propel. However, let us assume for a moment that this internal machinery is entirely reversible, including to the motor itself, so that reversing time only has the effect of making the motor change the direction of its rotation. 
We see that the run-and-tumble model is not capable of describing this time-reversal in which a run becomes a lengthy (rather than near-instantaneous) tumble and a tumble becomes an instantaneous (rather than extended) burst of forward motion.

Nevertheless, even for those models that are unfaithful to physical time-reversal, looking for a time-reversal-like symmetry can still be interesting and inform on properties of the statistics of trajectories, not necessarily as the real time is reversed but, for instance, as a video of the system is played backward.
Furthermore, if such a time-reversal symmetry is uncovered, then it often\footnote{We will see that this is not always the case for EMT-symmetry.} grants access to valuable information such as the stationary probability measure and current.

Thus, being able to systematically say whether a given system satisfies a particular time-reversal symmetry, or measuring the extent to which it breaks it, is a crucial step in the endeavour of extending statistical physics and thermodynamics to out-of-equilibrium systems.
While an extensive amount of work has been dedicated to T-reversal, less attention has been devoted to MT-reversal, and even less to its extended version, EMT-reversal~\cite{de1962non,graham1971generalized,risken1996fokker,van1992stochastic,gardiner1985handbook,haussmann1986time,jiang2004mathematical,maes1999fluctuation,seifert2005entropy,seifert2012stochastic,yang2021bivectorial}. 
The objective of this article is to fill this gap for stochastic processes in finite dimension. 
Note that, in a companion paper, we generalize the results of this article to the infinite-dimensional setting of field theory, which is often the best suited framework to describe collective behaviors.
\\

In section~\ref{sec:contextAndTRSconstruction}, we start by arguing that a choice of a reference measure -- which plays the role of either a gauge or a prior, depending on the context -- is required to describe the properties of the generic Markov process~\eqref{EDS} in a way
that is both coordinate–free and independent of the prescription used to define stochastic integrals. This idea builds upon several scattered pioneering works~\cite{graham1977covariant,polettini2012nonequilibrium,cates2022stochastic}.
Then, after having recalled the precise definition of MT-reversal, we give a general definition of EMT-reversal, uncover a commutativity constraint which is required for this time-reversal to be well defined, and give the EMT-reversed dynamics of the general stochastic process~\eqref{EDS}. 
In addition, we show that every dynamics is in fact invariant under at least one such EMT-reversal. One could then think that an EMT-reversal symmetry is too weak a property to have any interesting consequence. We show that, remarkably, uncovering such a symmetry in a certain explicit way allows to construct the stationary probability measure and current explicitly, just like in the cases of MT and T-reversal. 

In section~\ref{sec:EMT-reversibility}, we give a triad of conditions that the generic dynamics under study needs to satisfy in order to be EMT-reversible. 
In particular, we give conditions that do not appeal to the stationary measure -- which is out of reach in general -- and can thus be systematically tested, in contrast with many other conditions appearing in the literature. 
The first of these conditions is an Onsager-like symmetry of the diffusion tensor, the second corresponds to the vanishing of what we here call the irreversible cycle affinity, and the third and last one ensures the existence of a (certain kind of) stationary measure.
While the single condition to which our triad is reduced for T-reversibility was already known, our triplet of reversibility conditions is, we believe, new for (E)MT-reversal.

We then show that, assuming that the diffusion tensor is invertible and satisfies the Onsager-like symmetry, the entropy production rate~\cite{maes1999fluctuation,seifert2005entropy} can be expressed as the linear superposition of two independent sources, the first being proportional to the irreversible cycle affinity, while the second is proportional to the term whose cancellation is required by the third reversibility condition mentioned above.
In section~\ref{subsubsec:voritcityOperator}, we further show that the irreversible cycle affinity can be interpreted as the vorticity of the symmetric part of the drift under the mirror operation, for a Riemannian geometry imposed by the diffusion tensor.
In~\ref{subsec:linkDeterministicTRS}, we also discuss the link between the stochastic EMT-reversibility of dynamics~\eqref{EDS} and the deterministic reversibility of what we show to be interpretable as its Stratonovitch-average dynamics.

While all of these results are new for EMT-reversal, some of them were already known for MT-reversal on the Euclidean space $\mbR^d$. Nevertheless, we generalize, for the first time for (E)MT-reversal, all these results to a dynamics on a manifold in section~\ref{sec:generalization}. In particular, we describe how they are invariant under a change of reference measure, using the language of gauge theory in section~\ref{subsec:gaugeTheory}. We also interpret the various entropy production sources in terms of this gauge theory.

Finally, in section~\ref{sec:applications}, we illustrate the application of our results with a few pedagogical examples.

\newpage
\section{Covariant description of an SDE and construction of its EMT-reversal}
\label{sec:contextAndTRSconstruction}

\subsection{Choice of reference measure: the price of a prescription-free covariant description of a stochastic dynamics}
\label{subsec:generalContext}

In sections~\ref{sec:contextAndTRSconstruction} \&~\ref{sec:EMT-reversibility}, we consider the following stochastic differential equation (SDE) on $\mbR^d$ over the time interval $\mbT\equiv [0,\mcT]$:
\begin{equation}
\dot{\bx}_t = \bA(\bx_t) + \bs_{(\varepsilon)}(\bx_t) + \bb_\alpha(\bx_t)\eta^\alpha_t \ ,
\label{initialSDE}
\end{equation}
where $\varepsilon\in[0,1]$ is the prescription of the stochastic integral (spanning Ito \via Stratonovich to anti-Ito, see below), $\bb_1,\dots,\bb_n$ are $n$ vector fields over $\mbR^d$, $\boldeta=(\eta^1,\dots,\eta^n)^\top$ is a Gaussian white noise on $\mbR^n$ of zero mean and variance $\llangle \eta_t^\alpha\eta_{t'}^\beta \rrangle = 2\delta^{\alpha\beta}\delta(t-t')$, and we implicitly summed over repeated indices -- a convention we keep throughout this article. In eq.~\eqref{initialSDE}, following~\cite{cates2022stochastic}, we split the \textit{total $\varepsilon$-drift} $\bA_{(\varepsilon)}\equiv \bA + \bs_{(\varepsilon)}$ into the sum of the so-called \textit{spurious drift}, whose coordinate expression is given by
\begin{equation}
s^i_{(\varepsilon)} = \partial_j b^i_\alpha b^j_\alpha - 2\varepsilon b^j_\alpha\partial_jb^i_\alpha \ ,
\label{spuriousDrift}
\end{equation}
and what we here call the \textit{raw drift} $\bA$, which is a map from $\mbR^d$ to itself -- but not a vector field, as discussed below.
For all values of the prescription parameter $\varepsilon\in[0,1]$, the SDEs~\eqref{initialSDE} describe the same stochastic dynamics~\cite{cates2022stochastic}. Hence, the decomposition between raw and spurious drifts, where the former is fixed and the latter is given by~\eqref{spuriousDrift}, allows a prescription-free description of the underlying stochastic process (\ie that is independent of the choice of prescription). Although this is not the only decomposition having this property (for instance we could have erased the first term on the right-hand side of eq.~\eqref{spuriousDrift} and transferred it into the raw drift), it is the one that is most physically sound, at least in orthonormal coordinate frames, as discussed in~\cite{cates2022stochastic}. In this description, the spurious drift~\eqref{spuriousDrift} should thus be thought of as an inherent partner to the noise term $\bb_\alpha\eta^\alpha_t$ rather than part of any physical drift.

This separation between raw and spurious drifts allows to understand dynamics of the form~\eqref{initialSDE} in a more coherent way. Nevertheless, the raw drift $\bA$ (just like the total $\varepsilon$-drift $\bA_{(\varepsilon)}$, for any $\varepsilon\neq1/2$) suffers an annoying issue: the way it transforms under a change of coordinates -- which is imposed by the chain rule(s) of stochastic calculus -- does not follow that of a vector field~\cite{graham1977covariant}. This can be seen by considering the Stratonovitch $\varepsilon=1/2$ version of dynamics~\eqref{initialSDE}: according to the Stratonovitch chain rule, $\dot{\bx}_t$ transforms as a vector field, $\dot{x}^i_t\to \dot{\tilde{x}}^i_t= (\partial \tilde{x}^i/\partial x^j) \dot{x}^j_t$. Consequently, both the noise term $b^i_\alpha(\bx_t)\eta^\alpha_t\to \tilde{b}^i_\alpha(\bx_t)\eta^\alpha_t = (\partial \tilde{x}^i/\partial x^j) b^j_\alpha(\bx_t)\eta^\alpha_t$ and the deterministic $\varepsilon=1/2$-total drift $A^i_{(1/2)}\to \tilde{A}^i_{(1/2)}=(\partial \tilde{x}^i/\partial x^j) A^j_{(1/2)}$ also transform as vector fields, respectively. In contrast, the spurious drift $\bs_{(1/2)}$ does not transform as a vector field: 
\begin{equation}
s^i_{(1/2)}=b^i_\alpha\partial_j  b^j_\alpha  \to \tilde{s}_{(\varepsilon)}^i= \tilde{b}^i_\alpha\partial_j  \tilde{b}^j_\alpha \neq (\partial \tilde{x}^i/\partial x^j)s^j_{(1/2)} \ .
\end{equation}
In turn the raw drift $\bA$ is not a vector field either.
In this article, we argue that a good way to address this issue is to further split the raw drift $\bA$ as follows:
\begin{equation}
\bA = \ba_{\blambda} + \bh_{\blambda} \ ,
\label{lambdaDriftANDpriorSplit}
\end{equation}
where $\blambda$ is a reference measure over $\mbR^d$ that we assume to be \textit{smooth and positive}, a property that we here define as follows: $\blambda$ has a Radon-Nikodym derivative (\ie a density) with respect to the Lebesgue measure $\blambda_0$ over $\mbR^d$ and this density, denoted by $\rmd\blambda/\rmd\blambda_0$, is smooth and positive.
We respectively call $\blambda$ the \textit{gauge}, $\ba_{\blambda}$ the \textit{$\blambda$-drift} (or simply the \textit{drift}, when there is no ambiguity on the gauge), and $\bh_{\blambda}$ the \textit{$\blambda$-gauge drift}, names that are justified by several properties discussed after equation~\eqref{FPeq02} below. In any coordinate system, $\bh_{\blambda}$ reads 
\begin{equation}
h^i_{\blambda} = b^i_\alpha b^j_\alpha \partial_j \ln\lambda  \ ,
\label{priorDrift}
\end{equation}
where $\lambda(\bx)>0$ is \textit{the volume element} of the gauge measure $\blambda$ in the considered coordinate system, \ie $\blambda = \lambda(\bx) \rmd \bx \equiv \lambda (\bx) \rmd x^1\dots \rmd x^d$.
The resulting $\blambda$-drift $\ba_{\blambda}$ is now a proper contravariant vector field over $\mbR^d$. Indeed, for any gauge $\blambda$, 
\begin{equation}
s^i_{(1/2)}+h^i_{\blambda} = b^i_\alpha \partial_j b^j_\alpha +  b^i_\alpha b^j_\alpha \partial_j \ln\lambda = b^i_\alpha \lambda^{-1} \partial_j \lambda b^j_\alpha  \ ,
\end{equation}
\ie $\bs_{(1/2)} + \bh_{\blambda} = \Div_{\blambda} (\bb_\alpha) \bb_\alpha$, where the divergence operator with respect to the measure $\blambda$ reads $\Div_{\blambda} (\bb_\alpha) = \lambda^{-1}\partial_j (\lambda b^j_\alpha)$. Hence $\bs_{(1/2)} + \bh_{\blambda}$ is a proper vector field, and, in turn, so is $\ba_{\blambda}$ (for any $\varepsilon\in[0,1]$, since $\ba_{\blambda}$ does not depend on $\varepsilon$).
Note that $\bh_{\blambda}$ -- and hence $\ba_{\blambda}$ -- remain unchanged upon multiplying $\lambda$ by any non-zero real number (the same for all $\bx$).

Let us denote by $\bP_t$ the probability measure of the solution to dynamics~\eqref{initialSDE} at time $t$, starting from a given initial distribution $\bP_0$, and $p_t\equiv\frac{\rmd \bP_t}{\rmd \blambda}$ the corresponding probability density\footnote{Note that some authors, like~\cite{graham1977covariant}, define the probability density of $\bP_t$ in a different way: in any coordinate system $(x^1,\dots x^d)$, it is given by what multiplies $\rmd x^1\dots\rmd x^d$ in the local expression of $\bP_t$. With the notation adopted in our article, this notion of density reads $\lambda(\bx)\frac{\rmd \bP_t}{\rmd \blambda}(\bx)$, and is hence not invariant under a coordinate change, \ie it is not a proper scalar-valued function. Consequently, with such convention, the Fokker-Planck equation takes a slightly different form (see~\cite{graham1977covariant}), but the description remains equivalent, and their coordinate expression are identical in orthonormal coordinate frames.} with respect to the gauge $\blambda$ ($\rmd/\rmd\blambda$ denoting the Radon-Nikodym derivative). This probability density satisfies the Fokker-Planck equation:
\begin{equation}
\partial_t p_t = -\Div_{\blambda} \left[ \ba_{\blambda} p_t - \bD \cdot \rmd p_t \right] \ ,
\label{FPeq01}
\end{equation}
where $\Div_{\blambda}$ is the divergence operator\footnote{We recall that if $\blambda$ and $\bu$ are respectively a measure and a vector field over a smooth manifold, then $\Div_{\blambda}(\bu)$ is the rate of change of $\blambda$ along the flow lines of $\bu$.} with respect to $\blambda$, whose expression we already gave, $\bD$ is the diffusion tensor: $D^{ij}\equiv b^i_\alpha b^j_\alpha$, $\rmd p_t$ the differential of $p_t$, and $\cdot$ the full tensor contraction\footnote{By ``full tensor contraction'' we mean that we contract as many adjacent indices as possible. For instance if $\bT$ is a contravariant tensor of order $p$ and $\bS$ is a covariant tensor of order $q\leq p$, then $\bS\cdot \bT$ is a contravairant tensor of order $p-q$ with coordinates $[S\cdot T]^{i_{q+1}\dots i_p}=S_{i_1\dots i_q} T^{i_1\dots i_p}$, and \textit{vice versa} if $q\geq p$.}. The coordinate expression of eq.~\eqref{FPeq01} is thus
\begin{equation}
\partial_t p_t = -\lambda^{-1} \partial_i \lambda \left[ a_{\blambda}^i p_t - D^{ij}\partial_j p_t \right] \ .
\label{FPeq02}
\end{equation}
Throughout the paper we will denote by $\mcW_{\blambda}$ the so-called Fokker-Planck operator (in the gauge $\blambda$), whose application to $p_t$ is given by the right-hand side of eq.~\eqref{FPeq01}.
For a vanishing $\blambda$-drift $\ba_{\blambda}$, the space of positive measures that are invariant under dynamics~\eqref{initialSDE} is given by\footnote{Indeed, when $\ba_{\blambda}=0$, a uniform $\blambda$-density $p=constant$ is a stationary solution of eq.~\eqref{FPeq01}.} $c\blambda$, with $c$ any positive real number. In other words, as soon as $\ba_{\blambda}=0$, if the initial condition of dynamics~\eqref{initialSDE} is drawn from any measure $c\blambda$, where $c>0$, then the instantaneous distribution of the state $\bx_t$ is time invariant. In particular, when $\blambda$ has a finite integral over $\mbR^d$, the stationary probability measure is given by
 $\bP_{\rm ss} =\blambda/\int_{\mbR^d}\blambda$.

With given vector fields $(\bb_1,\dots,\bb_n)$, the process is entirely determined either by choosing $\bA$, or by choosing the pair $(\blambda,\ba_{\blambda})$. Hence, with a fixed raw drift $\bA$, changing $\blambda$ changes $\ba_{\blambda}$, and in this case, $\blambda$ is really a gauge choice: any formula expressing a physical observable, like \eg the entropy production, as a function of $\ba_{\blambda}$ will actually be gauge invariant (we discuss this in detail in section~\ref{subsec:gaugeTheory}). On the other hand, with a given vector field $\ba$, changing the measure $\blambda$ while always taking $\ba_{\blambda}\equiv\ba$ changes $\bA$, and hence the stochastic process itself. This means that $\blambda$ is no longer a mere gauge choice but it plays the role of a \textit{prior}: it is a part of the model specification that has to be done \textit{a priori} and that will impact the physics of the resulting model by fixing the set $\{c\blambda,c>0\}$ of measures that would be invariant under the chosen noise in the absence of any drift $\ba_{\blambda}=\ba=0$. 
Equivalently, in the absence of any physical drift, the prior is the measure with respect to which dynamics~\eqref{initialSDE} minimizes the relative entropy\footnote{Indeed, denoting $\mathrm{KL}[\bP|\bQ] \equiv \int \ln \frac{d\bP}{\rmd\bQ}\rmd \bP$ the relative entropy, a.k.a Kullback-Leibler divergence, the functional $F_{\blambda}[p_t]\equiv \mathrm{KL}[\bP_t|\blambda]= \int p_t\ln p_t \rmd \blambda$ is easily shown to satisfy $\rmd F /\rmd t = -\int (\partial_i p_t) D^{ij} (\partial_j p_t) p_t\rmd \blambda\leq 0$, and is hence minimized by dynamics~\eqref{initialSDE} with $\ba_{\blambda}=0$.} of $\bP_t$. 
Note that, when the reference measure plays the role of a prior, changing $\blambda\to\bmu$ changes the stationary measure by a factor of $\rmd \bmu/\rmd \blambda$. Further, if a given density $p_0$ is chosen as the initial condition whatever the prior, then changing the prior $\blambda\to\bmu$ changes the initial distribution $\bP_0=p_0\blambda\to \bP_0=p_0\bmu$ and the whole solution of the dynamics changes as $\bP_t=p_t\blambda \to \bP_t=p_t\bmu$, with the same $p_t$, \ie the whole solution $\bP_t$ of the dynamics is only scaled by a spatially-varying factor $\frac{\rmd\bmu}{\rmd\blambda}(\bx)$.

The split~\eqref{lambdaDriftANDpriorSplit} thus allows the stochastic-integral-prescription-free and  coordinate-free modeling of a given noisy physical system to be more transparent:
If one wants to describe a random process whose noise is driven by $n$ vector fields $\bb_\alpha$ and which is drifted by a physical vector field $\ba$, one also needs to choose a prior measure $\blambda$, the latter fixing the set of measures which would be invariant under the dynamics if the drift $\ba$ were to vanish. Once these geometrical objects are fixed, one can then make arbitrary gauge changes $\blambda\to\bmu$ to express the dynamics with respect to a different reference measure, without modifying the physical model.
The advantage of this modeling procedure lies both in the geometrical nature of $\blambda$ and $\ba_{\blambda}$ and on their clear physical meaning. 
Indeed, as the raw drift $\bA$ is not tensorial, it cannot be thought of as a vector field of forces for instance, in contrast with $\ba_{\blambda}$. Furthermore, even though the total Stratonovitch drift $\bA_{(1/2)}$ is tensorial, interpreting it as a physical force field  is not satisfying in general, as discussed in~\cite{cates2022stochastic}.


Note that, if one models a random process on $\mbR^d$ by writing down a SDE in a global coordinate system (\eg Euclidean or spherical) and thinks about the raw drift $\bA$ as a ``force field'' without taking care of the gauge measure, then this amounts to implicitly choosing the Lebesgue measure of the considered coordinate system as the reference measure. Besides, it is worth mentioning that some authors~\cite{de2023path,de2022discretized}, to describe things in a coordinate-free way, following~\cite{graham1977covariant}, implicitly use as the gauge the volume measure associated to the tensor $\bD^{-1}$, viewed as a Riemannian metric.

Throughout this article we will assume that $\bA$ is fixed, so that whatever the reference measure $\blambda$ in eq.~\eqref{EDS}, we always study the same given stochastic process. For this reason, we will keep calling $\blambda$ the gauge rather that the prior. Nevertheless, up until section~\ref{subsec:gaugeTheory}, the reference measure $\blambda$ will be fixed, so that all the results would be the same if $\blambda$ played the role of a prior.

To lighten notations, we will simply write $\ba$ instead of $\ba_{\blambda}$ when there is no ambiguity. Dynamics~\eqref{initialSDE} can thus be rewritten as
\begin{equation}
\dot{\bx}_t = \ba + \bh_{\blambda} + \bs_{(\varepsilon)} + \bb_\alpha\eta^\alpha_t
\label{EDS}
\end{equation}
where the noise is defined as in eq.~\eqref{initialSDE}, $\ba$ is a given vector field over $\mbR^d$, $\bs_{(\varepsilon)}$ is given by eq.~\eqref{spuriousDrift}, and $\bh_{\blambda}$ is as in eq.~\eqref{priorDrift} for a fixed gauge $\blambda$.
Except in section~\ref{sec:generalization} where we generalize our results to arbitrary smooth manifolds (and in section~\ref{subsec:ChiralABP2d} where we give an example on a manifold), dynamics~\eqref{EDS} takes place on $\mbR^d$. 
In section~\ref{subsec:gaugeTheory}, we shall further discuss the choice of reference measure through the lens of gauge theory.
Otherwise we will choose the Lebesgue measure, denoted by $\blambda_0$, as the gauge: $\blambda=\blambda_0$. Note that in this case, the coordinates of $\bh_{\blambda_0}$ vanish in any affine coordinate system. 
We further assume that the pair $(\ba,\bD)$ is such that dynamics~\eqref{EDS} has a unique stationary probability measure $\bP_{\rm ss}$ and that this measure is smooth and positive (see definition below eq.~\eqref{lambdaDriftANDpriorSplit}).
 The set of all pairs $(\ba,\bD)$ such that this holds is denoted by $\mkX$, \ie
\begin{eqnarray}
\mkX &\equiv & \{ (\ba,\bD) \text{ such that dynamics~\eqref{EDS} has a unique stationary measure} \\
& & \text{and this measure is smooth and positive} \} \ . 
\label{mkXdefinition}
\end{eqnarray}
Finally, we assume that the initial condition of eq.~\eqref{EDS} is drawn from $\bPss$, and we denote its $\blambda$-density by $\pss$. Hence, the corresponding stochastic process is stationary.

\subsection{Construction of EMT-reversal}
\label{sec:construction}

The aim of this section is to construct a proper notion of the EMT-dual process of dynamics~\eqref{EDS}, consisting in an inversion of the time variable, together with, possibly, a flip of some degrees of freedom and a direct modification of the drift $\ba$ and diffusion tensor $\bD$.
We start by recalling in section~\ref{subsec:MT-dual} the definition of what we call the MT-dual process, which, in contrast with the EMT-dual, does not include any direct modification of the pair $(\ba,\bD)$.
In section~\ref{subsec:EMT-dual}, we then turn to the construction of the full EMT-dual process -- which we will equally call the EMT-reversed or EMT-adjoint process. The operation of associating to a process its (E)MT-dual will be called \textit{(E)MT-reversal}.

\subsubsection{MT-reversal}
\label{subsec:MT-dual}

In dynamics~\eqref{EDS}, the vector $\bx_t\in\mbR^d$ contains all the degrees of freedom of the system at time $t$.
It is thus not necessarily a position variable only. It can possibly contains variables like the momentum or self-propulsion force, of one or several particles. Hence the various component of $\bx_t$ can have different parities under time reversal. 
We thus introduce a linear map $\mkm:\mbR^d\to\mbR^d$ -- the notation $\mkm$ standing for ``mirror'' -- that represents the parity of all the degrees of freedom under time-reversal.
As such, it must be involutive, \ie such that $\mkm\circ\mkm = id_{\mbR^d}$, where $\circ$ is the composition of functions and $id_{\mbR^d}$ is the identity map of $\mbR^d$.
For instance, if $\bx=(\br,\bp)$ is a vector containing the position and momentum vectors, then the natural involution to consider is $\mkm:(\br,\bp)\to (\br,-\bp)$.

The linear involution $\mkm$ induces two other, important, linear involutions: 
the first one, the so-called pullback, associates to any function $f:\mbR^d\to\mbR$ another function defined by:
\begin{equation}
(\mkm^\ast f)(\bx) \equiv f(\mkm(\bx)) \ ,
\end{equation}
while the second one, referred to as the pushforward, associates to any contravariant tensor $\bT$ a new tensor of the same type, denoted by $\mkm_*\bT$, whose coordinates read:
\begin{equation}
[\mkm_*T]^{j_1,\dots,j_k} (\bx) \equiv T^{i_1,\dots,i_k}(\mkm(\bx))[\partial_{i_1}\mkm^{j_1}(\mkm(\bx))]\dots [\partial_{i_k}\mkm^{j_k}(\mkm(\bx))] \ .
\label{pushforwardDef}
\end{equation}
These maps are respectively the right ones to take the mirror image of functions and tensor fields. For instance, the vector $[\mkm_*v]^i(\mkm(\bx))=\partial_j\mkm^i(\bx)v^j(\bx)$ is a tangent vector at point $\mkm(\bx)$ obtained by taking the mirror image of the tangent vector $\bv(\bx)$ at $\bx$ and gluing it to the mirrored base point $\mkm(\bx)$ (see figure~\ref{fig:ExampleMTreversible} in section~\ref{subsec:typicalMTreversibleProcess} below for an example).

Starting from the stationary dynamics~\eqref{EDS} for a given pair $(\ba,\bD)$, whose path probability is denoted by $\mcP_{(\ba,\bD)}$, its $\mkm\mkt$-dual is defined as the stochastic process whose steady-state path probability, denoted by $\mkT(\mcP_{(\ba,\bD)})$, reads
\begin{equation}
\mkT(\mcP_{(\ba,\bD)}) [(\bx_t)_{t\in\mbT}] =\mcP_{(\ba,\bD)}[(\mkm(\bx_{\mkt(t)}))_{t\in\mbT}] \ ,
\end{equation}
for all trajectories $(\bx_t)_{t\in\mbT}$, where we introduced the time-inversion map $\mkt:t\in\mbT\mapsto \mcT-t\in \mbT$.
 It can be shown (see appendix~\ref{app:MTreversal} or~\cite{risken1996fokker,van1992stochastic,gardiner1985handbook}) that the resulting MT-reversed dynamics is given by eq.~\eqref{EDS} where the pair drift-diffusion tensor $(\ba,\bD)$ is replaced by $\mkT(\ba,\bD)=(\mkT\ba,\mkT\bD)$, with 
\begin{eqnarray}
\mkT\bD &=& \mkm_\ast\bD  \ , \label{TrevD}\\
\mkT\ba &=& 2 (\mkm_\ast\bD) \cdot \rmd \ln (\mkm^\ast \pss) - \mkm_\ast\ba = \mkm_*\left[ 2\bD\cdot \rmd\ln \pss - \ba\right] \label{Treva} \ ,
\end{eqnarray}
the last equality stemming from the chain rule and the linearity of $\mkm_*$. 
Note that, by definition of $\mkT(\ba,\bD)$, the path probability of the $\mkm\mkt$-dual process, $\mkT(\mcP_{(\ba,\bD)})$, also reads $\mcP_{\mkT(\ba,\bD)}$.
Finally, let us specify that eqs.~\eqref{TrevD} \&~\eqref{Treva} are valid when the gauge is $\mkm$-invariant. This is indeed the case here since we choose the gauge to be the Lebesgue measure and $\mkm$ to be linear (see section~\ref{sec:generalization} for a more general setting).

In this section, we have (re)constructed a notion of time-reversal, namely the MT-reversal, that is appropriate both \eg for overdamped~\eqref{overdamped_Langevin} and underdamped~\eqref{underdamped_Langevin_01}-\eqref{underdamped_Langevin_02} Langevin particles in an external potential.
With the example of a Langevin particle in a the magnetic field~\eqref{Langevin_Magnetic_field_01}-\eqref{Langevin_Magnetic_field_02}, we nonetheless saw that it can sometimes be necessary to be able to describe a more general form of stochastic time-reversal. 

To account for the latter example, together with the similar one of a particle in a centrifuge, a generalized notion of time-reversal was introduced~\cite{onsager1931reciprocal2,de1962non,risken1996fokker,van1992stochastic,gardiner1985handbook} in which the drift depends on a finite set of parameters, each of which being either even or odd under time-reversal. We will refer to this notion as \textit{parameterized EMT-reversal}.
In this article, we prioritize a more general notion of EMT-reversal, defined in the next section,  which subsumes that of parameterized EMT-reversal but without referring to any parameter\footnote{In particular, we argue in section~\ref{subsec:EMT-dual} that, to account for certain time-reversal symmetries through the notion of parameterized EMT-reversal, some \textit{ad hoc} parameters need to be introduced, a superfluous circumvention our notion of EMT-reversal does not require.}.

%

\subsubsection{EMT-reversal}
\label{subsec:EMT-dual}

In this article, we consider this extended notion of time-reversal to amount to directly transforming the drift, and possibly the diffusion tensor of dynamics~\eqref{EDS}, in addition to the reversal of time and potentially that of some degrees of freedom.
To formalize this mathematically, we assume we have a new involution, $\mke=(\mke_\ba,\mke_\bD)$, directly defined on a subset $\mcY\subseteq\mkX$ of pairs (drift, diffusion tensor). 
This map also needs to be well defined on $\mke(\mcY)$ since we need to be able to re-apply $\mke$. This is the case \eg for the Langevin particle in a magnetic field if we take $\mcY$ to be the set $(\ba,\bD)$ appearing in eqs.~\eqref{Langevin_Magnetic_field_01} \&~\eqref{Langevin_Magnetic_field_02}, with, for instance, all parameters fixed but $\bB\in\mbR^3$, since in this case $\mke$ is well defined on $\mcY=\mke(\mcY)$. 
In what follows, to lighten the notations, we will omit the subscripts of the component maps $\mke_\ba$ and $\mke_\bD$ when there is no ambiguity. 

Denoting by $\overline{\mcP_{(\ba,\bD)}}$ the path probability of the EMT-reversed process we aim to construct, a natural approach could be to define it as
\begin{equation}
\overline{\mcP_{(\ba,\bD)}}[(\bx_t)_{t\in\mbT}]\equiv \mcP_{\mke(\ba,\bD)}[\mkm(\bx_{\mcT-t})_{t\in\mbT}] \ .
\end{equation} 
The issue is that this notation hides an important ambiguity in the construction. Indeed, using the notation of the previous section, the above definition reads 
\begin{equation}
\overline{\mcP_{(\ba,\bD)}} \equiv \mcP_{\mkT\mke(\ba,\bD)} \ ,
\label{def:EMT01}
\end{equation}
whereas we could have also set 
\begin{equation}
\overline{\mcP_{(\ba,\bD)}} \equiv \mcP_{\mke\mkT(\ba,\bD)} \ ,
\label{def:EMT02}
\end{equation}
which is \textit{a priori} different.

Nevertheless, a crucial constraint is that our EMT-reversal has to be involutive, since reversing time twice must amount to do nothing. Whether we choose~\eqref{def:EMT01} or~\eqref{def:EMT02}, this constraint reads $\mkT\mke\mkT\mke=id$.
Since both $\mkT$ and $\mke$ are involutive, composing this last equality from the left by $\mke\mkT$ gives
\begin{equation}
\mkT\mke=\mke\mkT  \ .
\label{EMT_commutativityConstraint}
\end{equation}
Thus, to be well defined as a form of time-reversal, our maps $\mke$ and $\mkm$ have to be such that $\mke$ and $\mkT$ commute. If this is the case, then the two definitions~\eqref{def:EMT01} and~\eqref{def:EMT02} coincide. 
Consequently, the $\mke\mkm\mkt$-reversed process is unambiguously defined and is solution of eq.~\eqref{EDS} where the pair $(\ba,\bD)$ is replaced by $(\reversed{\ba}, \reversed{\bD})$, the latter being defined through 
\begin{equation}
\mcP_{(\reversed{\ba},\reversed{\bD})}=\overline{\mcP_{(\ba,\bD)}} \ .
\label{equivarianceBar}
\end{equation}
It follows from eqs.~\eqref{def:EMT02} \&~\eqref{equivarianceBar} that the pair (drift, diffusion tensor) of the $\mke\mkm\mkt$-dual satisfies $(\reversed{\ba},\reversed{\bD})=\mke\mkT(\ba,\bD)$, \ie
\begin{eqnarray}
\reversed{\bD} &=& \mke(\mkm_\ast\bD)  \ , \label{eq:EMTdualD}\\
\reversed{\ba} &=&  \mke(\mkm_*\left[ 2\bD\cdot \rmd\ln \pss - \ba\right]) \label{eq:EMTduala}\ ,
\end{eqnarray}
or equivalently, from eqs.~\eqref{def:EMT01} \&~\eqref{equivarianceBar}, $(\reversed{\ba},\reversed{\bD})=\mkT\mke(\ba,\bD)$, \ie
\begin{eqnarray}
\reversed{\bD} &=& \mkm_\ast \mke(\bD)  \ , \label{eq:EMTdualDbis} \\
\reversed{\ba} &=& \mkm_*\left[ 2\mke(\bD)\cdot\rmd\ln \pss^\mke - \mke(\ba)\right]  \label{eq:EMTdualabis} \ ,
\end{eqnarray}
where the equivalence is a consequence of the commutativity~\eqref{EMT_commutativityConstraint}, and $\pss^\mke$ stands for the stationary probability density of dynamics~\eqref{EDS} with drift $\mke(\ba)$ and diffusion tensor $\mke(\bD)$.

Interestingly, this general construction implies that, for any stochastic process that is stationary solution of eq.~\eqref{EDS} for an arbitrary pair $(\ba,\bD)\in\mkX$, and for any linear involution $\mkm$ over $\mbR^d$ (hence, notably for  $\mkm=id_{\mbR^d}$), we can find a direct modification $\mke$ of the drift $\ba$ and diffusion tensor $\bD$ such that the process is $\mke\mkm\mkt$-symmetric, \ie such that
\begin{equation}
\overline{\mcP_{(\ba,\bD)}}[(\bx_t)_{t\in\mbT}]\equiv \mcP_{\mke(\ba,\bD)}[\mkm(\bx_{\mcT-t})_{t\in\mbT}] =\mcP_{(\ba,\bD)}[(\bx_t)_{t\in\mbT}] \ .
\label{EMTreversibiliyCond0}
\end{equation}
Previously we used the same notation, $\mkT$, to denote different maps, including one acting on the space of path probabilities and another on the space $\mkX$. Upon denoting the latter by $\mkT_\mkX$, we see that, indeed, taking $\mke=\mkT_\mkX$ satisfies the EMT-reversibility condition~\eqref{EMTreversibiliyCond0}, since $\mcP_{\mkT_\mkX(\ba,\bD)}[\mkm(\bx_{\mcT-t})_{t\in\mbT}]=\mcP_{\mkT_\mkX\mkT_\mkX(\ba,\bD)}[(\bx_t)_{t\in\mbT}]=\mcP_{(\ba,\bD)}[(\bx_t)_{t\in\mbT}]$ by definition of the involution $\mkT_\mkX$. 
Hence, \textit{any dynamics of the form~\eqref{EDS} admits at least one EMT-symmetry}\footnote{In fact, whenever $d>1$, it has an infinite number of EMT-symmetries, since one can take any arbitrary $\mkm$ in this construction.}. 
In words, these ``universal'' EMT-symmetries consist in the following construction: applying a $\mkm\mkt$-reversal in the path probability $\mcP_{(\ba,\bD)}$ amounts to transform the pair $(\ba,\bD)\mapsto \mkT_\mkX(\ba,\bD)$; the action of $\mke=\mkT_\mkX$ consists in re-inverting the pair: $\mkT_\mkX(\ba,\bD)\mapsto (\ba,\bD) $. This ensures that the probability of the initial process to travel a trajectory $(\bx_t)_{t\in\mbT}$, $\mcP_{(\ba,\bD)}[(\bx_t)_{t\in\mbT}]$, is equal to that of the process with the reversed pair $\mke(\ba,\bD)$ to generate the $\mkm\mkt$-reversed trajectories $\mkm(\bx_{\mcT-t})_{t\in\mbT}$, $\mcP_{\mke(\ba,\bD)}[\mkm(\bx_{\mcT-t})_{t\in\mbT}]$.

For a given $\mkm$, introducing the symmetric and skew-symmetric parts of $(\ba,\bD)$ under the map $\mkT_\mkX$,
\begin{equation}
(\ba_{\pm},\bD_{\pm})\equiv \frac{1}{2}\left[(\ba,\bD)\pm \mkT_{\mkX}(\ba,\bD)\right] \ ,
\end{equation}
allows reformulating the map $\mke\equiv\mkT_\mkX$ of the corresponding universal $\mke\mkm\mkt$-symmetry as
\begin{equation}
\mke:(\ba_+ +\ba_-, \bD_+ +\bD_-) \mapsto (\ba_+ -\ba_-, \bD_+ -\bD_-) \ .
\label{formulationUniversalEMT}
\end{equation}
Interestingly, such a universal EMT-symmetry were already spotted in~\cite{graham1980solution} in the case where $\mkm=id_{\mbR^d}$, \ie for even degrees of freedom. In order to frame this symmetry in the context of parameterized EMT-reversal (that we described at the end of section~\ref{subsec:MT-dual}), the author of~\cite{graham1980solution} introduced a real parameter $\alpha$ and set $\ba(\alpha)\equiv \ba_+ + \alpha\ba_-$. The initial dynamics corresponds to $\alpha=1$, while the T-reversed has a drift $\ba(\alpha=-1)$. Upon declaring $\alpha$ odd under time-reversal, one then obtains a \textit{parameterized} EMT-symmetry of dynamics~\eqref{EDS}. This is equivalent
 to our $\mke\mkm\mkt$-symmetry\footnote{We recall that, when $\mkm=id_{\mbR^d}$, then $\mkT\bD=\bD$ and hence $\bD_+=\bD$ and $\bD_-=0$.} with $\mke$ given by~\eqref{formulationUniversalEMT} but where, in contrast, our notion of EMT-reversal does not require the introduction of any \textit{ad hoc} parameter such as $\alpha$.


Note that in a given situation $\mcY\subset\mkX$, with given maps $\mkm$ and $\mke$, it can be quite hard to confirm explicitly that the map that sends a process to its $\mke\mkm\mkt$-dual is involutive, \ie to check that $\mke$ and $\mkT$ actually commute. Indeed, this amounts to show that the right-hand sides of eqs~\eqref{eq:EMTdualD} \&~\eqref{eq:EMTduala} coincide with that of eqs.~\eqref{eq:EMTdualDbis} \&~\eqref{eq:EMTdualabis}, respectively.
But if one finds a map $\mke$ such that the (subclass $\mcY$ of) dynamics under study turns out to be $\mke\mkm\mkt$-reversible, \ie satisfies $(\ba,\bD)=(\reversed{\ba},\reversed{\bD})$, with $(\reversed{\ba},\reversed{\bD})$ given by eqs.~\eqref{eq:EMTdualD} \&~\eqref{eq:EMTduala}, then we automatically know that the required commutativity holds since, at least on $\mcY\subset\mkX$, $\mke=\mkT$.

At first sight, the fact that an EMT-symmetry {\em always} exists for dynamics of the form~\eqref{EDS} seems to imply that there is no important consequence of finding such a symmetry for a given system.
In particular, for a given mirror map $\mkm$, finding an explicit expression for the map $\mke$ given above such that dynamics~\eqref{EDS} is $\mke\mkm\mkt$-reversible, $\mke=\mkT_\mkX$, generically requires knowledge of $\pss$. However, if one can spot a map $\mke$ that does not require the knowledge of $\pss$ such that dynamics~\eqref{EDS} is $\mke\mkm\mkt$-reversible, \eg by flipping a part of the drift as in the case of Langevin particle in a magnetic field, \textit{then one can use this to obtain the explicit expression of the stationary density} $\pss$.
%
%
%
Indeed, in this case, upon applying $[2\bD]^{-1}\mkm_*\mke$ to both sides of eq.~\eqref{eq:EMTduala}, and using the reversibility condition\footnote{Indeed, as we will see in the next section, the $\mke\mkm\mkt$-reversibility of dynamics~\eqref{EDS} is equivalent to $(\reversed{\ba},\reversed{\bD})=(\ba,\bD)$.} $\reversed{\ba}=\ba$, one gets $\rmd\ln\pss = \bD^{-1}(\mkm_*\mke(\ba)+\ba)/2$, which implies
\begin{equation}
\pss(\bx) = \frac{1}{Z}\exp\left\{\int^\bx_{\bx_0} \bD^{-1}\left[\frac{\mkm_*\mke(\ba)+\ba}{2}\right]\cdot \rmd \ell \right\} \ ,
\label{pssWhenEMTsym}
\end{equation}
where $\int_{\bx_0}^\bx\rmd \ell$ stands for integrating along any line joining an arbitrary chosen point $\bx_0\in\mbR^d$ to $\bx$, and $Z$ is a normalization constant. 
That is why, in the next section, despite the fact that the presented new results might mostly be important in practice for the study of MT-reversibility, we include in their description a possibly non-trivial $\mke$, \ie our result are stated for the more general notion of EMT-reversal.

Finally, note that in eq.~\eqref{pssWhenEMTsym}, we should have written the tensor contraction symbol `$\cdot$' between $\bD^{-1}$ and $\frac{\mkm_*\mke(\ba)+\ba}{2}$. Similarly, we will thereafter sometimes omit this symbol to lighten notations, especially when the contraction can be thought of as a matrix multiplication between a column vector and a square matrix, as in eq.~\eqref{pssWhenEMTsym}.
Besides, to get the expression~\eqref{pssWhenEMTsym} of $\pss$, we implicitly assumed $\bD$ to be invertible. But, even if it is not the case, the aforementioned situation can still grants an explicit expression of $\pss$ (see section~\ref{subsec:nonInvertibleD} for details).

\newpage
\section{$\mke\mkm\mkt$-reversibility or lack thereof on $\mbR^d$}
\label{sec:EMT-reversibility}

In this section, for a fixed choice of the maps $\mkm$ and $\mke$, we give a full procedure to check whether dynamics~\eqref{EDS} is $\mke\mkm\mkt$-symmetric and, if not, to what extent this symmetry is broken. 
The need to assess the $\mke\mkm\mkt$-(ir)reversibility of dynamics~\eqref{EDS} may arise in two distinct contexts. 

In the fist situation, one is interested in the effect of the ``real'' time-reversal on the system. This requires the model to be able to faithfully represent the real time-reversal of the system, which is not always possible as discussed in section~\ref{sec:intro} above for the case of a run-and-tumble model.
In this first situation, the maps $\mkm$ and $\mke$ corresponding to this real time-reversal are both imposed by the physics of the problem.

In the second situation, one is simply looking for a potential time-reversal-like symmetry (in the form of an EMT-symmetry) of the dynamics. A motivation for this could be simply the search for statistical symmetries of the dynamics, or for its stationary probability measure and current. In this case, one can adopt a trial and error strategy by guessing maps $\mkm$ and $\mke$ and testing whether dynamics~\eqref{EDS} is $\mke\mkm\mkt$-reversible or not. Success can yield substantial insights, including an explicit construction of the stationary measure itself via eq.~\eqref{pssWhenEMTsym} above.

\subsection{Reversibility conditions involving an unknown function}

\label{subsec:firstEMTcond}
For given maps $\mke$ and $\mkm$, a stochastic process, which is the solution of equation~\eqref{EDS} for some $(\ba,\bD)\in \mkX$, is said to be $\mke\mkm\mkt$-reversible or $\mke\mkm\mkt$-symmetric iff it is invariant under the $\mke\mkm\mkt$-reversal that we built in section~\ref{sec:construction}. This can be formulated in the space $\mkX$ as
\begin{equation}
(\ba,\bD) = (\reversed{\ba},\reversed{\bD}) \  ,
\label{emktRev-cond-01}
\end{equation}
where $\reversed{\ba}$ and $\reversed{\bD}$ are respectively given by eqs.~\eqref{eq:EMTduala} and~\eqref{eq:EMTdualD}.
Since $\mke\mkT$ is an involution, one can define the symmetric and skew-symmetric parts, \textit{under $\mke\mkm\mkt$-reversal}, of respectively the diffusion tensor:
\begin{equation}
{}^S\bD \equiv  \frac{1}{2}\left(\bD + \reversed{\bD}\right) \ \ \text{and} \  \ {}^A\bD \equiv  \frac{1}{2}\left(\bD - \reversed{\bD}\right) \ ,
\end{equation}
and the drift:
\begin{equation}
{}^S\ba \equiv  \frac{1}{2}\left(\ba + \reversed{\ba}\right) \ \ \text{and} \  \ {}^A\ba \equiv  \frac{1}{2}\left(\ba - \reversed{\ba}\right) \ .
\end{equation}
Naturally, the $\mke\mkm\mkt$-reversibility condition~\eqref{emktRev-cond-01} can be rephrased as $({}^A\ba,{}^A\bD)=(0,0)$ or $({}^S\ba,{}^S\bD)=(\ba,\bD)$. Similar symmetric and skew-symmetric object \textit{under time-reversal alone} have already been used to characterize the irreversibility of stochastic processes, as in~\cite{bertini2015macroscopic}.
In this article, we will rather use the symmetric and skew-symmetric parts, \textit{with respect to the map} $\mkm_*\mke$\footnote{Note that this is a slight abuse of terminology. Indeed, while the condition for $\mke\mkT$ to be involutive requires $\mkm_*\mke_\bD=\mke_\bD\mkm_*$, which implies that $\mkm_*\mke_\bD$ is an involution itself, it is not the case for $\mkm_*\mke_\ba$ in general. Hence, $\ba^S$ (respectively $\ba^A$) are not necessarily symmetric (respectively skew-symmetric) with respect to $\mkm_*\mke$. But the decomposition $\ba=\ba^S+\ba^A$ is still valid, and the invariance of $\ba$ with respect to $\mkm_*\mke$ still reads $\ba^S=\ba$ or equivalently $\ba^A=0$. When $\mke=id$, the symmetric and skew-symmetric part of $\ba$ (respectively of $\bD$) under the mirror operation $\mkm_*$ are sometimes respectively called the irreversible and reversible drift (respectively diffusion tensor) in the literature~\cite{risken1996fokker,graham1971generalized,dal2021fluctuation}.}, of $\bD$:
\begin{equation}
\bD^S\equiv \frac{1}{2}\left(\bD+\mkm_*\mke\bD\right) \ \ \text{and} \ \ \bD^A\equiv \frac{1}{2}\left(\bD-\mkm_*\mke\bD\right) \ ,
\end{equation}
and of $\ba$:
\begin{equation}
\ba^S\equiv \frac{1}{2}\left(\ba+\mkm_*\mke\ba\right) \ \ \text{and} \ \ \ba^A\equiv \frac{1}{2}\left(\ba-\mkm_*\mke\ba\right) \ ,
\end{equation}
which are always explicitly accessible. 
The reversibility condition~\eqref{emktRev-cond-01}, together with eqs.~\eqref{eq:EMTdualD} \&~\eqref{eq:EMTduala}, explicitly involves the stationary probability $\pss$, which is generically inaccessible. 
Hence in the general case it is not possible to explicitly check whether reversibility holds for any given choice of $\mke$ and $\mkm$.
In section~\ref{subsec:whenDisInvertible} (and in section~\ref{subsec:nonInvertibleD} in the case where $\bD$ is not invertible), we shall give reversibility conditions that only involve $\ba^{A,S}$ and $\bD^{A,S}$ and that can thus always be checked explicitly.

We start by noting that eqs.~\eqref{eq:EMTdualD} \&~\eqref{eq:EMTduala} can be used to reformulate the reversibility condition~\eqref{emktRev-cond-01} as follows: dynamics~\eqref{EDS}, with $(\ba,\bD)\in\mkX$, is $\mke\mkm\mkt$-reversible iff
\begin{eqnarray}
\bD^A &=& 0 \ , \label{TRS-D-cond-01} \\ 
\ba^S &=& \bD\cdot \rmd \ln \pss \ , \label{TRS-a-cond-01} 
\end{eqnarray}
\textit{where $\pss$ is the stationary density of dynamics~\eqref{EDS}} (which exists and is unique since we assumed $(\ba,\bD)\in\mkX$). To obtain eq.~\eqref{TRS-a-cond-01}, we have applied $\mkm_*\mke$ to both sides of the reversibility constraint $\ba=\mke\mkm_*(2\bD\cdot \rmd\log\pss-\ba)$.

Note that, by analogy between the probability and thermodynamical currents~\cite{graham1971fluctuations}, condition~\eqref{TRS-D-cond-01}, which can be reformulated as $\mkm_*\mke\bD=\bD$, can be seen as an \textit{Onsager-Casimir symmetry}.
We also observe that this reversibility condition can already be checked explicitly in a systematic way, since it does not require the knowledge of $\pss$. 

This is not the case of condition~\eqref{TRS-a-cond-01}, which explicitly involve $\pss$.
A first step towards getting reversibility conditions that can all be systematically checked consists in the following reformulation of~\eqref{TRS-D-cond-01} \&~\eqref{TRS-a-cond-01}:
dynamics~\eqref{EDS} is $\mke\mkm\mkt$-reversible iff \textit{there exists a smooth positive function $\tpss$ whose integral over $\mbR^d$ is finite and which is such that}
\begin{eqnarray}
\bD^A &=& 0 \ , \label{TRS-D-cond-02} \\ 
\ba^S &=& \bD\cdot \rmd \ln \tpss \ , \label{TRS-a-cond-02}  \\
\Div(\ba \tpss - \bD\cdot \rmd \tpss) &=& 0 \label{TRS-div-cond-02} \ ,
\end{eqnarray}
where $\Div$ is the divergence operator (with respect to our current gauge: the Lebesgue measure).  
Indeed, conditions~\eqref{TRS-D-cond-01}-\eqref{TRS-a-cond-01} and~\eqref{TRS-D-cond-02}-\eqref{TRS-div-cond-02} are equivalent. Firstly because conditions~\eqref{TRS-D-cond-01} and~\eqref{TRS-D-cond-02} are identical. Furthermore, if~\eqref{TRS-a-cond-01} is satisfied, then choosing the function $\tpss$ to be the stationary density, $\tpss \equiv \pss$, obviously ensures that condition~\eqref{TRS-a-cond-02} is also fulfilled since it is then identical to eq.~\eqref{TRS-a-cond-01}, while the additional condition~\eqref{TRS-div-cond-02}, which checked whether $\tpss$ lies in the kernel of the Fokker-Planck operator, also holds since $\pss$ is stationary.
Conversely, if condition~\eqref{TRS-div-cond-02} is satisfied, 
we first conclude that $\pss\propto\tpss$ since the finiteness of the overall integral of $\tpss$ ensures that the latter can be normalized to obtain the stationary probability density (the assumption $(\ba,\bD)\in\mkX$ ensuring the uniqueness of $\pss$). If we further assume that~\eqref{TRS-a-cond-02} holds, then so does~\eqref{TRS-a-cond-01} since it is invariant under the rescaling of $\pss$.
Finally, note that eq.~\eqref{TRS-a-cond-02} can be used to simplify~\eqref{TRS-div-cond-02} in such a way that the set of reversibility conditions~\eqref{TRS-D-cond-02}-\eqref{TRS-div-cond-02} can be rewritten as: 
there exists a smooth positive function $\tpss$ whose integral over $\mbR^d$ is finite and which is such that
\begin{eqnarray}
\bD^A &=& 0 \ , \label{TRS-D-cond-03} \\ 
\ba^S &=& \bD\cdot \rmd \ln \tpss \ , \label{TRS-a-cond-03}  \\
\Div(\ba^A \tpss ) &=& 0 \label{TRS-div-cond-03} \ .
\end{eqnarray}

The point of the slight reformulation of conditions~\eqref{TRS-D-cond-01} \&~\eqref{TRS-a-cond-01} as~\eqref{TRS-D-cond-03}-\eqref{TRS-div-cond-03} (or equivalently as~\eqref{TRS-D-cond-02}-\eqref{TRS-div-cond-02}) is that the former can be a bit misleading in practice.
To see this, let us
%
consider the simple example of a Langevin particle in $\mbR^3$ in an external fore field $\bF(\br)$:
\begin{eqnarray}
\dot{\br} &=& \bv \ , \label{UnderLangevinR}\\
\dot{\bv} &=& \bF(\br) - \gamma \bv + \sqrt{2\gamma kT}\boldeta \ , \label{UnderLangevinV}
\end{eqnarray}
with $\gamma$ the dumping coefficient, $\boldeta$ a centered, delta-correlated, Gaussian white noise, and where we set the particle's mass to unity for simplicity. Equations~\eqref{UnderLangevinR} \&~\eqref{UnderLangevinV} are written in the canonical basis of $\mbR^6$, so that, the noise being additive, and the gauge choice being the Lebesgue measure $\blambda_0=\Pi_{i=1}^3\rmd r^i\rmd v^i$, the spurious and $\blambda_0$-gauge drifts vanish.
In the case of dynamics~\eqref{UnderLangevinR}-\eqref{UnderLangevinV}, as already stated in the introduction, the natural involutions $\mke$ and $\mkm$ to consider are $\mke=id$ and $\mkm(\br,\bv)=(\br,-\bv)$. 
The drift $\ba=(\bv,\bF -\gamma \bv)^\top$ thus has $\mkm_*$-(anti)symmetric parts given by
\begin{equation}
\ba^S = \begin{pmatrix} 0 \\ -\gamma \bv \end{pmatrix} \quad \text{and} \quad \ba^A = \begin{pmatrix} \bv \\ \bF  \end{pmatrix} \ ,
\end{equation}
while the diffusion tensor 
\begin{equation}
\bD \equiv \begin{pmatrix}
 0 & 0 \\ 
 0 & 2\gamma k T \; \bI_3
\end{pmatrix}
\label{LangevinDiffusionTensor}
\end{equation}
is such that
\begin{equation}
\mkm_*\bD = \begin{pmatrix} \bI_3 & 0 \\  0 & -\bI_3 \end{pmatrix}
\begin{pmatrix} 0 & 0 \\  0 & 2\gamma k T \; \bI_3 \end{pmatrix}
\begin{pmatrix} \bI_3 & 0 \\  0 & -\bI_3 \end{pmatrix} = \bD
\label{LangevinOnsagerSym}
\end{equation}
and hence satisfies condition~\eqref{TRS-D-cond-01}. In eqs.~\eqref{LangevinDiffusionTensor} \&~\eqref{LangevinOnsagerSym}, $\bI_3$ is the $3\times 3$ identity matrix. 
The second $\mkm \mkt$-reversibility condition~\eqref{TRS-a-cond-01} then reads
\begin{equation}
-\gamma v^i =\gamma kT \frac{\partial}{\partial v^i} \ln \pss \ .
\label{treatressCondition}
\end{equation}
A careless interpretation of eq.~\eqref{treatressCondition} could lead to conclude that the stationary density is given by
\begin{equation}
\pss(\br,\bv) = \frac{1}{Z}\exp\left[-\frac{1}{kT}\left(|\bv|^2/2 +\psi(\br)\right)\right] \ ,
\label{UnderLangevinCondaS}
\end{equation}
where $Z$ is the normalization constant, $|\bv|^2\equiv v^iv^j\delta_{ij}$ the euclidean square norm of $\bv$, and $\psi(\br)$ a function of $\br$ such that $\int_{\mbR^d}\exp(-\psi(\br)/kT)\rmd\br < \infty$, and that, in turn, dynamics~\eqref{UnderLangevinR}-\eqref{UnderLangevinV} is reversible whatever the external force field $\bF(\br)$, which is obviously wrong!

The problem comes from the fact that, although one can define the probability density given by the right-hand side of eq.~\eqref{UnderLangevinCondaS}, nothing ensures that it is the stationary probability!
To know whether this holds, we need to check whether the right-hand side of eq.~\eqref{UnderLangevinCondaS}, that we now denote by $\tpss$, lies in the kernel of the Fokker-Planck operator, \ie that
\begin{equation}
\frac{\partial}{\partial r^i} (v^i\tpss) + \frac{\partial}{\partial v^i} \left(F^i\tpss - \gamma v^i \tpss - \gamma kT\frac{\partial}{\partial v^i} \tpss \right) = 0 \ .
\end{equation}
Using the definition of $\tpss$, this implies that
\begin{equation}
v^i\left[\frac{\partial}{\partial r^i} \psi(\br) +  F^i(\br)\right] = 0 \ .
\label{UnderLangevinIntermedia}
\end{equation}
In turn, since~\eqref{UnderLangevinIntermedia} must hold for all $\br$ and $\bv$, we conclude that $\tpss$ is the stationary density iff
\begin{equation}
\bF = - \nabla_\br \psi \ ,
\end{equation}
\ie iff the force field $\bF(\br)$ is conservative. 

The example of dynamics~\eqref{UnderLangevinR}-\eqref{UnderLangevinV} illustrates how the appropriate interpretation of conditions~\eqref{TRS-D-cond-01} \&~\eqref{TRS-a-cond-01} comes down to using conditions~\eqref{TRS-D-cond-02}-\eqref{TRS-div-cond-02} (or, equivalently, conditions~\eqref{TRS-D-cond-03}-\eqref{TRS-div-cond-03}) in practice.

\subsection{EMT-reversibility when $\bD$ is invertible}
\label{subsec:whenDisInvertible}

In this section, we further analyze the conditions for $\mke\mkm\mkt$-reversibility of dynamics~\eqref{EDS} in the case where $\bD(\bx)$ is invertible for all $\bx\in\mbR^d$, delaying the case where $\bD$ has a non-trivial null-space to section~\ref{subsec:nonInvertibleD}.

\subsubsection{EMT-reversibility conditions without any unknown function}
\label{subsubsec:EMTrevCondDinvert}

While eq.~\eqref{TRS-D-cond-03} is easily checked for a generic dynamics of the form~\eqref{EDS}, this is not the case for conditions~\eqref{TRS-a-cond-03} and~\eqref{TRS-div-cond-03}, since they involve an unknown function, $\tpss$, the explicit knowledge of which is equivalent to that of the stationary density, which is out of reach in general.
%
%
In the case where $\bD$ is invertible, the $\mke\mkm\mkt$-reversibililty conditions~\eqref{TRS-D-cond-03}-\eqref{TRS-div-cond-03} can be re-written as
\begin{gather}
\bD^A=0 \ , \label{TRS-D-cond-04} \\
\rmd(\bD^{-1}\ba^S) = 0  \quad \text{and} \quad \int_{\mbR^d} \exp \left[\int_{\bx_0}^\bx \bD^{-1}\ba^S \cdot \rmd \ell \right] \rmd\bx < \infty \ ,  \label{TRS-a-cond-04} \\
\Div( \ba^A) + \ba^A\cdot\bD^{-1} \ba^S = 0 \ . \label{TRS-div-cond-04}
\end{gather}
In eq.~\eqref{TRS-a-cond-04} $\bx_0$ is an arbitrary point in $\mbR^d$ and $\int_{\bx_0}^\bx \bD^{-1}\ba^S \cdot \rmd \ell$ stands for the line integral of $\bD^{-1}\ba^S$ along any arbitrary smooth path from $\bx_0$ to $\bx$.
In eq.~\eqref{TRS-a-cond-04} \&~\eqref{TRS-div-cond-04}, $\bD^{-1}$ is a covariant tensor field of order two which is the inverse of the contravariant tensor field $\bD$, in the sense that $D^{ij}[D^{-1}]_{jk}=\delta^i_j$ in any coordinate system. Hence $[D^{-1}a^S]_i=[D^{-1}]_{ij}[a^S]^j$ is a covariant vector field, \ie a differential one-form, and  $\rmd(\bD^{-1}\ba^S)$ stands for its exterior derivative, which reads
\begin{equation}
\rmd(\bD^{-1}\ba^S)= \sum_{1\leq i<j \leq d} \left( \partial_i [D^{-1}a^S]_j - \partial_j [D^{-1}a^S]_i \right) \rmd x^i \wedge \rmd x^j \ .
\label{ExtDeriv}
\end{equation}
In eq.~\eqref{ExtDeriv}, $(\rmd x^1,\dots,\rmd x^d)$ is the dual basis of an arbitrary (possibly local) basis $(\be_1,\dots,\be_d)$ of $\mbR^d$, \ie $\rmd x^i(\be_j)=\delta^i_j$, and the exterior product $\wedge$ of the basis co-vectors $\rmd x^i$ and $\rmd x^j$ is a two-form that associates to any pair $\bu,\bv$ of vectors the real number $\rmd x^i \wedge \rmd x^j(\bu,\bv)=u^iv^j-v^iu^j$. Furthermore, we recall that the exterior derivative $\rmd$ is essentially a generalization of the curl operator\footnote{Actually, as $\bD^{-1}$ is a symmetric, positive, covariant tensor of order two, we can use it as a Riemannian metric on $\mbR^d$. In dimension $d=3$, we would then have $\rmd(\bD^{-1}\ba^S)(\bu,\bv)=\bv \cdot [ (\nabla\times \ba^S)\times \bu]$, where `$\cdot$',  `$\nabla\times$', and `$\times$' respectively stand for the dot product, the curl operator, and the cross product, all associated to the metric $\bD^{-1}$.} to dimension $d>3$.
%
%
%
%

The reversibility condition~\eqref{TRS-D-cond-04} being identical to~\eqref{TRS-D-cond-03}, we now need to show that the remaining pair of conditions~\eqref{TRS-a-cond-04} \&~\eqref{TRS-div-cond-04} is equivalent to the pair~\eqref{TRS-a-cond-03} \&~\eqref{TRS-div-cond-03}.
Let us start by assuming the former and prove the latter. The fist part of condition~\eqref{TRS-a-cond-04} means that the one form $\bD^{-1}\ba^S$ is closed, and since $\mbR^d$ is simply connected, it is also exact, \ie $[D^{-1}a^S]_i= -\partial_i \psi$, where any primitive $\psi$ can be expressed as $\psi(\bx)=-\int^\bx_{\bx_0} \bD^{-1}\ba^S\cdot\rmd\ell$, up to an additive constant.
The second part of condition~\eqref{TRS-a-cond-04} implies that we can define the probability density $\tpss(\bx) \equiv e^{-\psi(\bx)}/\int_{\mbR^d} e^{-\psi}$, for which condition~\eqref{TRS-a-cond-03} holds. Condition~\eqref{TRS-div-cond-04} then ensures that the resulting $\tpss$ lies in the kernel of the Fokker-Planck operator. Indeed it reads $\lambda_0^{-1}\partial_i \lambda_0 [a^A]^i + [a^A]^i\partial_i\ln\tpss=0$ with the $\tpss$ we have just defined (we remind that $\lambda_0$ is the volume element of the Lebesgue measure in the considered coordinate system; see section~\ref{subsec:generalContext}). Multiplying by $\tpss$ this last equation and using the chain rule, we get $\Div[\tpss\ba^A]=0$, \ie that condition~\eqref{TRS-div-cond-03} is satisfied -- which in particular means that $\tpss$ is the stationary probability density $\pss$.

Conversely, if we now assume that conditions~\eqref{TRS-a-cond-03} and~\eqref{TRS-div-cond-03} are satisfied, then the first part of~\eqref{TRS-a-cond-04} holds because $\bD^{-1}\ba^S$ is an exact one-form\footnote{We recall that the exterior derivative of an exact form is always zero thanks to Schwarz theorem.}, while the second part stems from the fact that $\tpss$ has a finite overall integral on $\mbR^d$. Furthermore, we see that using the chain-rule on the left-hand side of eq.~\eqref{TRS-div-cond-03} gives $\tpss\Div(\ba^A)+\ba^A\cdot\rmd\tpss=0$. Dividing this last equality by $\tpss$ and and using eq.~\eqref{TRS-a-cond-03} finally gives~\eqref{TRS-div-cond-04}.

Thus, when $\bD$ is invertible, dynamics~\eqref{EDS} is $\mke\mkm\mkt$-reversible if and only if conditions~\eqref{TRS-D-cond-04}-\eqref{TRS-div-cond-04} are satisfied. And these conditions can be \textit{explicitly checked in a systematic way}, unlike conditions~\eqref{TRS-a-cond-03} \&~\eqref{TRS-div-cond-03}.
We note that the reversibility conditions~\eqref{TRS-D-cond-04}, \eqref{TRS-a-cond-04} and~\eqref{TRS-div-cond-03} (the latter hence still involving an unknown function) were already known in the case of parameterized EMT-reversibility~\cite{gardiner1985handbook}, but not the full triad~\eqref{TRS-D-cond-04}-\eqref{TRS-div-cond-04}, which we believe to be new.

Interestingly, while the obstruction to simple T-reversibility is entirely quantified by the cycle affinity~\cite{jiang2004mathematical,yang2021bivectorial} $\rmd (\bD^{-1}\ba)$, which is a 2-form, \ie a field of skew-symmetric bilinear maps, \textit{the obstruction to EMT-reversibility is composed of three fields of different natures}: a two form $\rmd (\bD^{-1}\ba^S)$, that we call the \textit{symmetric} (or \textit{irreversible}) \textit{cycle affinity} by comparison with the T-reversal case; a tensor field $\bD^A$; and a scalar field corresponding to the left-hand side of eq.~\eqref{TRS-div-cond-04}.

Let us finally comment on the normalizability condition in~\eqref{TRS-a-cond-04}. Everything that we did so far -- namely the construction of the EMT-reversal and the reversibility conditions -- can be straightforwardly extended to the case where the space $\mkX$ defined in~\eqref{mkXdefinition} is replaced by the larger set $\tilde{\mkX} \supset \mkX$ defined as
\begin{eqnarray}
\tilde{\mkX}&\equiv & \{(\ba,\bD) \text{ such that the kernel of $\mcW_{\blambda}$ is one-dimensional and that} \\
& & \text{all $\bnu\in\ker(\mcW_{\blambda})\backslash\{0\}$ are smooth and nowhere vanishing measures} \} \ , 
\label{tildeMKXdefinition}
\end{eqnarray}
where $\mcW_{\blambda}$ is the Fokker-Planck operator of dynamics~\eqref{EDS}. By ``smooth and nowhere vanishing'', we mean that $\bnu\in\ker(\mcW_{\blambda})\backslash\{0\}$ is assumed to have a Lebesgue density $\rmd \bnu/\rmd\blambda$ that is smooth and nonzero everywhere.
In this more general context, dynamics~\eqref{EDS} might not have any normalizable stationary measure, but still have a family of stationary measures parameterized by an overall constant multiplier. When it is the case, \ie when $(\ba,\bD)\in\tilde{\mkX}\setminus\mkX$, then all the probabilities appearing in the previous sections (and in appendix~\ref{app:MTreversal}), become non-normalized ``weights'' -- that can still be used to calculate the ``relative probability'' of any pair of events. The notion of an EMT-dual process can still be properly defined, in terms of these ``path-weights''. Furthermore, the drift and diffusion tensor of this EMT-dual process are still given by eqs.~\eqref{eq:EMTdualD} \&~\eqref{eq:EMTduala} (or equivalently by eqs.~\eqref{eq:EMTdualDbis} \&~\eqref{eq:EMTdualabis}), the latter equations being manifestly invariant upon rescaling $\pss$, hence they are independent of which non-normalized stationary $\lambda_0$-density $\pss$ that is chosen among the (positive) stationary ones. 
Consequently, the EMT-reversibility conditions~\eqref{TRS-D-cond-03}-\eqref{TRS-div-cond-03} are still valid in this generalized sense. Further, condition~\eqref{TRS-a-cond-04} deprived of the normalizability constraint simply reads
\begin{equation}
\rmd (\bD^{-1}\ba^S) = 0 \ .
\label{TRS-a-cond-05}
\end{equation}
Conditions~\eqref{TRS-D-cond-04} \&~\eqref{TRS-div-cond-04}, together with~\eqref{TRS-a-cond-05}, are now EMT-reversibility conditions for a possible not-normalizable stationary state of dynamics~\eqref{EDS}.  
We give an example of such a situation of ``non-normalized reversibility'' in section~\ref{subsec:chiralAOUP}.

\subsubsection{Entropy production}
\label{subsubsec:IEPRonRd}

In this section, we assume that condition~\eqref{TRS-D-cond-04} is satisfied, \ie that\footnote{Note that the condition $\mkm_*\mke\bD=\bD$ is automatically satisfied when $\mkm$ and $\mke$ are trivial, \ie in the case of T-symmetry. Similarly, in the case of $T$-symmetry, $\ba^A=0$ and condition~\eqref{TRS-div-cond-04} consequently always holds.} $\mkm_*\mke\bD=\bD$. We then show (as detailed in appendix~\ref{app:IEPR}) that the informatic entropy production rate\footnote{The IEPR is a usefully general measure of irreversibility in coarse-grained systems. In cases where the coarse-graining can be encapsulated by coupling to a heat bath, it  reduces to the entropy production as defined in (stochastic) thermodynamics~\cite{seifert2005entropy}.} (IEPR) with respect to the given $\mke\mkm\mkt$-reversal under study, which is defined as~\cite{fodor2022irreversibility}
\begin{equation}
\sigma \equiv \lim_{\mcT\to \infty}\ln  \frac{\mcP[(\bx_t)_{t\in\mbT}]}{\reversed{\mcP}[(\bx_t)_{t\in\mbT}]}\ ,
\end{equation}
with $\reversed{\mcP}$ as defined in section~\ref{sec:construction}, reads
\begin{equation}
\sigma = \int_{\mbR^d} \bJss \cdot \bD^{-1}\ba^S \lambda_0\rmd \bx - \int_{\mbR^d}\pss(\Div \; \ba^A + \ba^A\cdot\bD^{-1} \ba^S)\lambda_0\rmd\bx \ ,
\label{IEPR_01}
\end{equation}
where $\bJss\equiv \pss\ba-\bD\cdot \rmd\pss$ is the stationary probability current. 
Since the latter is divergence free and $\mbR^d$ is simply-connected, we know from Helmholtz-Hodge theory that there exists a contravariant, skew-symmetric, tensor of order two, $C^{ij}$, such that $J_{\rm ss}^j = -\lambda_0^{-1}\partial_i \lambda_0 C^{ij}$~\cite{fecko2006differential}.
The first term on the right-hand side of eq.~\eqref{IEPR_01} then reads
\begin{equation}
\int_{\mbR^d} \Jss^j [D^{-1}a^S]_j \lambda_0\rmd \bx = -\int (\lambda_0^{-1}\partial_i \lambda_0 C^{ij})[D^{-1}a^S]_j \lambda_0\rmd \bx \ .
\end{equation} 
Integrating by parts and using the skew-symmetry $C^{ij}=-C^{ji}$ gives
\begin{equation}
\int_{\mbR^d} \Jss^j [D^{-1}a^S]_j \lambda_0\rmd \bx = \frac{1}{2}\int C^{ij} ( \partial_i [D^{-1}a^S]_j - \partial_j [D^{-1}a^S]_i) \lambda_0\rmd \bx \ .
\end{equation}
Thus the IEPR~\eqref{IEPR_01} can be re-written as
\begin{equation}
\sigma = \int_{\mbR^d} \frac{1}{2}\bC\cdot \rmd (\bD^{-1}\ba^S) \lambda_0\rmd \bx - \int_{\mbR^d}\pss(\Div \; \ba^A + \ba^A\cdot\bD^{-1} \ba^S)\lambda_0\rmd\bx \ .
\label{IEPR_02}
\end{equation}
Dynamics~\eqref{EDS} is $\mke\mkm\mkt$-reversible iff $\sigma=0$. It is also $\mke\mkm\mkt$-reversible iff conditions~\eqref{TRS-a-cond-04} and~\eqref{TRS-div-cond-04} are both satisfied (we have already assumed condition~\eqref{TRS-D-cond-04} to be fulfilled). Hence, we conclude that $\sigma$ vanishes iff the two integrands on the right-hand side of eq.~\eqref{IEPR_02} vanish simultaneously. In particular, this means that the two integrals in the expression~\eqref{IEPR_02} for $\sigma$ cannot be non-zero and cancel each other out, nor they can both vanish without their integrands themselves being identically zero. The two terms $\rmd (\bD^{-1}\ba^S)$ and $\Div (\ba^A) + \ba^A\cdot\bD^{-1} \ba^S$ can thus be interpreted as two independent sources of entropy production.

The expression~\eqref{IEPR_02} generalizes to EMT-reversal a result previously obtained for T-reversal~\cite{yang2021bivectorial}. In the latter, only the first term remains on the right-hand side of eq.~\eqref{IEPR_02}, and $\rmd (\bD^{-1}\ba^S)=\rmd (\bD^{-1}\ba)$ is then the cycle affinity. In the more general case of EMT-reversal, we see that only the ``symmetric'' cycle affinity $\rmd (\bD^{-1}\ba^S)$ is responsible for producing entropy, while the ``skew-symmetric'' cycle affinity $\rmd (\bD^{-1}\ba^A)$ is not, so that its presence does not violate time-reversal symmetry. Thus $\rmd (\bD^{-1}\ba^S)$ and $\rmd (\bD^{-1}\ba^A)$ can be regarded respectively as the \textit{irreversible} and \textit{reversible cycle affinities}.

In the next section, we show that the irreversible cycle affinity, which is the first entropy-production source on the right-hand side of eq.~\eqref{IEPR_02}, can be geometrically viewed as (dual to) a vorticity. In contrast, we note that a good physical or geometrical interpretation of the second entropy-production source, $\Div (\ba^A) + \ba^A\cdot\bD^{-1} \ba^S$, is still lacking -- although in section~\ref{subsec:gaugeTheory} we give a partial interpretation of it in terms of gauge theory.

\subsubsection{Symmetric cycle affinity two-form in relation to vorticity of $\ba^S$}
\label{subsubsec:voritcityOperator}
In this section, we show that $\rmd(\bD^{-1}\ba^S)$ is dual to the vorticity operator of the vector field $\ba^S$ for a Riemannian metric given by $\bD^{-1}$.

To this end, let us start by considering a smooth vector field $\bv(\bx)$ over $\mbR^d$. It can be locally Taylor-expanded at fist order as $\bv(\bx+\bu)\simeq\bv(\bx)+\nabla \bv(\bx)\cdot \bu$, where $\nabla \bv$ is the Euclidean gradient of $\bv$, whose Euclidean coordinates\footnote{Here we call ``Euclidean coordinates'' any orthonormal coordinate system for the canonical Euclidean metric of $\mbR^d$.} read $[\nabla \bv]^{ij}=\partial_jv^i$, and `$\cdot$' here stands for the canonical scalar product of $\mbR^d$. We can then decompose the first order term into its symmetric and skew-symmetric parts, so that the Taylor expansion reads:
\begin{equation}
\bv(\bx+\bu) \simeq \bv(\bx) + \bE\cdot\bu + \bOmega\cdot \bu \ ,
\label{velocityFieldExp01}
\end{equation}
with  $E^{ij}=(\partial_j v^i + \partial_i v^j)/2$ and $\Omega^{ij}=(\partial_j v^i - \partial_i v^j)/2$ in Euclidean coordinates. If the vector field $\bv$ is seen as the velocity field of some fluid, $\bE$ is then called the strain-rate tensor and, in $d=3$ dimensions, $\bOmega\cdot\bu=\frac{1}{2}(\nabla\times\bv)\times\bu$, where $\times$ is the cross product, and $\nabla\times\bv$, which is the curl of $\bv$,  is called the vorticity vector. By extension, we will call $\bOmega$ the vorticity tensor.

The expansion~\eqref{velocityFieldExp01} can be extended to a more general setting, where the canonical Euclidean metric is replaced by an arbitrary Riemannian metric.
Since the diffusion tensor associated to eq.~\eqref{EDS} is defined as $\bD\equiv \bb_\alpha\bb_\alpha$, it is clearly positive semi-definite. Having assumed (see the beginning of section~\ref{subsec:whenDisInvertible}) that $\bD(\bx)$ is in addition everywhere invertible, we conclude that $\bD$ is then positive definite. In turn, so is $\bD^{-1}(\bx)$. The latter, which is a covariant tensor field of order two, can thus be promoted to a Riemannian metric on $\mbR^d$ that we denote by $g$, \ie for any pair $\bu,\bv$ of tangent vectors at $\bx\in\mbR^d$: $g(\bu,\bv)\equiv u^i[D^{-1}]_{ij}(\bx)v^j$.
Using this metric, we can associate to any vector field $\bu$ a dual one-form, traditionally denoted by $\bu^\flat$, which is such that $[\bu^\flat]_i=g_{ij}u^j=[D^{-1}]_{ij}u^j$.

Let us denote by $\widetilde{\nabla}$ the Levi-Civita connection associated to the metric $g$. Any vector field $\bv(\bx)$ can then be Taylor expanded in a covariant way as~\cite{chavel2006riemannian}:
\begin{equation}
\bv(\exp_{\bx}(\bu))\simeq \tau_{\bx\to\exp_{\bx}(\bu)}\left[\bv(\bx) + \wnabla_\bu\bv (\bx) \right] \ ,
\label{covariantTaylor}
\end{equation}
up to linear order in $\bu$, where $\exp_\bx$ is the Riemannian $g$-exponential map\footnote{The Riemannian exponential map $\exp_{\bx}$ sends a tangent vector $\bu$ at $\bx$ to the point $\exp_{\bx}(\bu)$ reached at time one by the geodesic that passes at time zero through $\bx$ with speed $\bu$.}, and $\tau_{\bx\to\exp_{\bx}(\bu)}$ is the parallel transport from $\bx$ to $\exp_{\bx}(\bu)$ along the geodesic joining these two points.

Since $\wnabla \bv(\bx)$ linearly maps any vector $\bu$ at $\bx$ to another vector $\wnabla_\bu\bv(\bx)$ at $\bx$, we can consider the adjoint $(\wnabla \bv)^\top$ with respect to $g$, whose action on a vector $\bu$ we denote by $\wnabla_\bu\bv^\top$.
It is easy to show that in coordinates $[(\wnabla \bv)^\top]^\ell_k=g^{\ell j}[\wnabla\bv]^i_jg_{ik}$. More importantly, we can now split $\wnabla\bv$ into its $g$- symmetric and skew-symmetric part in the usual way, that we respectively denote by $\wbE$ and $\wbOmega$: $\wnabla_\bu\bv (\bx) = \wbE(\bx)\cdot \bu + \wbOmega(\bx)\cdot \bu$, the dot `$\cdot$' here denoting the tensor contraction again. The covariant Taylor expansion~\eqref{covariantTaylor} then reads
\begin{equation}
\bv(\exp_{\bx}(\bu))\simeq \tau_{\bx\to\exp_{\bx}(\bu)}\left[\bv(\bx) + \wbE(\bx)\cdot\bu + \wbOmega(\bx)\cdot \bu \right] \ ,
\end{equation}
where, in this Riemannian context, the tensors $\wbE$ and $\wbOmega$ play similar roles to that of $\bE$ and $\bOmega$ in the Euclidean context.

We then show in appendix~\ref{app:vorticity_operator} that for any vector fields $\bu,\bw,$: $g(\wbOmega\cdot\bu , \bw) = \rmd \bv^\flat (\bu,\bw)/2$, where $\rmd$ is the exterior derivative. In particular, this is true for the symmetric part $\bv\equiv \ba^S$ of the drift of dynamics~\eqref{EDS}:
\begin{equation}
g(\wbOmega\cdot\bu , \bw) = \frac{1}{2} \rmd [\ba^S]^\flat (\bu,\bw)=\frac{1}{2} \rmd [\bD^{-1}\ba^S](\bu,\bw) \ .
\end{equation}
This can be reformulated as:
\begin{equation}
\wbOmega = -\frac{1}{2} \bD\cdot \rmd [\bD^{-1}\ba^S] \ ,
\label{vorticityOperatorVS2form}
\end{equation}
or,  in coordinates, $\widetilde{\Omega}^i_k = D^{ij} (\partial_k [D^{-1}a^S]_j - \partial_j [D^{-1}a^S]_k)/2$.
This means that the symmetric cycle affinity $\rmd [\bD^{-1}\ba^S]$ is dual to the $\bD^{-1}$-skew-symmetric part of the linear term in the $\bD^{-1}$-covariant Taylor expansion of $\ba^S$. In other words, the symmetric cycle affinity -- which we identified in section~\eqref{subsubsec:IEPRonRd} as a source of entropy production~\eqref{IEPR_02} -- measures the infinitesimal rotations (for the metric $\bD^{-1}$) locally generated by the symmetric part $\ba^S$ of the drift.
For this reason, $\rmd[\bD^{-1}\ba^S]$ will also be referred to as the \textit{vorticity two-form}, while we call the operator $\wbOmega$ on the left-hand side of eq.~\eqref{vorticityOperatorVS2form} the \textit{vorticity operator}.

\subsection{The case of a non-invertible $\bD$}
\label{subsec:nonInvertibleD}

In this section, we further study the EMT-reversibility conditions and give a formula for the IEPR in a case where $\bD(\bx)$ is not everywhere invertible. 
We do not investigate the fully general situation but rather focus on a context, akin to that of the underdamped Langevin process, where the vector fields $\{\bb_{\alpha}\}_\alpha$ and the map $\mke$ satisfy certain constraints described below.

The mirror map $\mkm$ being a linear involution, there exists a split of the state space $\mbR^d=\mbR^{d_1}\oplus \mbR^{d_2}$ such that $\mkm =(id_{\mbR^{d_1}},-id_{\mbR^{d_2}})$. In this section we assume that, with this decomposition of the state space, all the $\bb_\alpha$'s have vanishing components along $\mbR^{d_1}$. In other words, in an adapted basis, the matrix $\bB\equiv (\bb_1,\dots,\bb_n)$ takes the form
\begin{equation}
\bB=\begin{pmatrix} 0 & 0 \\ 0 & \bB_2 \end{pmatrix} \ .
\end{equation}
In particular, this means that the spurious drift obeys $\bs^{(\varepsilon)}=(0, \bs^{(\varepsilon)}_2)^\top$ -- where we write the prescription parameter $\varepsilon\in[0,1]$ as a superscript for convenience -- and that the diffusion tensor reads
\begin{equation}
\bD=\begin{pmatrix} 0 & 0 \\ 0 & \bD_2 \end{pmatrix} \ ,
\end{equation}
with $\bD_2=\bB_2\bB_2^\top$.
We further assume that $\bD_2$ is invertible and that the map $\mke$ acts on the $\mbR^{d_2}$-component only, \ie $\mke(\ba_1,\ba_2)=(\ba_1,\mke\ba_2)^\top$ and  
\begin{equation}
\mke\bD=\begin{pmatrix} 0 & 0 \\ 0 & \mke\bD_2 \end{pmatrix} \ .
\label{eq:eStructure}
\end{equation} 

\subsubsection{EMT-reversibility conditions}
In a similar way to what we did in section~\ref{subsec:whenDisInvertible}, the $\mke\mkm\mkt$-reversibility conditions~\eqref{TRS-D-cond-02}-\eqref{TRS-div-cond-02} are equivalent to the existence of a smooth positive function $\tpss$ whose integral over $\mbR^d$ is finite, and which is such that
\begin{gather}
\bD^A_2 = 0 \ , \label{TRS-D-cond-sing}\\
\ba^S_1 = 0 \ , \label{TRS-a1-cond-sing} \\
\rmd_2 (\bD_2^{-1}\ba^S_2) = 0 \ , \label{TRS-a2-cond-sing}\\
\Div_1 (\tpss\ba_1) + \tpss\left[\Div_2(\ba^A_2) + \ba^A_2\cdot\bD_2^{-1}\ba_2^S \right]  = 0 \label{TRS-div-cond-sing} \ ,
\end{gather}
where the symmetric and skew-symmetric parts with respect to $\mkm_*\mke$ are defined just as in section~\eqref{subsec:firstEMTcond}, $\Div_1$ and $\Div_2$ are respectively the divergence operators\footnote{More precisely, $\Div_1$ and $\Div_2$ are the divergence operators with respect to the Lebesgue measures in $\mbR^{d_1}$ and $\mbR^{d_2}$, respectively. Upon denoting these Lebesgue measures respectively by $\blambda_1$ and $\blambda_2$, we have $\lambda_0(\bx_1,\bx_2)\rmd x_1^1\dots\rmd x_1^{d_1}\rmd x_2^1\dots\rmd x_2^{d_2}=\lambda_1(\bx_1)\lambda_2(\bx_2)\rmd x_1^1\dots\rmd x_1^{d_1}\rmd x_2^1\dots\rmd x_2^{d_2}$ in any coordinate system which is adapted to the decomposition $\mbR^d=\mbR^{d_1} \oplus\mbR^{d_2}$.} in $\mbR^{d_1}$ and $\mbR^{d_2}$, and $\rmd_2$ stands for the exterior derivative in $\mbR^{d_2}$: 
\begin{equation}
\rmd_2 (\bD_2^{-1}\ba^S_2) = \sum_{1\leq i\leq j<d_2} \left[ \frac{\partial [D_2^{-1}a^S_2]_i}{\partial x_2^j} - \frac{\partial [D_2^{-1}a^S_2]_j}{\partial x_2^i} \right]\rmd x_2^j\wedge\rmd x_2^i \ .
\end{equation}
Indeed, condition~\eqref{TRS-D-cond-sing} directly stems from eq.~\eqref{TRS-D-cond-02} together with eq.~\eqref{eq:eStructure} and the action of $\mkm$ on the factors of $\mbR^{d_1}\oplus\mbR^{d_2}$. In this setting, condition~\eqref{TRS-a-cond-02} reads $(\ba^S_1,\ba^S_2)=(0,\bD_2\cdot\rmd_2\ln\tpss)$. The first component of this last equality is eq.~\eqref{TRS-a1-cond-sing} while the second component is equivalent to eq.~\eqref{TRS-a2-cond-sing}, just as in the case of an invertible $\bD$ discussed in section~\ref{subsec:whenDisInvertible}. 
Here, the subtlety lies in the fact that eq.~\eqref{TRS-a2-cond-sing} involves $\rmd_2$ and not $\rmd$. Indeed, while eq.~\eqref{TRS-a2-cond-sing} implies the existence of a potential $\psi(\bx_1,\bx_2)$ such that $[D^{-1}_2a_2^S]_i=-\partial\psi/\partial x_2^i$, imposing $\rmd(\bD_2^{-1}\ba^S_2)=0$ would imply the existence of a potential $\psi(\bx_2)$, independent of $\bx_1$, such that $[D^{-1}_2a_2^S]_i=-\partial\psi/\partial x_2^i$, which is much stronger than what time-reversal requires.
Finally, eq.~\eqref{TRS-div-cond-sing} stems from using the Leibniz rule in eq.~\eqref{TRS-div-cond-02} and replacing $\rmd_2\ln\tpss$ by $\bD_2^{-1}\ba^S_2$, the justification of the equivalence of eqs.~\eqref{TRS-div-cond-03} and~\eqref{TRS-div-cond-sing} being the same as in section~\ref{subsec:whenDisInvertible}.

Contrary to condition~\eqref{TRS-div-cond-04} for the case of an invertible $\bD$, condition~\eqref{TRS-div-cond-sing} still involves an unknown function $\tpss$, akin to the stationary probability $\pss$. The latter condition can be reformulated by noting that the reversibility condition $\ba_2=\bD_2\cdot\rmd_2\ln\tpss$ imposes that there exists a real-valued function $c(\bx_1)$ such that $\tpss = \exp(-\psi_c)$ where
\begin{equation}
\psi_c(\bx_1,\bx_2) = -\int_0^1 [D^{-1}_2a_2^S(\bx_1,t\bx_2)]_i x_2^i \rmd t + c(\bx_1) \ .
\label{eq:singDexplPotential}
\end{equation}
This implies that condition~\eqref{TRS-div-cond-sing} can be replaced by\footnote{Just as conditions~\eqref{TRS-D-cond-04}, \eqref{TRS-div-cond-04}, \&~\eqref{TRS-a-cond-05} in the case where $\bD$ is invertible, equations~\eqref{TRS-D-cond-sing}-\eqref{TRS-a2-cond-sing} \&~\eqref{TRS-div-cond-sing-bis} are actually conditions of non-normalized $\mke\mkm\mkt$-reversibility. To include the normalizability requirement, conditions~\eqref{TRS-a2-cond-sing} and~\eqref{TRS-div-cond-sing} should together be replaced by the single condition: $\rmd_2(\bD^{-1}_2\ba^S_2)=0$ and $\exists \; c(\bx_1)$ such that $\int_{\mbR^d}\exp(-\psi_c)<\infty$ and the equality in~\eqref{TRS-div-cond-sing-bis} is satisfied, with $\psi_c$ given by eq.~\eqref{eq:singDexplPotential}.}:
\begin{equation}
\exists \ c(\bx_1) \ \ \text{such that} \ \ \ \ba_1\cdot\rmd c = \Div_1(\ba_1) - \ba_1\cdot \rmd_1 \psi_0 + \Div_2(\ba^A_2) + \ba^A_2\cdot\bD_2^{-1}\ba_2^S  \ .
\label{TRS-div-cond-sing-bis}
\end{equation}
Unfortunately, this condition still requires determining whether a certain PDE admits a solution. 
In a case by case situation, one can try to solve eq.~\eqref{TRS-div-cond-sing-bis} by \eg the method of characteristics. But we are not aware of any general theorem allowing us to replace condition~\eqref{TRS-div-cond-sing-bis} with one verifying that an analytically calculable quantity equally vanishes.

\subsubsection{Entropy production}
Let us now turn to the inspection of the entropy production. With the current hypothesis on the structure of $\bD$, the path-wise probability of dynamics~\eqref{EDS} is proportional to  $\delta[\dot{\bx}_1(t)-\ba_1]\times \exp(-\mcS_2)$, where $\mcS_2$ here denotes the Onsager-Machlup action associated to the dynamics of the $\bx_2$ component. In particular, this implies that the IEPR of dynamics~\eqref{EDS} diverges to infinity as soon as condition~\eqref{TRS-a1-cond-sing} is violated. From now on, we consider the case where the two $\mke\mkm\mkt$-reversibility conditions~\eqref{TRS-D-cond-sing} \&~\eqref{TRS-a1-cond-sing} are satisfied. While the fulfillment of the former condition, as in section~\ref{subsubsec:IEPRonRd}, simplifies the computation of the IEPR, that of the latter, which also reads $\ba_1(\bx_1,-\bx_2)=-\ba_1(\bx_1,\bx_2)$, prevents the entropy production from diverging. The computation of appendix~\ref{app:IEPR} can then be straightforwardly adapted to show that the IEPR reads 
\begin{equation}
\sigma = \int_{\mbR^d} \bJ_{\mathrm{ss},2} \cdot \bD_2^{-1}\ba^S_2\lambda_0\rmd \bx - \int_{\mbR^d} \pss \left[\Div_2 \ba^A_2 + \ba^A_2\cdot\bD_2^{-1}\ba_2^S \right] \lambda_0\rmd\bx \ ,
\label{IEPR-sing0}
\end{equation}
where $ \bJ_{\mathrm{ss},2}$ is the $\mbR^{d_2}$-component of the stationary probability current: $\bJss=(\bJ_{\mathrm{ss},1}, \bJ_{\mathrm{ss},2})$. As in section~\ref{subsubsec:IEPRonRd}, since $\Div_{\blambda_0}\bJss=0$, there exists a field $\bC(\bx_1,\bx_2)$ of skew-symmetric contravariant tensors of order two such that $J_{\rm ss}^j=-\lambda_0^{-1}\partial_i \lambda_0 C^{ij}$. Note that, if we simply take the convention of looking at $\bD_2^{-1}\ba_2^S$ as an element of (the dual of) $\mbR^d$ with its $d_1$ first components being equal to zero, rather that as an element of $\mbR^{d_2}$,  the first integrand on the right-hand side of eq.~\eqref{IEPR-sing0} reads $\bJ_{\mathrm{ss},2} \cdot \bD_2^{-1}\ba^S_2 =\sum_{i=d_1+1}^{d_2} \Jss^i [D_2^{-1}a_2^S]_i=\sum_{i=1}^{d_2} \Jss^i [D_2^{-1}a_2^S]_i$. Just as we did in section~\ref{subsubsec:IEPRonRd}, we can then replace $\Jss^i$ by $-\lambda_0^{-1}\partial_i \lambda_0 C^{ij}$, integrate by parts, and rearrange the terms using the skew-symmetry of $C^{ij}$ to get
\begin{equation}
\sigma = \frac{1}{2}\int_{\mbR^d} \bC \cdot \rmd (\bD_2^{-1}\ba^S_2)\lambda_0\rmd \bx - \int_{\mbR^d} \pss \left[\Div_2 \ba^A_2 + \ba^A_2\cdot\bD_2^{-1}\ba_2^S \right] \lambda_0\rmd\bx \ .
\label{IEPR-sing}
\end{equation}
On the one hand, demanding condition~\eqref{TRS-div-cond-sing} for $\mke\mkm\mkt$-symmetry leads to the cancellation of the second integral in the expression~\eqref{IEPR-sing} of $\sigma$, since this integral then reads $\int_{\mbR^{d_2}}\int_{\mbR^{d_1}} \Div_1 (\ba_1\pss) \lambda_0\rmd\bx_1\rmd\bx_2$, which vanishes thanks to Stokes' theorem. Interestingly, contrary to the invertible-$\bD$ case, reversibility does not impose the cancellation of the integrand, but only that of the integral with respect to $\bx_1$, for all $\bx_2$.

On the other hand, we know that imposing the last remaining reversibility condition~\eqref{TRS-a2-cond-sing} must result in the cancellation of the last remaining part of $\sigma$, \ie of the first integral on the right-hand side of eq.~\eqref{IEPR-sing}.
But contrary to the invertible-$\bD$ case, it is now not obvious at all to see from eq.~\eqref{IEPR-sing} how this happens, since the first integral in eq.~\eqref{IEPR-sing} involves the exterior derivative $\rmd$ on the full $\mbR^d$, while it is the exterior derivative $\rmd_2$ on $\mbR^{d_2}$ that appears in reversibility condition~\eqref{TRS-a2-cond-sing}, the difference between the two being
\begin{equation}
\rmd(\bD_2^{-1}\ba^S_2)-\rmd_2(\bD_2^{-1}\ba^S_2)= \sum_{i=1}^{d_2}\sum_{j=1}^{d_1}\frac{\partial [D_2^{-1}a^S_2]_i}{\partial x_1^j} \rmd x_1^j\wedge\rmd x_2^i \ .
\end{equation}
This difference vanishes iff $\bD^{-1}_2\ba_2^S$ does not depend on $\bx_1$, in which case the connection between eqs.~\eqref{TRS-a2-cond-sing} and~\eqref{IEPR-sing} is much clearer.

This is the case \eg for the underdamped Langevin dynamics~\eqref{UnderLangevinR} \&~\eqref{UnderLangevinV}, with $\mke=id$ and $(\bx_1,\bx_2)\equiv(\br,\bv)$, and hence $\ba^S=(0,\ba_2^S)=(0,-\gamma\bv)$, $\ba^A=(\ba_1,\ba_2^A)=(\bv,\bF(\br))$, $D_2^{ij}=\delta^{ij}\gamma kT$. Indeed, this implies that $\rmd(\bD_2^{-1}\ba^S_2)=\rmd_2(\bD_2^{-1}\ba^S_2)=0$. Furthermore, in this case, it is straightforward to see that conditions~\eqref{TRS-D-cond-sing}-\eqref{TRS-a2-cond-sing} are fulfilled, while condition~\eqref{TRS-div-cond-sing-bis} reads:
\begin{equation}
\exists \ c(\br) \ \ \text{such that} \ \ \ \bv\cdot \left[ \nabla_\br c - \frac{1}{kT}\bF(\br) \right] = 0 \ ,
\end{equation}
which means that the MT-reversibility of dynamics~\eqref{UnderLangevinR} \&~\eqref{UnderLangevinV} requires the force field to be conservative, as is well known. Finally, a direct application of formula~\eqref{IEPR-sing} gives the IEPR which is, as expected:
\begin{equation}
\sigma= \frac{1}{kT}\int \bv\cdot\bF(\br) \pss(\br,\bv)\rmd\br\rmd\bv \ .
\end{equation}
\\

To conclude, we have focused in this section~\ref{subsec:nonInvertibleD} on a situation where $\bD$ is not invertible and satisfies certain constraints. We have obtained conditions for EMT-reversiblity~\eqref{TRS-D-cond-sing}-\eqref{TRS-a2-cond-sing} \&~\eqref{TRS-div-cond-sing-bis} that do not require the knowledge of (a function akin to) $\pss$ and formulas for the IEPR~\eqref{IEPR-sing0}-\eqref{IEPR-sing} in the case where conditions~\eqref{TRS-D-cond-sing} \&~\eqref{TRS-a1-cond-sing} are fulfilled. But as opposed to the case of invertible $\bD$ studied in section~\ref{subsec:whenDisInvertible}, not all the reversibility conditions~\eqref{TRS-D-cond-sing}-\eqref{TRS-a2-cond-sing} \&~\eqref{TRS-div-cond-sing-bis} involve verifying the vanishing of an analytically calculable quantity, since condition~\eqref{TRS-div-cond-sing-bis} consists in figuring out whether a certain PDE admits a solution. It would be interesting to know whether this latter condition can also be turned into verifying that an analytically computable quantity vanishes.

In the next section, we study the relation between the EMT-reversibility of dynamics~\eqref{EDS} and that of two associated deterministic dynamics.

\subsection{Stochastic \textit{vs} deterministic $\mke\mkm\mkt$-reversibility}
\label{subsec:linkDeterministicTRS}

Let us consider the deterministic dynamics
\begin{equation}
\dot{\bx}_t = \bvss(\bx_t) \ ,
\label{Strato-average-dyn}
\end{equation}
where $\bvss\equiv\bJss/\pss = \ba-\bD\cdot \rmd\ln\pss$ is the so-called stationary mean velocity of dynamics~\eqref{EDS}. By definition of $\bvss$, the flow of dynamics~\eqref{Strato-average-dyn} leaves the stationary probability $\bPss$ invariant. In this section, we give a statistical interpretation of this dynamics and discuss the relation between its deterministic EMT-reversibility and the stochastic EMT-reversibility of dynamics~\eqref{EDS}.

\paragraph{Stratonovitch average dynamics.} Let us start by showing that dynamics~\eqref{Strato-average-dyn} can be interpreted as the Stratonovitch average of dynamics~\eqref{EDS}. A known result (which, for completeness, we prove in appendix~\ref{app:StratoAverageDyncs}), is that, in steady state, the velocity $\dot{\bx}_t$ and the probability current $\bJss=\pss\ba - \bD\cdot\rmd \pss$ are related through\footnote{In this paragraph, for simplicity, we make the proof by writing integrals in an affine coordinate system, so that the Lebesgue measure reads $\blambda_0=\rmd\bx$.}
\begin{equation}
\llangle \alpha(\bx_t)  \dot{\bx}_t \rrangle = \int_{\mbR^d} \alpha(\bx) \bJss(\bx) \rmd\bx \ ,
\label{relationAverageSpeedCurrent01}
\end{equation}
where the left-hand side is defined as
\begin{equation}
\llangle \alpha(\bx_t)  \dot{\bx}_t \rrangle\equiv \lim_{\Delta t\to 0} \llangle \alpha\left(\bx_t +\frac{\Delta\bx_t}{2}\right) \frac{\Delta\bx_t}{\Delta t} \rrangle \ .
\label{relationAverageSpeedCurrent02}
\end{equation}
We now show that this means $\bvss = \bJss/\pss$ can be interpreted as a Stratonovitch average velocity, in the sense that
\begin{equation}
\bv_t (\by) = \lim_{\Delta t\to 0} \llangle \frac{\Delta\bx_t}{\Delta t} \left| \bx_t + \frac{\Delta \bx_t}{2} = \by \right. \rrangle \ ,
\end{equation}
where $\Delta \bx_t \equiv \bx_{t+\Delta t} - \bx_t$ and the vertical bar stands for conditioning.
Indeed, in general, if $f(X,Y)$ is a function of two random variables $X$ and $Y$, then 
\begin{equation}
\llangle f(X,Y)\left| Y=y_0 \right. \rrangle = \int f(x,y) p(x|Y=y_0) \rmd x \ .
\end{equation}
But $p(x|Y=y_0)= p(x,y_0)/p(y_0) = \int \rmd y \; p(x,y)\delta(y-y_0)/p(y_0)$, hence
\begin{equation}
\llangle f(X,Y)\left| Y=y_0 \right. \rrangle = \int f(x,y) \frac{\delta(y-y_0)}{p(y_0)} p(x,y) \rmd x \rmd y = \llangle f(X,Y) \frac{\delta(Y-y_0)}{p(y_0)}\rrangle \ .
\end{equation}
Hence, we have
\begin{equation}
 \llangle \frac{\Delta\bx_t}{\Delta t} \left| \bx_t + \frac{\Delta \bx_t}{2} = \by \right. \rrangle = \llangle \frac{\Delta\bx_t}{\Delta t} \frac{\delta (\bx_t + \Delta\bx_t/2 - \by)}{p(\bx_t + \Delta\bx_t/2 = \by)} \rrangle \ .
\end{equation}
Using the latter equation, together with eqs.~\eqref{relationAverageSpeedCurrent01} \&~\eqref{relationAverageSpeedCurrent02}, we finally conclude that
\begin{equation}
\lim_{\Delta t \to 0} \llangle \frac{\Delta\bx_t}{\Delta t} \left| \bx_t + \frac{\Delta \bx_t}{2}=\by \right. \rrangle = \int \frac{\delta(\bx-\by)}{\pss(\by)}\bJss(\bx) \rmd \bx = \bvss(\by) \ .
\end{equation}

\paragraph{From stochastic to deterministic EMT-reversibility.} It is known that if stochastic dynamics~\eqref{EDS} is MT-reversible, then the deterministic dynamics~\eqref{Strato-average-dyn}, that we have just shown to be the Stratonovitch average dynamics, is MT-reversible in a deterministic sense~\cite{van1992stochastic}. We here show that this generalizes to EMT-reversibility.

We first need to define the $\mke\mkm\mkt$-reversal of the deterministic dynamics~\eqref{Strato-average-dyn}. A symmetry of an ordinary differential equation (ODE) is define as a map that send solutions to solutions. Hence, we must associate to the maps $\mke, \mkm$, and $\mkt$ a map that associates other trajectories to solutions of eq.~\eqref{EDS} on the time interval $\mbT$. 
As for the stochastic case, we start by defining how $\mkm$ and $\mkt$ together act on~\eqref{Strato-average-dyn}. This is done be taking a solution $(\bx_t)_{t\in\mbT}$ of eq.~\eqref{Strato-average-dyn} and mapping it to the $\mkm\mkt$-reversed path:
$(\by_t)_{t\in\mbT}= (\mkm(\bx_{\mcT-t}))_{t\in\mbT}$. Time-differentiating the resulting path, we get
\begin{equation}
\frac{\rmd}{\rmd t} \by_t = \frac{\partial\mkm}{\partial x^i} \frac{\rmd x^i_{\mcT-t}}{\rmd t} =- \frac{\partial\mkm}{\partial x^i}  v_{\rm ss}^i(\bx_{\mcT-t}) = -\mkm_*\bvss(\by_t) \ ,
\end{equation}
where we have used the fact that $(\bx_t)_{t\in\mbT}$ is a solution of eq.~\eqref{Strato-average-dyn} together with the definition~\eqref{pushforwardDef} of $\mkm_*$. Hence, the $\mkm\mkt$-reversed solutions of eq.~\eqref{Strato-average-dyn} satisfy the $\mkm\mkt$-reversed dynamics:
\begin{equation}
\dot{\bx}_t = -\mkm_*\bvss(\bx_t) \ .
\end{equation}
%

Next, we define $\mke_{\bv}(\bvss)$ to be the stationary velocity of the process~\eqref{EDS} after the map $\mke$ has been applied to $\ba$ and $\bD$, \ie
\begin{equation}
\mke_{\bv}(\bvss)\equiv \mke_{\ba}(\ba) - \mke_{\bD}(\bD)\cdot\rmd\ln\pss^{\mke} \ .
\label{mkeVssDef}
\end{equation}
%
Then, we simply set the ODE~\eqref{Strato-average-dyn} to be mapped by $\mke$ onto the dynamics:
\begin{equation}
\dot{\bx}_t = \mke_\bv(\bvss)(\bx_t) \ .
\end{equation}
%
Finally, we can define the $\mke\mkm\mkt$-reversal of dynamics~\eqref{Strato-average-dyn} as
\begin{equation}
\dot{\bx}_t = -\mkm_*\mke_\bv(\bvss)(\bx_t) \ ,
\end{equation}
and the Stratonovitch average dynamics~\eqref{Strato-average-dyn} is then $\mke\mkm\mkt$-reversible in a deterministic sense iff
\begin{equation}
\bvss=-\mkm_*\mke_\bv(\bvss) \ .
\label{determEMTrevCond}
\end{equation}

We now show that, assuming dynamics~\eqref{EDS} is $\mke\mkm\mkt$-reversible (in the stochastic sense), then its Stratonovitch average dynamics~\eqref{Strato-average-dyn} is deterministically $\mke\mkm\mkt$-reversible.
First, reversibility of dynamics~\eqref{EDS} implies that $\reversed{\ba}=\ba$, where $\reversed{\ba}$ is given by eq.~\eqref{eq:EMTduala}. Applying the map $\mkm_*\mke_{\ba}$ to this last equality, we get
\begin{equation}
 \mkm_*\mke_{\ba}(\ba)=2\bD\cdot\rmd\ln\pss-\ba \ .
 \label{tempEMTrev}
\end{equation}
Then, using the definition~\eqref{mkeVssDef}, we get 
\begin{equation}
\bvss+\mkm_*\mke_\bv(\bvss) = \ba - \bD\cdot\rmd\ln\pss + \mkm_*\mke_\ba(\ba) - \mkm_*\mke_\bD(\bD)\cdot\rmd \ln \mkm^*\pss^{\mke} \ .
\end{equation}
But $\mkm^*\pss^\mke$ is the probability density of the $\mke\mkm\mkt$-reversed process. Having assumed reversibility of dynamics~\eqref{EDS}, we have that $\mkm^*\pss^\mke=\pss$. Similarly, reversibility implies that $\mkm_*\mke_\bD(\bD)=\bD$. Hence, we can conclude that $\bvss+\mkm_*\mke_\bv(\bvss) = \ba +\mkm_*\mke_\ba(\ba) - 2\bD\cdot\rmd\ln\pss=0$, where the last equality is given by~\eqref{tempEMTrev}. The Stratonovitch average dynamics~\eqref{Strato-average-dyn} is thus deterministically $\mke\mkm\mkt$-reversible.

\paragraph{Zero-noise limit.} The zero-noise limit of equation~\eqref{EDS} is the following deterministic dynamics:
\begin{equation}
\dot{\bx}_t = \ba(\bx_t) \ .
\label{zeroNoiseLimit}
\end{equation}
Similarly to what has been done in the previous paragraph for the Stratonovitch average dynamics, the zero-noise dynamics~\eqref{zeroNoiseLimit} is $\mke\mkm\mkt$-reversible iff $\ba = -\mkm_*\mke_\ba(\ba)$, which also reads
\begin{equation}
\ba^S = 0 \ .
\label{condZeroNoiseEMTrev}
\end{equation}
We immediately see that condition~\eqref{condZeroNoiseEMTrev} is much stronger than the stochastic $\mke\mkm\mkt$-reversibility condition $\ba^S = \bD\cdot\rmd \ln\pss$. In fact, the latter implies the former iff the stationary probability is uniform\footnote{More precisely, when $\bD$ is singular $\pss$ only needs to be independent of the variables lying outside of the kernel if $\bD$.}. And condition~\eqref{condZeroNoiseEMTrev} does not even involve the stochastic reversibility since conditions~\eqref{TRS-D-cond-03}-\eqref{TRS-div-cond-03} have no reason to be satisfied.

Finally, note that a zero-noise equation of the form~\eqref{zeroNoiseLimit} is sometimes regarded as reversible iff $\ba$ is gradient~\cite{fruchart2021non}. Such a statement in fact needs to be made more precise: the condition that $\ba$ is a gradient can be seen as a reversibility condition only in the context of the noisy dynamics~\eqref{EDS} with a diffusion matrix that is uniform and proportional the identity: $\bD = D \bI_{d}$.
\\

To summarize, stochastic and deterministic reversibility are two distinct notions of time-reversal symmetry. Nevertheless, they can be related in certain peculiar situations. We have shown that the stochastic EMT-reversibility of dynamics~\eqref{EDS} implies the deterministic EMT-reversibility of dynamics~\eqref{Strato-average-dyn}, and we have identified the latter dynamics as the Stratonovitch average of the former. Finally, we have briefly discussed the case of the zero-noise limit~\eqref{zeroNoiseLimit}. In particular, we have highlighted that requiring $\ba$ in eq.~\eqref{zeroNoiseLimit} to be a gradient can be understood as a reversibility condition (something which is sometimes asserted in an imprecise way in the literature) in one precise situation only.

\newpage

\section{Generalization: SDE on manifolds \& alternative gauges}
\label{sec:generalization}

In sections~\ref{sec:contextAndTRSconstruction} \&~\ref{sec:EMT-reversibility}, we restricted ourselves for simplicity to cases where
the stochastic dynamics~\eqref{EDS} was defined on $\mbR^d$, the mirror symmetry $\mkm$ was linear, and we chose the Lebesgue measure $\blambda_0$ as the gauge measure.
In this section, we generalize all our results to the following more general context: dynamics~\eqref{EDS} takes place on a smooth, oriented, connected, manifold $\mcM$ without boundary, the mirror symmetry $\mkm:\mcM\to\mcM$ is an arbitrary smooth involution, and the gauge $\blambda$ is an arbitrary smooth, positive, measure\footnote{This means that the gauge $\blambda$ is such that, in any coordinate chart, $\blambda$ has a smooth positive density with respect to the Lebesgue measure of the chart. \label{smoothMeasDef}} on $\mcM$.

Because we took great care to present our previous results in a coordinate-free way, their adaptation to the current more general context will be fairly straightforward.
There are essentially two ingredients that will affect our previous results. 
The first one is the fact that $\mcM$ is not necessarily simply connected, as opposed to $\mbR^d$. 
The second one is that our arbitrary gauge $\blambda$ could now be modified by the mirror symmetry $\mkm$. Mathematically, this means that the pushforward measure, denoted by $\mkm_*\blambda$, that reads in local coordinates:
\begin{equation}
[\mkm_*\blambda] = [\mkm_*\lambda] (\bx)\rmd \bx = \lambda(\mkm(\bx)) \left|\mathrm{det} \left[(\partial_j\mkm^i)(\bx)\right] \right|\rmd \bx \ ,
\label{pushforwardMeasure}
\end{equation}
with $\rmd\bx\equiv \rmd x^1\dots\rmd x^d$, has no reason \textit{a priori} to be equal to $\blambda=\lambda(\bx)\rmd\bx$, whereas we had $\mkm_*\blambda_0=\blambda_0$, since linear involutions always preserve Euclidean volume (their determinant is always equal to 1, in absolute value). 

Let us choose a smooth positive measure $\blambda$ on $\mcM$, and consider the stochastic dynamics~\eqref{EDS} on $\mcM$, whose expression in $\varepsilon$-prescription\footnote{To the best of our knowledge, stochastic integrals on manifolds have only been globally defined so far for Ito ($\varepsilon=0$) and Stratonovich ($\varepsilon=1/2$) prescriptions. Nevertheless, one can still use any prescription $\varepsilon\in[0,1]$ in a given coordinate chart.} we write again here for convenience:
\begin{equation}
\dot{\bx}_t = \ba_{\blambda} + \bh_{\blambda} + \bs_{(\varepsilon)} + \bb_\alpha \eta_t^\alpha \ .
\label{EDSmanifold}
\end{equation}
The local coordinate expressions of $\bh_{\blambda}$ and $\bs_{(\varepsilon)}$ are as in section~\ref{sec:contextAndTRSconstruction}. The noises $\eta_t^\alpha$ are also as in section~\ref{sec:contextAndTRSconstruction}.
The $\bb_\alpha$, $\alpha=1\dots n$, and $\ba_{\blambda}$ are now vector fields over $\mcM$. Note that we now explicitly write down the gauge $\blambda$ as a subscript of the $\blambda$-drift, $\ba_{\blambda}$. In section~\ref{subsec:manifolds}, the gauge will be fixed, whereas in section~\ref{subsec:gaugeTheory}, we will study what happens as we vary the gauge, with a fixed stochastic dynamics, \ie with a fixed raw drift $\bA$. In the latter section, the drift $\ba_{\blambda}$ will thus vary with the gauge and it is to emphasize this dependency that we write $\blambda$ as a subscript of $\ba_{\blambda}$.

Note that the Fokker-Planck equation of the $\blambda$-density of the solution to dynamics~\eqref{EDSmanifold} is still\footnote{This can be readily checked by noting that the generator of dynamics~\eqref{EDS} is still given~\cite{barp2021unifying,bect2006generalized,elworthy1998stochastic,hsu2002stochastic} in local coordinate by $\mcL=A^i_{(1/2)}\partial_i + b^i_{\alpha}\partial_i b^j_\alpha \partial_j$, with $\bA_{(1/2)}$ the total drift of the Stratonovitch version of~\eqref{EDSmanifold}. The Fokker-Planck operator for the $\lambda$-density is then given by the $L^2(\blambda)$-adjoint of $\mcL$, which coincides with the right-hand side of eq.~\eqref{FPeq01}.} given by eq.~\eqref{FPeq01}. Also to emphasize its dependency to the gauge measure $\blambda$, we now denote by $\pss^{\blambda}$ the $\blambda$-density of the stationary measure $\bPss$ of the solution to~\eqref{EDS}, \ie $\pss^{\blambda}\equiv \frac{\rmd \bPss}{\rmd\blambda}$.

\subsection{EMT-reversal for stochastic dynamics on a manifold}
\label{subsec:manifolds}

\subsubsection{(E)MT-dual dynamics}
Let $\mkm$ be a smooth involution on $\mcM$. The $\mkm\mkt$-reversed dynamics of eq.~\eqref{EDSmanifold}, as defined in section~\ref{subsec:MT-dual}, has a diffusion tensor $\mkT\bD$ and a $\blambda$-drift $\mkT\ba_{\blambda}$ respectively given by (see appendix~\ref{app:MTreversal}):
\begin{eqnarray}
\mkT\bD &=& \mkm_*\bD \label{manifoldMTrev-D} \ ,\\
\mkT\ba_{\blambda} &=& \mkm_*\left[2\bD\cdot \rmd\ln \pss^{\blambda} - \ba_{\blambda} - \bD\cdot\rmd\ln\frac{\mkm_*\lambda}{\lambda}\right] \ , \label{manifoldMTrev-a}
\end{eqnarray}
In eq.~\eqref{manifoldMTrev-a}, the function $\frac{\mkm_*\lambda}{\lambda}(\bx)$, given by the ratio between the volume element\footnote{We recall that in this article, a measure is always written in bold font, while its volume element in a local chart is denoted by the same letter written in regular font.} $\mkm_*\lambda$ of $\mkm_*\blambda$ (given in eq.~\eqref{pushforwardMeasure}) and the volume element $\lambda$ of $\blambda$, coincides with the Radon-Nikodym derivative of $\mkm_*\blambda$ with respect to $\blambda$: $\frac{\mkm_*\lambda}{\lambda}=\frac{\rmd \mkm_*\blambda}{\rmd\blambda}$. 
Note that, just like in the Euclidean case, using the linearity of $\mkm_*$ together with the chain-rule and the fact that $\mkm^*\left[\frac{\mkm_*\lambda}{\lambda}\right] = \frac{\lambda}{\mkm_*\lambda}$, it can be easily shown that eq.~\eqref{manifoldMTrev-a} is equivalent to 
\begin{equation}
\mkT\ba_{\blambda} = 2[\mkm_*\bD]\cdot \rmd[\mkm^*\ln \pss^{\blambda}] - \mkm_*\ba_{\blambda} - [\mkm_*\bD]\cdot\rmd\ln\frac{\lambda}{\mkm_*\lambda} \ .
\end{equation}

Let us consider an involutive map $\mke$ that sends a pair of vector and second-order contravariant tensor fields $(\ba_{\blambda},\bD)$ to another one $\mke(\ba_{\blambda},\bD)\equiv(\mke^{\blambda}_\ba(\ba_{\blambda}),\mke_\bD(\bD))$. Note that we add the superscript $\blambda$ to the map $\mke_{\ba}$ to emphasize that it is defined for a given gauge $\blambda$ (in section~\ref{subsec:gaugeTheory}, we will discuss how it changes with the gauge). As we did previously, when there is no ambiguity, we will just write $\mke$ in place of $\mke_{\ba}^{\blambda}$ and $\mke_\bD$. 
The diffusion tensor $\reversed{\bD}$ and $\blambda$-drift $\reversed{\ba}_{\blambda}$ of the $\mke\mkm\mkt$-reversed dynamics of~\eqref{EDSmanifold} are then easily obtained, as in section~\ref{subsec:EMT-dual}:
\begin{eqnarray}
\reversed{\bD} &=& \mke(\mkm_*\bD) \label{manifoldEMTrev-D-01}\\
\reversed{\ba}_{\blambda} &=& \mke\left(\mkm_*\left[2\bD\cdot \rmd\ln \pss^{\blambda} - \ba_{\blambda} - \bD\cdot\rmd\ln\frac{\mkm_*\lambda}{\lambda}\right]\right) \label{manifoldEMTrev-a-01}
\end{eqnarray}
or, alternatively, using the commutativity constraint~\eqref{EMT_commutativityConstraint},
\begin{eqnarray}
\reversed{\bD} &=& \mkm_*\mke(\bD) \label{manifoldEMTrev-D-02}\\
\reversed{\ba}_{\blambda} &=& \mkm_*\left[2\mke(\bD)\cdot \rmd\ln (\pss^\mke)^{\blambda} - \mke(\ba_{\blambda}) - \mke(\bD)\cdot\rmd\ln\frac{\mkm_*\lambda}{\lambda}\right] \label{manifoldEMTrev-a-02} \ .
\end{eqnarray}
 In eq.~\eqref{manifoldEMTrev-a-02}, $(\pss^\mke)^{\blambda}\equiv \frac{\rmd \bPss^\mke}{\rmd\blambda}$, \ie it is the $\blambda$-density of the stationary measure $\bPss^\mke$ of the dynamics obtained from~\eqref{EDSmanifold} by replacing $\ba_{\blambda}$ by $\mke(\ba_{\blambda})$ and $\bD$ by $\mke(\bD)$.

\subsubsection{Reversibility conditions and entropy production}
We now introduce the vector fields
\begin{eqnarray}
\mcA^+_{\blambda} \equiv\ba^S_{\blambda} +\frac{1}{2}\bD\cdot \rmd \ln \frac{\mkm_*\lambda}{\lambda} \ , \label{curlyAvectField1}\\
\mcA^- \equiv\ba^A_{\blambda} -\frac{1}{2}\bD\cdot \rmd \ln \frac{\mkm_*\lambda}{\lambda} \ .
\label{curlyAvectField2}
\end{eqnarray}
Note that we have not used $\blambda$ as a subscript to $\mcA^-$ because the latter is actually gauge-invariant, as we will see in the next section.
Just as we did for dynamics~\eqref{EDS} on $\mbR^d$ in section~\ref{sec:EMT-reversibility}, but using formula~\eqref{manifoldEMTrev-D-01} \&~\eqref{manifoldEMTrev-a-01} for the $\mke\mkm\mkt$-dual drift and diffusion tensor, the conditions for dynamics~\eqref{EDSmanifold} to be $\mke\mkm\mkt$-reversible can be shown to be: there exists a smooth positive function $\tpss$ on $\mcM$ whose integral over $\mcM$ is finite and which is such that
\begin{gather}
\bD^A = 0 \ , \label{manifoldEMTrev-D-03}\\
\mcA^+_{\blambda} = \bD \cdot \rmd \ln \tpss\ , \label{manifoldEMTrev-a-03}\\
\Div_{\blambda}\left(\tpss\mcA^-\right) = 0 \ .\label{manifoldEMTrev-div-03}
\end{gather}
From now on we assume, as in section~\ref{subsec:whenDisInvertible}, that $\bD(\bx)$ \textit{is everywhere invertible}. Then, similarly to what we did on $\mbR^d$ in section~\ref{subsec:whenDisInvertible}, we can give $\mke\mkm\mkt$-reversibility conditions that do not involve any unknown function:
\begin{gather}
\bD^A = 0 \ , \label{manifoldEMTrev-D-04}\\
\rmd (\bD^{-1}\mcA_{\blambda}^+) = 0 \quad \text{and,} \quad \ \forall i=1\dots \dime(H_1(\mcM)), \ \oint_{\mcC_i}\bD^{-1}\mcA^+_{\blambda}\cdot \rmd\ell = 0 \label{manifoldEMTrev-a-04} \ ,\\
\Div_{\blambda} \left( \mcA^-\right) + \mcA^-\cdot\bD^{-1} \mcA^+_{\blambda} = 0 \ . \label{manifoldEMTrev-div-04}
\end{gather}
In eq.~\eqref{manifoldEMTrev-a-04}, the set of oriented loops $\{\mcC_i\}_{i=1\dots\dime(H_1(\mcM))}$ generates the first homology group $H_1(\mcM)$ of $\mcM$ (with real coefficients), \ie for each hole in $\mcM$ such that there exists a loop surrounding it that cannot be obtained as the boundary of a surface lying in $\mcM$, one has to choose such a loop\footnote{For example, if $\mcM$ is the two-dimensional torus, then $\dime(H_1(\mcM))=2$ and a pair of loops generating $H_1(\mcM)$ is then obtained by choosing one loop that surrounds the hole that is ``inside the torus'' (the interior of the ``donut'') and another loop that surrounds the ``central hole''.} $\mcC_i$.
The first equation in condition~\eqref{manifoldEMTrev-a-04} means that the one-form $\bD^{-1}\mcA^+_{\blambda}$ is closed. Provided this is true, the set of equations indexed by $i=1\dots\dime(H_1(\mcM))$ in condition~\eqref{manifoldEMTrev-a-04} means that the one-form $\bD^{-1}\mcA^+_{\blambda}$ is exact.

Conditions~\eqref{manifoldEMTrev-D-04}-\eqref{manifoldEMTrev-div-04} differ from conditions \eqref{TRS-D-cond-04},\eqref{TRS-div-cond-04} \&~\eqref{TRS-a-cond-05} in two ways. First, the vector fields $\ba^S_{\blambda}$ and $\ba^A_{\blambda}$ appearing in the latter are replaced in the former by $\mcA^+_{\blambda}$ and $\mcA^-$, respectively. This difference arises because the gauge $\blambda$ is now not necessarily $\mkm$-invariant. When this is actually the case, \ie when $\mkm_*\blambda=\blambda$, then we recover $\mcA^{+/-}_{\blambda}=\ba^{S/A}_{\blambda}$.
The second difference is the set of equations indexed by $i=1\dots \dime(H_1(\mcM))$ in condition~\eqref{manifoldEMTrev-a-04}.
When $\mcM$ is simply connected, like when $\mcM=\mbR^d$ for instance, then $\dime(H_1(\mcM))=0$ and these additional equations disappear. In this case, the first equation in~\eqref{manifoldEMTrev-a-04}, which in general only ensures that the one-form $\bD^{-1}\mcA^+_{\blambda}$ is closed, then also implies that it is exact.

Note that, just like condition~\eqref{TRS-a-cond-05} in $\mbR^d$, equation~\eqref{manifoldEMTrev-a-04} is actually a condition of $\mke\mkm\mkt$-reversibility for dynamics~\eqref{EDSmanifold} whose steady state is not necessarily normalizable. In order to require the existence of a normalizable stationary state, condition~\eqref{manifoldEMTrev-a-04} must be completed with the normalizability condition:
\begin{equation}
\int_{\mcM} \exp \left[\int_{\bx_0}^\bx \bD^{-1}\mcA^+_{\blambda} \cdot \rmd \ell \right] \lambda\rmd\bx < \infty \ ,
\label{manifoldEMTrev-a-04-addendum}
\end{equation}
where $\bx_0$ is an arbitrary point in $\mcM$, and $\int_{\bx_0}^\bx \bD^{-1}\mcA^+_{\blambda} \cdot \rmd \ell$ is then the integral of the exact one-form $\bD^{-1}\mcA^+_{\blambda}$ on a arbitrary smooth line joining $\bx_0$ to $\bx$ in $\mcM$.
When conditions~\eqref{manifoldEMTrev-D-04}-\eqref{manifoldEMTrev-div-04}, and~\eqref{manifoldEMTrev-a-04-addendum} are satisfied, dynamics~\eqref{EDSmanifold} is $\mke\mkm\mkt$-reversible and its stationary probability measure has a $\blambda$-density given by
\begin{equation}
\pss^{\blambda}(\bx) = Z^{-1}\exp \left[\int_{\bx_0}^\bx \bD^{-1}\mcA^+_{\blambda} \cdot \rmd \ell \right] \ ,
\label{manifoldSSproba}
\end{equation}
where the normalization factor $Z$ is given by the left-hand side of~\eqref{manifoldEMTrev-a-04-addendum}.

As in section~\ref{subsubsec:IEPRonRd}, we now focus on cases where condition~\eqref{manifoldEMTrev-D-04} is satisfied, \ie $\mkm_*\mke(\bD)=\bD$, and study the corresponding entropy production. Upon denoting by $\bJss^{\blambda}\equiv \ba_{\blambda}\pss^{\blambda}-\bD\cdot\rmd\pss^{\blambda}$ the stationary probability $\blambda$-current of dynamics~\eqref{EDSmanifold}, the IEPR of this dynamics reads (see appendix~\ref{app:IEPR}):
\begin{equation}
\sigma = \int_{\mcM} \bJss^{\blambda}\cdot\bD^{-1}\mcA^+_{\blambda} \; \lambda \rmd\bx - \int_{\mcM} \left[\Div_{\blambda}\left(\mcA^-\right) + \mcA^-\cdot\bD^{-1}\mcA^+_{\blambda} \right] \pss^{\blambda} \lambda\rmd \bx \ ,
\label{manifoldIEPR01}
\end{equation}
directly generalizing eq.~\eqref{IEPR_01} to the case of a manifold.

\subsubsection{Reformulating the reversibility conditions and IEPR using a Riemannian metric}
\label{subsubsec:RiemMetricReformulation}
Let us consider a Riemannian metric $g$ on $\mcM$ such that the gauge $\blambda$ is the corresponding Riemannian volume measure. Note that it is always possible to find such a metric because, starting from an arbitrary metric $g'$ on $\mcM$ (which always exists), we only need to rescale it as $g\equiv g' [\lambda/\sqrt{\mathrm{det}(g')}]^{2/d}$ to get a metric $g$ with the desired property.
%
%
Together with this metric $g$ comes a differential operator on differential forms, the so-called Hodge Laplacian~\cite{warner1983foundations}. We denote by $\Delta_1$ the corresponding Hodge Laplacian on one-forms. 
The one-forms on which $\Delta_1$ vanishes are called \textit{harmonic}, and their space is denoted by $\mcH_{\Delta_1}(\mcM)$, \ie $\mcH_{\Delta_1}(\mcM)\equiv \ker(\Delta_1)$.
According to Hodge-de Rham theory~\cite{warner1983foundations}, the spaces $\mcH_{\Delta_1}(\mcM)$ and $H_1(\mcM)$ are isomorphic, a property that allows us to reformulate both the $\mke\mkm\mkt$-reversibility condition~\eqref{manifoldEMTrev-a-04} and the expression~\eqref{manifoldIEPR01} of the IEPR. 
To this purpose, let $(\bgamma^i)_{i=1\dots\dime(H_1)}$ be an orthonormal basis of $\mcH_{\Delta_1}(\mcM)$.

As we said earlier, while the first part of condition~\eqref{manifoldEMTrev-a-04} ensures that $\bD^{-1}\mcA^+_{\blambda}$ is closed, the second one (\ie the set of loop integrals that have to vanish) ensures that $\bD^{-1}\mcA^+_{\blambda}$ is then exact. If this one-form is closed, \ie if $\rmd (\bD^{-1}\mcA^+_{\blambda})=0$, then the only way for it to be non-exact would be to have a non-zero harmonic component in its (orthonormal) Helmholtz-Hodge decomposition\footnote{Note that, while the Hodge decomposition is known to be valid on compact manifolds, when the manifold is noncompact, some additional hypothesis might be required to ensure the decomposition still holds. In particular, assuming that the differential form $\bD^{-1}\mcA_{\blambda}^+$ has a finite $L^2$-norm is sufficient. But it could be not quite satisfying in a physical context where $\bD^{-1}\mcA_{\blambda}^+$ is \eg a force field that has a confining part that prevents the ``particle from going to infinity''. Indeed, in this case, the confining force diverges at infinity and will hence have an infinite $L^2$-norm. See~\cite{chan2018hodge} and references therein. Hence, while our results involving Hodge decomposition are strictly valid when $\mcM$ is compact, a more careful treatment might be necessary in other cases.}. In other words, if it is closed, it is exact iff $\bD^{-1}\mcA^+_{\blambda}$ is orthogonal to $\mcH_{\Delta_1}(\mcM)$. Hence, the $\mke\mkm\mkt$-reversibility condition~\eqref{manifoldEMTrev-a-04} can be reformulated as follows:
\begin{equation}
\rmd (\bD^{-1} \mcA_{\blambda}^+) = 0 \quad \text{and,} \quad \ \forall \alpha=1\dots \dime(H_1(\mcM)), \ \int_{\mcM}\gamma^\alpha_ig^{ij}[D^{-1}\mcA^+_{\lambda}]_j \lambda\rmd\bx= 0  \label{manifoldEMTrev-a-04-bis} \ ,
\end{equation}
where each integral indexed by $\alpha$ corresponds to the global Hodge inner product between the one-form $\bgamma^\alpha$ and $\bD^{-1}\mcA^+_{\blambda}$.

Note that in the next section, we show all the results of this section to be gauge invariant. In particular this implies that instead of choosing $g$ such that its associated Riemannian volume measure is a given $\blambda$, we can choose whatever metric $g$ we want a take its Riemannian volume form as our gauge $\blambda$. For instance if we choose $g_{ij}=[D^{-1}]_{ij}$, just as in section~\ref{subsubsec:voritcityOperator}, then $g^{ij}=D^{ij}$ and $\gamma^\alpha_ig^{ij}[D^{-1}\mcA^+_{\lambda}]_j$ is simply $\gamma^\alpha_i[\mcA^+_{\lambda}]^i$.

Let us now turn to the IEPR. 
We recall that the Riemannian metric $g$ allows us to go from vector fields to one-forms and back, respectively by ``lowering and raising indices''. For instance, if $\bu$ is a vector field over $\mcM$ whose local coordinate expression reads $\bu=u^i\frac{\partial}{\partial x^i}$, then the one-form associated to it through $g$ is $\bu^\flat=u^ig_{ij}\rmd x^j$, where $(\rmd x^j)_j$ is the dual basis of $(\frac{\partial}{\partial x^j})_j$, \ie which is such that $\rmd x^j(\frac{\partial}{\partial x^i})=\delta^j_i$. Similarly, if $\balpha=\alpha_i\rmd x^i$ is a one-form on $\mcM$, its $g$-associated vector field is $\balpha^\sharp = \alpha_ig^{ij}\frac{\partial}{\partial x^j}$.
Then, in the language of Helmholtz-Hodge-de Rham theory, $\Div^{\blambda}(\bJss^{\blambda})=0$ means that the one form $[\bJss^{\blambda}]^\flat$ is co-closed. In turn, this implies that there exists a skew-symmmetric contravariant tensor $\bC_{\blambda}$, and a harmonic one-form $\bGamma_{\blambda}=[\Gamma_{\lambda}]_\alpha\bgamma^\alpha$ (note that the $[\Gamma_{\lambda}]_\alpha$, $\alpha=1\dots\dime H_1(\mcM)$, are constant coefficients, independent of $\bx\in\mcM$) such that the $\blambda$-probability current splits as follows:
\begin{equation}
\bJss^{\blambda}=-\Div_{\blambda}(\bC_{\blambda}) + \bGamma_{\blambda}^\sharp \ ,
\label{currentManifoldDecomp}
\end{equation}
where $\Div_{\blambda}(\bC_{\blambda})$ and $\bGamma_{\blambda}^\sharp=[\Gamma_\lambda]_\alpha (\bgamma^\alpha)^\sharp$ are vector fields whose coordinates are respectively given by $[\Div_{\blambda}(\bC_{\blambda})]^j\equiv \lambda^{-1}\partial_i \lambda C_{\lambda}^{ij}$  and $[\Gamma^\sharp_\lambda]^j=[\Gamma_\lambda]_\alpha[(\gamma^\alpha)^\sharp]^j\equiv [\Gamma_\lambda]_\alpha g^{ji}\gamma^\alpha_i$.
We can now inject expression~\eqref{currentManifoldDecomp} into that of $\sigma$~\eqref{manifoldIEPR01}. Integrating by parts, just as in section~\eqref{subsubsec:IEPRonRd}, we finally get
\begin{eqnarray}
\sigma &=& \frac{1}{2}\int_{\mcM} \bC_{\blambda}\cdot\rmd \left(\bD^{-1}\mcA^+_{\blambda}\right) \; \lambda \rmd\bx + [\Gamma_\lambda]_\alpha\int_{\mcM}(\bgamma^\alpha)^\sharp \cdot \bD^{-1}\mcA^+_{\blambda} \lambda \rmd \bx  \notag \\
& & -\int_{\mcM} \left[\Div_{\blambda}\left(\mcA^-\right) + \mcA^-\cdot\bD^{-1}\mcA^+_{\blambda} \right] \pss^{\blambda} \lambda\rmd \bx \ . 
\label{manifoldIEPR02}
\end{eqnarray}
Note that, similarly to what we had in section~\eqref{subsubsec:IEPRonRd}, having assumed condition~\eqref{manifoldEMTrev-D-04} fulfilled, the remaining $\mke\mkm\mkt$-reversibility conditions~\eqref{manifoldEMTrev-div-04} \&~\eqref{manifoldEMTrev-a-04-bis} appear as independent\footnote{We recall that the term ``independent'' is justified by the fact that for $\sigma$ to vanish, the integrand of each integral on the right-hand side of eq.~\eqref{manifoldIEPR02} must be identically zero.} sources of entropy production in~\eqref{manifoldIEPR02}.
While the interpretation of the first and last integrals on the right-hand side of eq.~\eqref{manifoldIEPR02} is similar to that of the Euclidean case, the other ones, indexed by $\alpha=1\dots \dime(H_1(\mcM))$, are related to the non-simply connectedness of $\mcM$. Each of them quantifies the entropy that is produced by the winding of $\bD^{-1}\mcA^+_{\blambda}$ (and $\bJss^{\blambda}$) around the holes of $\mcM$. This generalizes to (E)MT-reversal previous results obtained for T-reversal on a manifold~\cite{jiang2004mathematical}.
While the irreversible cycle affinity $\rmd\bD^{-1}\mcA^+_{\blambda}$ quantifies the local rotations around points on the manifold, the projection of $\bD^{-1}\mcA^+_{\blambda}$ on the space of harmonic one-forms, given by
\begin{equation}
\left[\bD^{-1}\mcA^+_{\blambda}\right]^{\rm harm}\equiv \sum_{\alpha=1}^{\dime H_1(\mcM)}\left[\int_{\mcM}(\bgamma^\alpha)^\sharp \cdot \bD^{-1}\mcA^+_{\blambda} \lambda \rmd \bx \right] \bgamma^\alpha \ ,
\label{harmonicCycleAff}
\end{equation}
quantifies rotations around holes. Consequently the former could be called ``local (or differential) irreversible cycle affinity'' and the latter ``topological irreversible cycle affinity'' -- or ``harmonic irreversible cycle affinity'' to emphasize its dependence on the choice of Riemannian metric.

Finally, note that all the other results obtained in $\mbR^d$ in the sections~\ref{subsec:whenDisInvertible} \&~\ref{subsec:linkDeterministicTRS} are easily shown to remain unchanged in this manifold context. This includes the interpretation of $\rmd (\bD^{-1}\mcA_{\blambda}^+)$ as the dual of the vorticity operator of $\mcA_{\blambda}^+$, and the link between the stochastic $\mke\mkm\mkt$-reversibility of~\eqref{EDSmanifold} and the deterministic ones of its Stratonovitch average dynamics and of its zero-noise limit.

\subsection{The gauge theory of the reference measure}
\label{subsec:gaugeTheory}


It has been shown that, in the case of the T-reversal of Markov chains, the cycle affinity can be interpreted as the curvature of a certain connection~\cite{polettini2012nonequilibrium}. The author of~\cite{polettini2012nonequilibrium} interprets the gauge transformation as a change of the reference measure with respect to which
the (relative) entropy is measured. Then rules for how the probability density and jump rates transform under a change of reference measure are postulated. The author shows that although the relative entropy changes with the prior, the entropy production is gauge invariant.

In this section, we revisit this interpretation in the light of the decomposition~\eqref{lambdaDriftANDpriorSplit} of the raw drift into $\blambda$-gauge drift and $\blambda$-drift, and generalize it to the case of EMT-reversal of Markov processes in continuous space-time with multiplicative (Gaussian) noise\footnote{Note that this gauge-theoretic interpretation in the case of a Langevin equation is also discussed in~\cite{polettini2012nonequilibrium}, but the authors focus on coordinate change. Indeed, in their case, the Fokker-Planck equation is not invariant under coordinate change because their definition of the ``probability density'', which is the same as in~\cite{graham1977covariant},  corresponds to the Radon-Nikodym derivative with respect to \textit{the Lebesgue measure of the coordinate chart}. In this article, the reference measure $\lambda$ is not define with respect to any coordinate system. The Fokker-Planck equation~\eqref{FPeq01} is hence invariant under coordinate change, but not under gauge transformation of $\blambda$. In addition, note that the generalization of this ``thermodynamic curvature'' to the case of the MT-reversal of Markov processes in continuous space-time is briefly mentioned in~\cite{dal2021fluctuation}.}. In particular, the way the various mathematical objects transform under a change of reference measure does not need to be postulated in our framework, but is a direct consequence of the decomposition~\eqref{lambdaDriftANDpriorSplit}.

As discussed in section~\ref{subsec:generalContext}, as we change the reference measure $\blambda\to\bmu$, we can consider that either $\ba$ or $\bA$ is fixed in relation~\eqref{lambdaDriftANDpriorSplit} -- in the former, $\blambda$ is a prior, while in the latter it is a gauge. 
Apart from~\ref{subsubsec:commentPrior}, where we briefly discuss the case where $\blambda$ plays the role of a prior, we consider in this section that choosing different $\blambda$ amounts to perform gauge transformations under which we prove our EMT-reversibility conditions and entropy production formula are invariant.
\\

\subsubsection{Gauge group, connection, and covariant derivative}
Let $\mcE$ be the set of all smooth measures on $\mcM$, and $\mcE_+$ the set of possible reference measures, \ie the subset of $\mcE$ made of positive measures -- we defined what we mean by ``smooth'' and ``positive'' measure in the footnote~\ref{smoothMeasDef} in section~\ref{subsec:manifolds}. 
For any $\blambda\in\mcE_+$, any measure $\bP\in\mcE$ can be ``decomposed'' as $\bP=p^{\blambda}\blambda$, with $p^{\blambda}=\frac{\rmd\bP}{\rmd\blambda}$, in a similar way a vector field can be decomposed in a coordinate frame.
Changing the gauge $\blambda\to\bmu =(\rmd \bmu/\rmd\blambda)\blambda$ amounts to multiply the initial gauge $\blambda$ by the positive function $\frac{\mu}{\lambda}(\bx)= \frac{\rmd \bmu}{\rmd\blambda}(\bx)$, while the density of $\bP$ changes as $p^{\blambda}\to p^{\bmu}=\frac{\lambda}{\mu} p^{\blambda}$. In other words, the multiplicative group of positive real numbers, $(\mbR_+,\times)$, here plays the role of a gauge group\footnote{In the language of differential geometry, the space $\mcE_+$ is a principal $(\mbR_+,\times)$-bundle over $\mcM$, while $\mcE$ is an associated vector bundle.}~\cite{bleecker2005gauge,hamilton2017mathematical}.

Relation~\eqref{lambdaDriftANDpriorSplit} then imposes that, upon changing the gauge $\blambda\to\bmu$ with a fixed raw drift $\bA=\ba_{\blambda}+ \bh_{\blambda}=\ba_{\bmu}+ \bh_{\bmu}$, the drift $\ba_{\blambda}$ becomes
\begin{equation}
\ba_{\bmu}=\ba_{\blambda}+ \bh_{\blambda}-\bh_{\bmu}=\ba_{\blambda} + \bD\cdot \rmd \ln \frac{\lambda}{\mu} \ .
\end{equation}
As mentioned in the previous section, the map $\mke_{\ba}^{\blambda}$, which is initially defined in a given gauge $\blambda$, changes so as to ensure the commutativity of applying the map $\mke$ to the drift and diffusion tensor of dynamics~\eqref{EDSmanifold} and making a gauge transformation, \ie
\begin{equation}
\mke_{\ba}^{\bmu}(\ba_{\bmu}) = \mke^{\bmu}_{\ba}\left(\ba_{\blambda} + \bD\cdot \rmd \ln \frac{\lambda}{\mu}\right) = \mke_{\ba}^{\blambda}(\ba_{\blambda}) + \mke_{\bD}(\bD)\cdot \rmd \ln \frac{\lambda}{\mu} \ .
\end{equation}
One can then easily show that, while the vector field $\mcA^-$ defined in~\eqref{curlyAvectField2} is gauge invariant, $\mcA_{\blambda}^+$, defined in eq.~\eqref{curlyAvectField1}, is transformed as 
\begin{equation}
\mcA^+_{\bmu} = \mcA^+_{\blambda} + \bD \cdot \rmd \ln \frac{\lambda}{\mu} \ .
\label{gaugeChangeCurlyA}
\end{equation}
Thanks to the way $\mcA^+_{\blambda}$ changes under a gauge transformation, we can use $-\bD^{-1}\mcA^+_{\blambda}$ as a gauge field (a.k.a a connection one-form) to define a $(\mbR_+,\times)$-connection on the space $\mcE$. 

Before showing this, let us first give a short explanation of what is the purpose of a $(\mbR_+,\times)$-connection in our context. While choosing a reference measure $\blambda$ allows associating to a given $\bP$ in $\mcE$ a number at each $\bx\in\mcM$, $\frac{\rmd \bP}{\rmd\blambda}(\bx)$, in a coordinate-free way, there is no canonical way of comparing these numbers at two distinct points $\bx, \by\in\mcM$. 
Being able to do such a comparison requires a way to ``parallel-transport'' the value of the density at $\bx$ to another point $\by$ (along any path connecting these two points), which is something that is allowed by the introduction of a so-called connection.
Although any explicit computation will still refer to a given gauge $\blambda$, we want a way of connecting the points $\bx,\by$ in $\mcM$ that is independent of the gauge choice. This is exactly the purpose of a so-called $(\mbR_+,\times)$-connection in the case where going from a gauge $\blambda$ to another $\bmu=\frac{\rmd\bmu}{\rmd\blambda}(\bx)\blambda$ requires the action of the gauge group $(\mbR_+,\times)$.

The connection whose gauge field is $-\bD^{-1}\mcA^+_{\blambda}$ has an infinitesimal counterpart: a $(\mbR_+,\times)$-covariant derivative, denoted by $\mcD$, which is defined by setting
\begin{equation}
\mcD \blambda \equiv -\bD^{-1}\mcA^+_{\blambda} \blambda \ .
\end{equation}
The Leibniz rule then gives the covariant derivative of an arbitrary measure $\bP\in\mcE$:
\begin{equation}
\mcD \bP = \mcD (p^{\blambda}\blambda)= (\rmd p^{\blambda})\blambda + p^{\blambda} \mcD\blambda = \left(\rmd p^{\blambda} -\bD^{-1}\mcA^+_{\blambda} p^{\blambda}\right)\blambda \ .
\end{equation}
This derivative associates to a measure\footnote{One could have expected that we focus on non-negative normalized measures, since we are ultimately working with probability measures. The point of studying the bigger space $\mcE$ is that a covariant derivative can only be define on ``vector bundles'', \ie we need our space to be stable under linear superposition.} $\bP\in\mcE$ and a vector field $\bu$ over $\mcM$ another measure: $\mcD_{\bu} \bP = (u^i \partial_i p^{\blambda} - u^i [D^{-1}]_{ij} [\mcA^+_{\blambda}]^jp^{\blambda})\blambda$, that also belongs to $\mcE$. The latter quantifies the infinitesimal variation of $\bP$ in the direction $\bu$, where the notion of spatial variation is defined by the connection.
The operator $\mcD$ defined above is a proper covariant derivative in the sense that, in addition to be linear and to satisfy the Leibniz rule (by construction), it is gauge invariant. Indeed,
\begin{eqnarray*}
\mcD (p^{\blambda}\blambda) &=& \blambda\left(\rmd - \bD^{-1}\mcA^+_{\blambda} \right) p^{\blambda} \\
&=& \frac{\lambda}{\mu}\bmu\left(\rmd - \bD^{-1}\mcA^+_{\bmu} - \rmd\ln \frac{\mu}{\lambda}\right) p^{\bmu}\frac{\mu}{\lambda} \\
&=&\bmu \left( \rmd p^{\bmu} -\bD^{-1}\mcA^+_{\bmu} \right)p^{\bmu} \\
&=& \mcD (p^{\bmu}\bmu) \ ,
\end{eqnarray*}
where the passage from the second line to third follows from the chain rule, while the rest is given by the way $\blambda$, $p^{\blambda}$, and $\mcA^+_{\blambda}$ change under gauge transformation.

%

\subsubsection{Reversibility conditions through the lens of gauge theory}
We can now easily prove that the $\mke\mkm\mkt$-reversibility conditions~\eqref{manifoldEMTrev-D-04}-\eqref{manifoldEMTrev-div-04} are gauge-invariant, something we expect since changing the gauge $\blambda$ does not change the dynamics. 
First, because $\bD$, $\mkm$, and $\mke_{\bD}$, are manifestly gauge-invariant, so is condition~\eqref{manifoldEMTrev-D-04}. 
Then, since the gauge field $-\bD^{-1}\mcA^+_{\blambda}$ changes by an exact one-form under a gauge transformation: $-\bD^{-1}\mcA^+_{\blambda} + \bD^{-1}\mcA^+_{\bmu}= \rmd \ln (\lambda/\mu)$, the reversibility condition~\eqref{manifoldEMTrev-a-04} (or equivalently condition~\eqref{manifoldEMTrev-a-04-bis}), which requires the gauge field to be an exact one-form, is also gauge invariant\footnote{This can be directly seen by noting that the exterior derivative of an exact form vanishes: $\rmd(\rmd\ln(\lambda/\mu))=0$, and that the integral of the exact one-form $\rmd\ln(\lambda/\mu)$ vanishes on any loop (or, for condition~\eqref{manifoldEMTrev-a-04-bis}, that the harmonic component of an exact form is identically zero).}.
Finally, conditions~\eqref{manifoldEMTrev-div-04} is also gauge invariant because the gauge change of the divergence operator $\Div_{\blambda}$, given by the chain rule:
\begin{equation}
\Div_{\blambda}\mcA^-=\lambda^{-1}\partial_i(\lambda \mu^{-1}\mu[\mcA^-]^i) = \Div_{\bmu} \mcA^- + \mcA^-\cdot \rmd\ln \frac{\lambda}{\mu} \ ,
\end{equation}
is compensated by the way $\mcA^+_{\blambda}$ changes.

Beyond their gauge-invariance, the $\mke\mkm\mkt$-reversibility conditions~\eqref{manifoldEMTrev-a-04} \&~\eqref{manifoldEMTrev-div-04} (or equivalently~\eqref{manifoldEMTrev-a-04-bis} \&~\eqref{manifoldEMTrev-div-04}) can be interpreted in terms of the gauge theory we have just described. (We have seen that condition~\eqref{manifoldEMTrev-D-04} has nothing to do with the gauge, so it is natural that it does not bear any gauge-theoretic interpretation.)

The full condition~\eqref{manifoldEMTrev-a-04} ensures that their exists a measure which is globally constant in the sense of our connection, \ie which is parallel-transported to itself along any path\footnote{This is analogous, in the case of a Levi-Civita connection on the tangent bundle of a manifold, to the existence of a global basis of tangent vector fields, each element of this basis being a parallel vector field, \ie a vector field that is invariant under parallel transport along any path.}. Indeed, the first part of condition~\eqref{manifoldEMTrev-a-04} requires the curvature of the connection defined above, which is given by $-\rmd(\bD^{-1}\mcA^+_{\blambda})$, to vanish. But the curvature quantifies the obstruction to the existence of a local measure that is $\mcD$-constant, \ie such that its covariant derivative vanishes -- which is the infinitesimal counterpart of being invariant under parallel transport. 
Note that if such a $\mcD$-constant measure $\bP=p^{\blambda}\blambda$ exists, it must be such that $\mcD\bP=(\rmd \ln p^{\blambda} - \bD^{-1}\mcA^+_{\blambda})\bP=0$, and hence satisfy  $\rmd \ln p^{\blambda} = \bD^{-1}\mcA^+_{\blambda}$.
The second part of eq.~\eqref{manifoldEMTrev-a-04} (or, equivalently, that of eq.~\eqref{manifoldEMTrev-a-04-bis}) requires the monodromy of the connection to vanish. But here the monodromy measures the obstruction to a locally $\mcD$-constant measure to be globally $\mcD$-constant. Hence, if the curvature and the monodromy vanish, then global $\mcD$-constant measures exist and must take the form given by the left-hand side of eq.~\eqref{manifoldSSproba} -- without being necessarily normalized.

Let us now give a gauge-theoretic interpretation of the reversibility condition~\eqref{manifoldEMTrev-div-04}.
Performing an integration by parts in the spatial integral of $\mcD_{\bu}\bP$, where $\bP=p^{\blambda}\blambda \in\mcE$ and $\bu$ is a vector field over $\mcM$, gives
\begin{eqnarray}
\int_{\mcM} \mcD_{\bu}\bP = \int_{\mcM} u^i\left( \partial_i p^{\blambda} -[D^{-1}\mcA^+_{\blambda}]_i p^{\blambda} \right)\blambda = -\int \left[\Div_{\blambda} \bu +\bu\cdot \bD^{-1}\mcA^+_{\blambda} \right] \bP \ .
\label{covariantAdj}
\end{eqnarray}
Equation~\eqref{covariantAdj} means that the adjoint $\mcD^*$ of our covariant derivative, which can be thought of as a ``covariant divergence'', and which associates a scalar valued function to a vector-field $\bu$, reads
\begin{equation}
\mcD^*\bu = -\Div_{\blambda} \bu -\bu\cdot \bD^{-1}\mcA^+_{\blambda} \ .
\label{covariantDiv}
\end{equation}
Hence the $\mke\mkm\mkt$-condition~\eqref{manifoldEMTrev-div-04} can be re-written as
\begin{equation}
\mcD^*\mcA^-=0 \ ,
\label{TRS-cond-cov-div}
\end{equation}
and therefore be interpreted as imposing that $\mcA^-$ has a vanishing covariant-divergence.
Interestingly, if condition~\eqref{manifoldEMTrev-a-04} (or equivalently~\eqref{manifoldEMTrev-a-04-bis}) is fulfilled, then, as stated above, there exists a measure $\bP$ that is $\mcD$-constant. Such a measure can then be chosen as the reference measure, a gauge in which $\mcA^+_{\bP}=0$ since $-\bD^{-1}\mcA^+_{\bP}\bP=\mcD\bP=0$. In particular, this implies that the covariant divergence~\eqref{covariantDiv} is simply the usual divergence with respect to $\bP$: $\mcD^*=\Div_{\bP}$.
In turn, the reversibility condition~\eqref{manifoldEMTrev-div-04} (or, equivalently, eq.~\eqref{TRS-cond-cov-div}) reduces to $\Div_{\bP}\mcA^-$ in that gauge, which is a reformulation of the eq.~\eqref{manifoldEMTrev-div-03} that requires $\mcA^-$ to preserve the stationary measure $\bPss\propto\bP$.
%
 Furthermore, as we have seen at the beginning of section~\ref{subsubsec:RiemMetricReformulation}, we can always find a Riemannian metric $g$ on $\mcM$ whose volume measure is $\bP$. Hodge theory then gives that the divergence-free condition $\Div_{\bP}\mcA^- = 0$ is equivalent to $\mcA^-$ being the superposition of a pure curl and of a harmonic vector field -- both types of vector fields being defined with respect to the metric $g$ whose volume measure is $\bP\propto\bPss$.
%
\\

\subsubsection{Entropy production and gauge theory}
\label{subsubsec:gaugeTheoryIEPR}
Let us now turn to the IEPR, and consider the case where the Onsager-Casimir condition~\eqref{manifoldEMTrev-D-04} is satisfied.
First, we have just shown that $\Div_{\blambda}(\mcA^-) + \mcA^-\cdot\bD^{-1}\mcA^+_{\blambda}$ is gauge invariant. Since, in addition, $\pss^{\blambda}\lambda=\pss^{\bmu}\mu$, the second integral on the right-hand side of the expression~\eqref{manifoldIEPR01} of $\sigma$ is gauge invariant itself.
Then, from the way $\pss^{\blambda}$ and $\ba_{\blambda}$ change under a gauge transformation, we straightforwardly conclude that $\bvss\equiv \bJss^{\blambda}/p^{\blambda}=\ba_{\blambda}-\bD\cdot\rmd\ln \pss^{\blambda}$ is also gauge invariant. Hence $\bJss^{\blambda}$ changes as $\pss^{\blambda}$ under a gauge transformation, \ie $\bJss^{\blambda}\to\bJss^{\bmu}=\bJss^{\blambda}\lambda/\mu$. Therefore, using~\eqref{gaugeChangeCurlyA}, we have
\begin{eqnarray}
\int_{\mcM} \bJss^{\blambda} \cdot \bD^{-1}\mcA^+_{\blambda} \lambda\rmd \bx
&=& \int_{\mcM}  \left[\bJss^{\bmu}\cdot \bD^{-1}\mcA^+_{\bmu} + \bJss^{\bmu}\cdot\rmd\ln\frac{\mu}{\lambda}\right] \mu\rmd\bx  \\
&=& \int_{\mcM}  \bJss^{\bmu}\cdot \bD^{-1}\mcA^+_{\bmu} \mu \rmd\bx - \int_{\mcM} \Div_{\bmu}(\bJss^{\bmu})\ln\frac{\mu}{\lambda} \mu \rmd\bx \ ,
\end{eqnarray} 
where we performed an integration by parts to go from the first line to the second. Because $\Div_{\bmu} \bJss^{\bmu} = 0$, the second integral on this second line vanishes. We conclude that the first term on the right-hand side of~\eqref{manifoldIEPR01} is also gauge invariant and, in turn, so is $\sigma$. Once again, this was to be expected since changing the gauge with a fixed $\bA$ does not change the stochastic process we are studying.

Consequently, the expression~\eqref{manifoldIEPR02} of $\sigma$ is also gauge invariant. Nonetheless, it is worth detailing some aspects of this invariance. To get eq.~\eqref{manifoldIEPR02} from eq.~\eqref{manifoldIEPR01}, we chose a Riemannian metric $g$ whose volume measure coincided with the gauge $\blambda$. Since the gauge $\blambda$ was arbitrary in the derivation of expression~\eqref{manifoldIEPR02} from~\eqref{manifoldIEPR01}, the former expression is valid in any gauge $\bmu$ under the condition of choosing a metric $g'$, like \eg $g'\equiv (\mu/\lambda)^{2/d}g$, whose volume measure is $\bmu$.
But this change of metric also changes the Hodge Laplacian and the corresponding harmonic one-forms (while still preserving the isomorphism $\mcH_{\Delta_1}(\mcM)\cong H_1(\mcM)$). 
Hence, while the last integral on the right-hand side of eq.~\eqref{manifoldIEPR02} is gauge invariant on its own, it is not generically the case for each of the other integrals: changing the gauge mixes them up despite leading to a similar expression in the new gauge and new metric.


The gauge-theoretic interpretation of the $\mke\mkm\mkt$-reversibility conditions~\eqref{manifoldEMTrev-a-04-bis} \&~\eqref{manifoldEMTrev-div-04}, which we discussed above, can be directly translated to each component of $\sigma$ in the expression~\eqref{manifoldIEPR02}.
The first integral on the right-hand side of eq.~\eqref{manifoldIEPR02} is an average of the curvature of our connection. The following terms, involving the harmonic fields $\bgamma^\alpha$, measure an average monodromy of this connection. Then, integrating by parts in the last integral on the right-hand side of eq.~\eqref{manifoldIEPR02} allows to rewrite it as
\begin{equation}
-\int_{\mcM} \left[\Div_{\blambda}\left(\mcA^-\right) + \mcA^-\cdot\bD^{-1}\mcA^+_{\blambda} \right] \pss^{\blambda} \lambda\rmd \bx  =  \int_{\mcM} \mcD_{\mcA^-}\bPss \ ,
\end{equation}
which means that this last term of $\sigma$ is the total covariant variation of the stationary measure $\bPss$ along $\mcA^-$ on $\mcM$.

Besides, note that starting from an arbitrary gauge $\blambda$, we can built an $\mkm$-invariant measure as $\blambda^S\equiv (\blambda + \mkm_*\blambda)/2$. Indeed, thanks to the linearity of the pushforward of measures~\eqref{pushforwardMeasure}, $\mkm_*\blambda^S=\blambda^S$. In particular, in such $\mkm$-invariant gauge, we have $\mcA^{+/-}_{\blambda}=\ba^{S/A}_{\blambda}$. 
Consequently, the expressions~\eqref{manifoldIEPR01} and~\eqref{manifoldIEPR02} of $\sigma$ and the reversibility conditions~\eqref{manifoldEMTrev-D-04}-\eqref{manifoldEMTrev-a-04} written in the gauge $\blambda_S$ only involve the vector fields $\ba^{S/A}_{\blambda}$, without any Radon-Nykodim derivative of $\mkm_*\blambda_S$ with respect to $\blambda_S$, so that we recover exactly the expressions\footnote{Up to the topological terms, that always vanish in $\mbR^d$.} we had derived in section~\ref{subsec:whenDisInvertible} where $\mcM=\mbR^d$ and the mirror map $\mkm$ was linear and hence preserving of the Lebesgue gauge $\blambda_0$.

%
%

\subsubsection{When the reference measure is a prior}
\label{subsubsec:commentPrior}

In this section, we briefly analyse the case where $\blambda$ plays the role of a prior, \ie where it is $\ba$ and not $\bA$ that is held constant in eq.~\eqref{lambdaDriftANDpriorSplit} as we vary the reference measure $\blambda$. Recall that in this case, changing $\blambda$ changes the physical model under study.

Let us consider a given vector field $\ba$ and two reference measures $\blambda$ and $\tilde{\blambda}$.
We are going to compare the $\mke\mkm\mkt$-reversibility conditions of the two different dynamics that are obtained by choosing $\ba$ as respectively the $\blambda$-drift and the $\tilde{\blambda}$-drift in eq.~\eqref{EDSmanifold}.
To distinguish these two process, every quantity associated to the second dynamics will be topped with a tilde.
In particular, in the first dynamics, the $\blambda$-drift is $\ba_{\blambda}\equiv \ba$, while in the second dynamics, $\widetilde{\ba_{\tblambda}}\equiv \ba$.

In order to compare the reversibility conditions of these dynamics, we need to relate key quantities that are involved in eq.~\eqref{manifoldEMTrev-D-04}-\eqref{manifoldEMTrev-div-04} and~\eqref{manifoldIEPR01}.

First, the Onsager-Casimir symmetry condition~\eqref{manifoldEMTrev-D-04} is obviously the same for the two processes since they share the same diffusion tensor $\bD$.
Then, the relation between the vector field $\widetilde{\mcA^+_{\tblambda}}=\ba^S+\frac{1}{2}\bD\cdot \rmd \ln \frac{\mkm_*\tilde{\lambda}}{\tilde{\lambda}}$ of the second process in the gauge $\tblambda$ and the vector field $\mcA^+_{\blambda}=\ba+\frac{1}{2}\bD\cdot\rmd \ln \frac{\mkm_*\lambda}{\lambda}$ of the first process in the gauge $\blambda$ reads:
\begin{equation}
\widetilde{\mcA^+_{\tblambda}} = \mcA^+_{\blambda} + \frac{1}{2}\bD\cdot\rmd \ln \frac{\lambda\mkm_*\tlambda}{\tlambda \mkm_*\lambda} \ .
\label{relationAplus}
\end{equation}
In particular,we see that $\bD^{-1}\widetilde{\mcA^+_{\tblambda}}$ and $\bD^{-1}\mcA^+_{\blambda}$ differ by an exact one-form. Therefore, the reversibility condition~\eqref{manifoldEMTrev-a-04} (or, equivalently, condition~\eqref{manifoldEMTrev-a-04-bis}) is satisfied in the first dynamics if and only if it is so in the second one.

Let us now turn to the last $\mke\mkm\mkt$-reversibility condition~\eqref{manifoldEMTrev-div-04}. Using definition~\eqref{curlyAvectField2}, it is easy to show that the gauge-invariant vector fields $\mcA^-$ and $\widetilde{\mcA^-}$ of the two dynamics are related by
\begin{equation}
\widetilde{\mcA^-} = \mcA^- + \frac{1}{2}\bD\cdot \rmd \ln\frac{\tlambda \mkm_*\lambda}{\lambda \mkm_*\tlambda} \ .
\label{relationAminus}
\end{equation}
Then, a straightforward calculation using the chain rule together with eqs.~\eqref{relationAplus} \&~\eqref{relationAminus}, give the difference between the left-hand side of eq.~\eqref{manifoldEMTrev-div-04} of the two dynamics:
\begin{eqnarray}
\Div_{\tblambda} \widetilde{\mcA^-} + \widetilde{\mcA^-}\cdot \bD \widetilde{\mcA^+_{\tblambda}} - \Div_{\blambda} \mcA^- + \mcA^- \cdot \bD \mcA^+_{\blambda} &=& \frac{1}{2}\Div_{\blambda} \left[ \bD \cdot \rmd \ln \frac{\tlambda \mkm_*\lambda}{\lambda \mkm_*\tlambda}\right] + \frac{1}{2}\mcA^-\cdot \rmd \ln \frac{\tlambda \mkm_*\tlambda}{\lambda \mkm_*\lambda} \notag\\
& & + \frac{1}{2} \left[\rmd \ln \frac{\tlambda \mkm_*\lambda}{\lambda \mkm_*\tlambda} \right] \cdot \left[ \mcA^+_{\blambda} + \frac{1}{2}\bD \rmd \ln \frac{\tlambda\mkm_*\tlambda}{\lambda \mkm_*\lambda} \right] \ . \qquad  \quad 
\label{diffLastRevCond}
\end{eqnarray}
Since the right-hand side of eq.~\eqref{diffLastRevCond} is generically non-zero, we conclude that even if one of the two processes satisfies the last $\mke\mkm\mkt$-reversibility condition~\eqref{manifoldEMTrev-div-04}, the other one will not in general.

Thus, when $\blambda$ plays the role of a prior, choosing two different $\blambda$ gives two distinct processes whose $\mke\mkm\mkt$-reversibility status differs in general. But note that, since this difference is only due to the third $\mke\mkm\mkt$-reversibility condition~\eqref{manifoldEMTrev-div-04} of our triad, which is automatically satisfied in the case of simple T-symmetry, \ie when $\mke$ and $\mkm$ are the identity, varying the prior does not actually affect the T-symmetry status of dynamics~\eqref{EDSmanifold}.

\newpage
	
\section{Applications and examples}
\label{sec:applications}

Although we believe the results obtained above to be powerful in many contexts, we do not attempt to demonstrate this here. Instead, this section is devoted to give a few pedagogical illustrations of the usage of these results to study the time-reversal properties of various systems, going from some simple particular examples to more generic situations.


%
Sections~\ref{subsec:chiralAOUP} \&~\ref{subsec:ChiralABP2d} are dedicated to the study of time-reversal properties of two different models of chiral active particles. Active particles are ``agents'' that are able to exert non-conservative forces on their environment to self-propel. In some cases, the dynamics of such particles can acquire an intrinsic chirality, which means that their self propulsion mechanism can break the (spatial) mirror symmetry. The study of such chiral active agents have attracted a lot of attention recently, in particular due their relevance in biological systems such as \eg sperm cells~\cite{riedel2005self,friedrich2007chemotaxis}, malaria parasites~\cite{patra2022collective}, or bacteria swimming near interfaces~\cite{diluzio2005escherichia,lauga2006swimming,di2011swimming}.

In section~\ref{subsec:chiralAOUP} we exhibit an EMT-symmetry of a free chiral active Ornstein-Uhlenbeck particle in $\mbR^3$ that, in turn, allows us to explicitly construct the stationary measure.

In section~\ref{subsec:ChiralABP2d} , we perform a similar analysis for a chiral active Brownian particle moving on a two-dimensional plane with periodic boundary conditions, \ie on a two-dimensional torus. In particular, the non-trivial topology of the configuration space constitutes an instructive playground to implement the tools we developed in section~\ref{subsec:manifolds} for the quantification of topological obstructions to time-reversibility.

Section~\ref{subsec:Kuramoto} is dedicated to the analysis of a class of EMT-transformation of the non-reciprocal Kuramoto model~\cite{acebron2005kuramoto,fruchart2021non,hong2011kuramoto,sompolinsky1986temporal}.
This section goes further in complexity in the sense that it studies a collection of interacting oscillators rather that a single agent. In particular, we exhibit an EMT-symmetry for the case of two non-reciprocally coupled oscillators and deduce the stationary measure and entropy production rate.

Finally, in section~\ref{subsec:typicalMTreversibleProcess} , we use the reversibility conditions~\eqref{manifoldEMTrev-D-03}-\eqref{manifoldEMTrev-div-03} to determine the structure of the most generic MT-reversible dynamics on a simply connected manifold.

\subsection{Chiral active Ornstein-Uhlenbeck particle}
\label{subsec:chiralAOUP}

This section is an inquiry into whether a particular instance of chiral active particle has an $\mke\mkm\mkt$-symmetry for suitable maps $\mke$ and $\mkm$. This is achieved by using the reversibility conditions~\eqref{TRS-D-cond-04}-\eqref{TRS-div-cond-04} which apply for the configuration space of this particle which is $\mbR^6$.

Consider a free chiral active Ornstein-Uhlenbeck particle~\cite{caprini2019active} in dimension $d=3$, whose dynamics in the canonical basis reads:
\begin{eqnarray}
\dot{\br} &=& v_0\bu + \sqrt{2D_t}\boldeta \label{ChiralAOUP01}\\ 
\dot{\bu} &=& -\frac{\bu}{\tau} + \bw\times\bu + \sqrt{\frac{2}{3\tau}}\bxi \label{ChiralAOUP02}
\end{eqnarray}
where $D_t$ is a translational diffusion coefficient, $\bu$ and $v_0$ are respectively the direction and typical amplitude of the self propulsion, $\tau$ is the persistence time, $\bw$ a rotation vector accounting for the chirality of the particle, `$\times$' the cross product, $\boldeta$ and $\bxi$ are independent, zero-mean, delta-correlated, Gaussian white noises, and the Lebesgue measure $\blambda_0$ has been chosen as the reference measure.
The drift and diffusion tensor thus respectively reads
\begin{equation}
\ba = \begin{pmatrix} v_0\bu \\ -\tau^{-1}\bu + \bw\times\bu \end{pmatrix} \ , \ \ \text{and} \ \ \bD = \begin{pmatrix} D_t \bI_3 & 0 \\ 0 & (3\tau)^{-1}\bI_3 \end{pmatrix} \ ,
\end{equation}
with $\bI_3$ the identity $3\times 3$ matrix.
Let us consider the linear involution
\begin{equation}
\mkm(\br,\bu)=(\br,\alpha\bu)^\top \ , \ \ \text{with} \ \ \alpha=\pm 1 \ ,
\end{equation}
together with the map $\mke$ that possibly reverses the chirality vector $\bw$, \ie
\begin{equation}
\mke(\ba,\bD) = \left[ \begin{pmatrix} v_0\bu \\ -\tau^{-1}\bu + \beta \bw\times \bu \end{pmatrix} , \bD  \right] \ , \ \ \text{with} \  \ \beta=\pm 1 \ .
\end{equation}
Let us compute the symmetric and skew-symmetric parts of $\ba$ under the map $\mkm_*\mke$. To this end, we first compute $\mkm_*\mke\ba$:
\begin{equation}
\mkm_*\mke\ba(\bx) = \begin{pmatrix} \bI_3 & 0 \\ 0 & \alpha\bI_3 \end{pmatrix} \begin{pmatrix} \alpha v_0 \bu \\  -\alpha\tau^{-1}\bu + \alpha\beta \bw\times\bu \end{pmatrix} = \begin{pmatrix} \alpha v_0 \bu \\ -\tau^{-1}\bu + \beta \bw\times\bu  \end{pmatrix} \ .
\end{equation}
Hence the symmetric part reads
\begin{equation}
\ba^S \equiv \frac{1}{2}(\ba + \mkm_*\mke\ba) = \begin{pmatrix} \frac{1+\alpha}{2}v_0\bu \\ -\frac{\bu}{\tau} +\frac{1+\beta}{2}\bw\times\bu \end{pmatrix} \ ,
\end{equation}
while the skew-symmetric is
\begin{equation}
\ba^A\equiv \frac{1}{2}(\ba - \mkm_*\mke\ba) = \begin{pmatrix} \frac{1-\alpha}{2}v_0\bu \\ \frac{1-\beta}{2}\bw\times\bu \end{pmatrix} \ .
\end{equation}
This implies, on the one hand that
\begin{equation}
\nabla\cdot \ba^A = \frac{1-\beta}{2} \nabla_\bu \cdot (\bw\times\bu) = 0 \ ,
\end{equation}
while on the other hand
\begin{eqnarray*}
\ba^A\bD^{-1}\ba^S &=& \begin{pmatrix} \frac{1-\alpha}{2}v_0\bu \\  \frac{1-\beta}{2}\bw\times\bu \end{pmatrix} \begin{pmatrix} D_t^{-1}\bI_3 & 0 \\ 0 & 3\tau\bI_3 \end{pmatrix}  \begin{pmatrix} \frac{1+\alpha}{2}v_0\bu \\ -\frac{\bu}{\tau} +\frac{1+\beta}{2}\bw\times\bu \end{pmatrix} \\
&=& \frac{v_0 |\bu|^2}{D_t} (1-\alpha^2) + \frac{3\tau | \bw\times\bu|^2}{4}(1-\beta^2) - \frac{3}{2}(1-\beta)\bu\cdot \bw\times\bu \ .
\end{eqnarray*}
Since $\alpha$ and $\beta$ are both equal to $\pm 1$, and $\bu\cdot \bw\times\bu=0$ by property of the cross product, we have
\begin{equation}
\ba^A\bD^{-1}\ba^S = 0 \ .
\end{equation}
Hence two of the three $\mke\mkm\mkt$-reversibility conditions~\eqref{TRS-D-cond-04} \&~\eqref{TRS-div-cond-04} are fulfilled whatever the values of $\alpha$ and $\beta$.
Let us now have a look at the last condition~\eqref{TRS-a-cond-04}. For this purpose, we compute $\rmd (\bD^{-1}\ba^S)$:
\begin{eqnarray*}
\rmd (\bD^{-1}\ba^S) &=& \rmd \left[ \frac{v_0}{2D_t}(1+\alpha) u^i\rmd r^i - 3 u^i\rmd u^i + \frac{3\tau}{2}(1+\beta) \epsilon_{ijk}w^ju^k\rmd u^i \right] \\
&=& \frac{v_0}{2D_t}(1+\alpha) \rmd u^i\wedge \rmd r^i - 3 \rmd u^i\wedge \rmd u^i + \frac{3\tau}{2}(1+\beta) \epsilon_{ijk}w^j \rmd u^k \wedge \rmd u^i \ .
\end{eqnarray*}
where $\epsilon_{ijk}$ the Levi-Civita symbol. Because of the skew-symmetry of the wedge product `$\wedge$', $\rmd u^i\wedge \rmd u^i$ vanishes. Hence we finally have
\begin{equation}
\rmd (\bD^{-1}\ba^S)  = \frac{v_0}{2D_t}(1+\alpha) \rmd u^i\wedge \rmd r^i + \frac{3\tau}{2}(1+\beta) \epsilon_{ijk}w^j \rmd u^k \wedge \rmd u^i \ .
\label{ChiralAOUPclosedFormCond}
\end{equation}
We immediately see that taking $\alpha=\beta=-1$ implies $\rmd (\bD^{-1}\ba^S)=0$, \ie $\mke\mkm\mkt$-reversibility condition~\eqref{TRS-a-cond-04} is satisfied. Thus, for $\mkm$ and $\mke$ such that $\alpha=-1$ and $\beta=-1$, respectively, dynamics~\eqref{ChiralAOUP01} \&~\eqref{ChiralAOUP02} is reversible.
Note that if $D_t$ was equal to zero, then the dynamics of our chiral active particle would map onto that of an underdamped Langevin particle in a magnetic field~\eqref{Langevin_Magnetic_field_01} \&~\eqref{Langevin_Magnetic_field_02} with a vanishing external potential. But here, the satisfying of the reversibility conditions~\eqref{TRS-D-cond-04}-\eqref{TRS-div-cond-04} shows this $\mke\mkm\mkt$-reversibility to hold even with a nonvanishing translational noise $D_t\neq 0$.

A similar property hold when the particle is not chiral: $\bw=0$. Indeed, despite the fact that the well known mapping between a free AOUP and a free underdamped Langevin particle does not hold any longer as soon as $D_t\neq 0$, dynamics~\eqref{ChiralAOUP01} \&~\eqref{ChiralAOUP02} is still $\mkm\mkt$-reversible, since eq.~\eqref{ChiralAOUPclosedFormCond} then gives $\rmd (\bD^{-1}\ba^S)=0$ for $\bw=0$ and $\alpha=-1$.

Finally, the $\mke\mkm\mkt$-symmetry of dynamics~\eqref{ChiralAOUP01} \&~\eqref{ChiralAOUP02} that we uncovered for $\mke$ and $\mkm$ such that $\alpha=\beta=-1$ grants a direct access to the explicit expression of all stationary measures:these are of the form
\begin{equation}
\bPss \propto \exp(-3|\bu|^2/2) \Pi_{i=1}^3\rmd r^i\rmd u^i\ .
\end{equation}
Because of their translation-invariance in the position variable $\br$, these stationary measures are not normalizable. Consequently, dynamics~\eqref{ChiralAOUP01} \&~\eqref{ChiralAOUP02} constitutes an instance of what we called ``non-normalized $\mke\mkm\mkt$-reversibility'' in section~\ref{subsubsec:EMTrevCondDinvert}.


\subsection{Chiral active Brownian particle in 2d}
\label{subsec:ChiralABP2d}
In this section, we study some EMT-reversal properties of another model of chiral active particle, namely that of the so-called chiral active Brownian particle~\cite{liebchen2017collective,volpe2014simulation}, whose dynamics is given by:
\begin{eqnarray}
\dot{\br} &=& v_0\bu + \sqrt{2D_t}\boldeta \ , \label{chiralABPdyn01}\\
\dot{\theta} &=& \omega + \sqrt{2D_r}\xi \ , \label{chiralABPdyn02}
\end{eqnarray}
where $\bu\equiv(\cos(\theta),\sin(\theta))$ is the direction of the self propulsion, $v_0$ its amplitude, $\omega$ the angular speed of the rotation of the chiral particle, $D_t$ and $D_r$ the translational and rotational diffusion constant, respectively, and $\boldeta$ and $\xi$ independent Gaussian white noises of zero mean and respective correlations $\langle \eta^i_t\eta^j_{t'} \rangle = 2\delta^{ij}\delta(t-t')$ and $\llangle \xi_t\xi_{t'}\rrangle = 2\delta(t-t')$. We assume that the position variable $\br$ is restricted to the unit square of the plane $\mbR^2$, with periodic boundary conditions, a space denoted by $\mbR^2/\mbZ^2$, while the angle $\theta$ parametrizes the unit circle $\mbS^1$.

As our gauge, we choose the usual Lebesgue measure $\blambda_0$ on the state space $(\mbR^2/\mbZ^2)\times\mbS^1$.
Although equation~\eqref{chiralABPdyn01} looks coordinate-free, it is actually only valid in Cartesian coordinates, where the $\blambda_0$-gauge drift $\bh_{\blambda_0}$ vanishes, together with the spurious drift $\bs_{(\varepsilon)}$ -- the notion of ``additive noise'' not being coordinate-free. In an arbitrary coordinate system, one would need to add $\bs_{(\varepsilon)}$ and $\bh_{\blambda_0}$ to the right-hand side of eq.~\eqref{chiralABPdyn01}, whose coordinate expression are respectively given by eqs.~\eqref{spuriousDrift} and~\eqref{priorDrift}.

In what follows, we will write all our equations in the orthonormal\footnote{We consider the metrics on the position and orientational spaces induced by the canonical Euclidean metric of $\mbR^2$.} frame $(\be_x,\be_y,\be_\theta)$, so that $\blambda_0=\rmd x\rmd y \rmd \theta$, the gauge drift vanishes, and the $ \blambda_0$-drift reads $\ba = (v_0\bu,\omega)$ while the diffusion tensor $\bD$ is 
\begin{equation}
\bD = \begin{pmatrix} D_t \bI_2 & 0 \\ 0 & D_r \end{pmatrix} \ .
\end{equation}

We consider the maps $\mkm(\br,\theta)\equiv (\br, \theta + \alpha \pi)$, where $\alpha$ is either $0$ or $1$, $\mke_\bD(\bD)\equiv\bD$, and $\mke_{\ba}(\ba)\equiv (v_0\bu, \beta\omega)$, with $\beta=\pm 1$. We then have that
\begin{equation}
[\mkm_*\mke(\ba)] (\br,\theta) = \begin{pmatrix} \bI_2 & 0 \\ 0 & 1 \end{pmatrix} \mke( \ba(\mkm(\br,\theta) ) = \begin{pmatrix} \hat{\alpha} v_0 \bu \\  \beta\omega \end{pmatrix} \ ,
\end{equation}
where $\hat{\alpha}\equiv \exp(i\alpha\pi)=\pm 1$, while $\mkm_*\mke(\bD) = \bD$.
Hence the symmetric and skew-symmetric parts of $\ba$ read
\begin{equation}
\ba^S = \begin{pmatrix} v_0\bu(1+\halpha)/2 \\ \omega (1+\beta)/2 & \end{pmatrix} \quad \text{and} \quad \ba^A = \begin{pmatrix} v_0\bu(1-\halpha)/2 \\ \omega (1-\beta)/2 & \end{pmatrix} \ ,
\end{equation}
while those of $\bD$ are simply $\bD^S=\bD$ and $\bD^A=0$. In particular, we see that the Onsager-like symmetry is satisfied, \ie the first $\mke\mkm\mkt$-reversibility condition~\eqref{manifoldEMTrev-D-04} is fulfilled, whatever $\alpha$ and $\beta$.
Note that our gauge is $\mkm$-invariant, \ie $\mkm_*\blambda_0=\blambda_0$. In particular, this implies that $\mcA^{+/-}_{\blambda_0}=\ba^{S/A}$, where $\mcA^{+/-}_{\blambda_0}$ are as defined in section~\ref{subsec:manifolds}.
In turn, we have that
\begin{equation}
\ba^A\cdot\bD^{-1}\ba^S = D^{-1}_tv_0^2 (1-\halpha^2)/4 + D_r^{-1} \omega^2(1-\beta^2)/4 = 0 \ ,
\end{equation}
where the last equality follows since $\halpha=\pm 1$ and $\beta\pm 1$, and 
\begin{equation}
\Div(\ba^A) = \frac{v_0(1-\halpha)}{2}\left[\partial_x \cos(\theta) + \partial_y \sin(\theta)\right] + \frac{1-\beta}{2}\partial_\theta \omega = 0 \ .
\end{equation}
This means that $\Div(\ba^A) + \ba^A\cdot\bD^{-1}\ba^S=0$, \ie that the third $\mke\mkm\mkt$-reversibility condition~\eqref{manifoldEMTrev-div-04} is always satisfied.
\\

Lets us now turn to the remaining reversibility condition~\eqref{manifoldEMTrev-a-04}, which requires the one-form
\begin{equation}
\bD^{-1}\ba^S=\frac{v_0(1+\halpha)}{2D_t}[\cos\theta \, \rmd x + \sin\theta \, \rmd y] + \frac{\omega(1+\beta)}{2D_r} \rmd\theta
\end{equation}
to be exact\footnote{Note that, even though the notation $(\rmd x,\rmd y, \rmd \theta)$ is usual to denote the dual coordinate frame of $(\be_x,\be_y,\be_\theta)$, it could be misleading because these one-forms are only closed, not exact.}. Let us first compute its exterior derivative:
\begin{equation}
\rmd \bD^{-1}\ba^S = \frac{v_0(1+\halpha)}{2D_t} \left[ -\sin\theta \,\rmd\theta \wedge \rmd x + \cos\theta\, \rmd\theta \wedge \rmd y \right] \ .
\end{equation}
We see that $\rmd \bD^{-1}\ba^S$ vanishes iff we flip the self-propulsion as time is reversed, \ie iff $\halpha=-1$ (or, equivalently, $\alpha=1$). 

As discussed in section~\ref{subsec:manifolds}, the local irreversible cycle affinity, $\rmd \bD^{-1}\ba^S$, quantifies only the local obstruction for $\bD^{-1}\ba^S$ to be exact. The global (or topological) obstruction remains to be considered.
To quantify the latter, as detailed in section~\ref{subsec:manifolds}, a first method  consists in computing the integral of $\bD^{-1}\ba^S$ around a set of loops generating the first homology group with real coefficients: $H_1(\mbR^2/\mbZ^2\times\mbS^1)$. Since $\mbR^2/\mbZ^2\cong \mbS^1\times\mbS^1$, we have that $H_1(\mbR^2/\mbZ^2\times\mbS^1)\cong H_1(\mbS^1\times\mbS^1\times\mbS^1) \cong\mbR^3$. As our generating set of loops, we choose $\{\mcC_x,\mcC_y,\mcC_\theta \}$, a parametrization of which is
\begin{eqnarray}
\mcC_x &:& t\in [0,1]\mapsto (x=t,y=0,\theta=\pi/2) \ , \\
\mcC_y &:& t\in [0,1]\mapsto (x=0,y=t,\theta=0) \ , \\
\mcC_\theta &:& t\in [0,2\pi]\mapsto (x=0,y=0,\theta=t) \ .
\end{eqnarray} 
While\footnote{Note that, unless we take, $\halpha=-1$, the one-form $\bD^{-1}\ba^S$ is not closed, in which case its contour integral is not necessarily the same for each element of a given homology class of $H_1(\mbR^2/\mbZ^2\times \mbS^1)$. For example, considering $\mcC'_x : t\in [0,1]\mapsto (x=t,y=0,\theta=0)$, which is in the same class as $\mcC_x$ (\ie it circles around the same hole of the state-space), we have $\int_{\mcC_x'}\bD^{-1}\ba^S = \frac{v_0(1+\halpha)}{2D_t}$, which does not vanish if $\halpha\neq -1$.}
\begin{equation}
\int_{\mcC_x} \bD^{-1}\ba^S = \int_{\mcC_y} \bD^{-1}\ba^S = 0 \ ,
\end{equation}
we have that
\begin{equation}
\int_{\mcC_\theta} \bD^{-1}\ba^S = \frac{\pi\omega(1+\beta)}{D_r} \ .
\label{oneTopoTRSobstr}
\end{equation}

As stated in section~\eqref{subsec:manifolds}, there is an alternative way of quantifying the topological obstruction for $\bD^{-1}\ba^S$ to be exact, which consists in computing its harmonic component $[\bD^{-1}\ba^S]^{\rm harm}$. In order to illustrate this second method, let us compute this harmonic component, that we called the topological irreversible cycle affinity.
We first choose as a Riemannian metric on $\mbR^2/\mbZ^2\times \mbS^1$ the one which is such that the frame $(\be_x,\be_y,\be_\theta)$ is everywhere orthonormal. In particular, it is such that the associated Riemannian volume measure is $\blambda_0$, as we assumed in the general setting of section~\ref{subsubsec:RiemMetricReformulation}.
Moreover, note that the dual frame $(\rmd x,\rmd y, \rmd \theta)$ is actually a basis of the space $\mcH_{\Delta_1}(\mbR^2/\mbZ^2\times \mbS^1)$ of harmonic one-forms. For the (global) Hodge inner product on one-forms, the latter basis is orthogonal but not normalized, as each element has a squared norm equal to $\int_{\mbR^2/\mbZ^2\times \mbS^1} \rmd \blambda_0 =2\pi$.
 To compute $[\bD^{-1}\ba^S]^{\rm harm}$ through formula~\eqref{harmonicCycleAff}, we hence take $(\bgamma_1,\bgamma_2,\bgamma_3)\equiv\frac{1}{\sqrt{2\pi}}(\rmd x,\rmd y, \rmd \theta)$, which implies that $(\bgamma_1^\sharp,\bgamma_2^\sharp,\bgamma_3^\sharp)=\frac{1}{\sqrt{2\pi}}(\be_x,\be_y, \be_\theta)$. The coordinates of $\bD^{-1}\ba^S$ along the $\bgamma_i$'s are thus 
\begin{eqnarray}
\int \gamma_1^\sharp\cdot \bD^{-1}\ba^S \rmd x\rmd y \rmd \theta &=& \frac{1}{\sqrt{2\pi}}\int \frac{v_0(1+\halpha)}{2D_t}\cos \theta \rmd x\rmd y \rmd \theta = 0 \ , \\
\int \gamma_2^\sharp\cdot \bD^{-1}\ba^S \rmd x\rmd y \rmd \theta &=& \frac{1}{\sqrt{2\pi}} \int \frac{v_0(1+\halpha)}{2D_t}\sin \theta \rmd x\rmd y \rmd \theta = 0 \ ,
\end{eqnarray}
and
\begin{equation}
\int \gamma_3^\sharp\cdot \bD^{-1}\ba^S \rmd x\rmd y \rmd \theta = \frac{1}{\sqrt{2\pi}}\int \frac{\omega(1+\beta)}{2D_r} \rmd x\rmd y \rmd \theta = \sqrt{\frac{\pi}{2}}\frac{\omega(1+\beta)}{D_r} \ .
\end{equation}
This gives the topological irreversible cycle affinity:
\begin{equation}
[\bD^{-1}\ba^S]^{\rm harm} = \frac{\omega(1+\beta)}{2D_r} \rmd \theta \ .
\end{equation}

Like eq.~\eqref{oneTopoTRSobstr}, this allows us to conclude that dynamics~\eqref{chiralABPdyn01} \&~\eqref{chiralABPdyn02} is $\mke\mkm\mkt$-reversible iff $\mkm$ and $\mke$ are such that $\halpha=\beta=-1$, \ie iff we flip $\bu$ and reverse the chirality of the particle as we reverse time.

Uncovering this explicit $\mke\mkm\mkt$-symmetry immediately grants access to the explicit expression of the stationary probability measure, just as in section~\ref{subsec:chiralAOUP}. It is such that the differential of its $\blambda_0$-density is equal to $\bD^{-1}\ba^S$, where $\ba^S$ corresponds to the $\mkm_*\mke$-symmetric part of $\ba$ with $\mkm$ and $\mke$ such that $\halpha=\beta=-1$: $\rmd \pss = \bD^{-1}\ba^S=0$. Hence $\pss=1/2\pi$, \ie the stationary measure is uniform.
Although this follows from the rotational symmetry of the problem, this symmetry was not used in the above derivation. This exemplifies how a well chosen ansatz for $\mke$ and $\mkm$ can, in favourable cases, uncover the stationary measure.

Since we know $\pss$, we also know the stationary probability $\blambda_0$-current: $\bJss=\ba\pss-\bD\cdot\rmd\pss= (v_0\bu,\omega)/2\pi $.
In turn, we can compute the IEPR, whatever $\alpha$ and $\beta$, directly from the expression~\eqref{manifoldIEPR01}:
\begin{equation}
\sigma = \int \bJss\cdot \bD^{-1}\ba^S \rmd x \rmd y\rmd\theta = \frac{v_0^2(1+\halpha)}{2D_t} + \frac{\omega^2(1+\beta)}{2D_r} \ ,
\end{equation}
which, as expected, vanishes iff $\halpha=\beta=-1$.

\subsection{Non-reciprocal Kuramoto model}
\label{subsec:Kuramoto}

In this section, we seek an EMT--symmetry of a non-reciprocal Kuramoto model \cite{acebron2005kuramoto,fruchart2021non,hong2011kuramoto,sompolinsky1986temporal} among a certain class of EMT-transformations.
The dynamics of the $N$ coupled oscillators reads
\begin{equation}
\partial_t\theta_i = \omega + \sum_j K_{ij}\sin(\theta_i-\theta_j) + \xi_i(t) \ ,
\label{Kuramoto_dynamics}
\end{equation}
where $\theta_i$ is the phase of the $i^{th}$ oscillator, $\omega$ the natural frequency of all oscillators, $\bK=(K_{ij})_{i,j=1\dots N}$ the matrix of coupling constants which is neither symmetric nor skew-symmetric \textit{a priori}, and $\xi_i(t)$ are Gaussian white noises of zero mean and correlations $\langle \xi_i(t)\xi_j(t') \rangle = 2D\delta_{ij}\delta(t-t')$. 
The dynamics~\eqref{Kuramoto_dynamics} takes place on the $N$-torus $\mbT_N\equiv \mbS^1\times\dots\times\mbS^1$. In the coordinate system $\btheta\equiv(\theta_1,\dots,\theta_N)$, its noise is additive and the spurious drift hence vanishes: $\bs_{(\varepsilon)}=0$.
Furthermore, choosing the Lebesgue measure $\blambda = \rmd \theta_1\dots\rmd \theta_N$ on $\mbT_N$ gives, in this coordinate system, a vanishing $\blambda$-spurious drift, $\bh_{\blambda}=0$, and a $\blambda$-drift $\ba=(a^i)_{i=1\dots N}=(\omega + \sum_j K_{ij}\sin(\theta_i-\theta_j))_{i=1\dots N}$

We consider the class of EMT-transformations corresponding to the mirror map $\mkm(\btheta)\equiv m \btheta$ together with the extension maps $\mke_\bD=id$ and $\mke(\ba)=(\alpha\omega + \sum_j (\beta K_{ij}^S + \gamma K_{ij}^A)\sin(\theta_i-\theta_j))_{i=1\dots N}$, with $m,\alpha,\beta,\gamma \in\{-1,1\}$.
Taking each of the parameters $m, \alpha, \beta, \gamma$ to $-1$ amounts to respectively reversing all the phases, the natural frequencies, the symmetric, and the antisymmetric coupling constants.

The mirror map leaves the gauge measure $\blambda$ invariant, $\mkm_*\blambda=\blambda$. In particular, this implies that $\mcA^{+/-}_{\blambda} = \ba^{S/A}_{\blambda}$.
Further, the diffusion tensor of dynamics~\eqref{Kuramoto_dynamics}, $\bD = D \bI_N$, is invariant under any of these EMT-transformations since $\mkm_*\mke(\bD) = m^2 \bD = \bD$. The Onsager-Casimir $\mke\mkm\mkt$-reversibility condition~\eqref{manifoldEMTrev-D-04} is thus satisfied, whatever the values of $m,\alpha,\beta,\gamma \in\{-1,1\}$.
As for the drift, its pushforward by $\mkm$ reads
\begin{equation}
[\mkm_*\ba(\btheta)]_i=m a_i(m\btheta) = m\omega + \sum_jK_{ij}\sin(\theta_i-\theta_j) \ ,
\end{equation}
and hence 
\begin{equation}
[\mkm_*\mke\ba (\btheta)]_i = m\alpha \omega + \sum_j  (\beta K^S_{ij} + \gamma K^A_{ij}) \sin(\theta_i-\theta_j) \ .
\end{equation}
Consequently, the symmetric and skew-symmetric parts of $\ba$ respectively read 
\begin{eqnarray}
[\ba^S]^i(\btheta) = \frac{1+m\alpha}{2}\omega + \sum_j \left( \frac{1+\beta}{2}K^S_{ij} + \frac{1+\gamma}{2}K^A_{ij} \right)\sin(\theta_i -\theta_j)
\end{eqnarray}
and 
\begin{eqnarray}
[\ba^A]^i(\btheta) = \frac{1-m\alpha}{2}\omega + \sum_j \left( \frac{1-\beta}{2}K^S_{ij} + \frac{1-\gamma}{2}K^A_{ij} \right)\sin(\theta_i -\theta_j) \ .
\end{eqnarray}

The second $\mke\mkm\mkt$-reversibility condition requires the one-form $\bD^{-1}\ba^S$ to be exact. Let us first compute its exterior derivative:
\begin{eqnarray}
\rmd (\bD^{-1}\ba^S) &=& D^{-1} \sum_{i,j} \left( \frac{1+\beta}{2}K^S_{ij} + \frac{1+\gamma}{2}K^A_{ij} \right) \cos(\theta_i-\theta_j) \rmd \theta_i \wedge \rmd \theta_j \notag \\ 
&=& D^{-1}\sum_{i<j} \left[ \left( \frac{1+\beta}{2}K^S_{ij} + \frac{1+\gamma}{2}K^A_{ij} \right)\cos(\theta_i-\theta_j) - \left( \frac{1+\beta}{2}K^S_{ji} + \frac{1+\gamma}{2}K^A_{ji} \right)\cos(\theta_j-\theta_i) \right] \rmd \theta_i\wedge\rmd\theta_j \notag \ ,
\end{eqnarray}
where we rearranged the sum to go from the first to the second line. Using the symmetry of the matrices $\bK^S$ and $\bK^A$ and of the cosine function, we finally get
\begin{equation}
\rmd (\bD^{-1}\ba^S) = D^{-1}(1+\gamma) \sum_{i<j} K^A_{ij}\cos(\theta_i-\theta_j) \rmd\theta_i\wedge\rmd\theta_j \ . \label{Kuramoto_ext_der} 
\end{equation}
From eq.~\eqref{Kuramoto_ext_der}, we conclude that $\mke\mkm\mkt$-reversibility within this class of EMT-transformations requires $\gamma=-1$ as soon as $\bK^A\neq 0$. The latter condition ensures the one-form $\bD^{-1}\ba^S$ to be closed, but in order to be exact, it must also satisfy the second part of condition~\eqref{manifoldEMTrev-a-04}.
The fist homology group of $\mbT_N$, $H_1(\mbT_N)$, being generated by the N loops $\mcC_i= \{(0,\dots,\theta_i,\dots,0) , \theta_i\in [0,2\pi[\}$, along which we have
\begin{eqnarray}
\int_{\mcC_i} \bD^{-1}a^S &=& \frac{1}{D}\int_0^{2\pi} \left[\frac{1+m\alpha}{2}\omega + \sum_j \left( \frac{1+\beta}{2}K^S_{ij} + \frac{1+\gamma}{2}K^A_{ij} \right)\sin(\theta_i -\theta_j)\right]\rmd \theta_i \\
&=& \frac{\pi(1+m\alpha)\omega}{D} \ ,
\end{eqnarray}
exactness also requires $m\alpha=-1$.

Let us now turn to the third reversibility condition~\eqref{manifoldEMTrev-div-04}. A straightforward calculation leads to:
\begin{eqnarray}
\Div_{\blambda}(\ba^A) + \ba^A\cdot \bD^{-1}\ba^S &=& \frac{1-\beta}{2}\sum_{i\neq j} K^S_{ij} \cos(\theta_i-\theta_j) + (1-m\alpha\gamma)\frac{\omega}{2D} \sum_{i,j}K^A_{ij} \sin(\theta_i-\theta_j) \\
&+& \frac{1}{D}\sum_{i,j,k} \left( \frac{1+\beta}{2}K^S_{ij} + \frac{1+\gamma}{2}K^A_{ij} \right) \left( \frac{1-\beta}{2}K^S_{ik} + \frac{1-\gamma}{2}K^A_{ik} \right) \sin(\theta_i-\theta_j)\sin(\theta_i-\theta_k)   \ . \notag
\end{eqnarray}
Injecting the previous conditions for $\mke\mkm\mkt$-reversibility, $\gamma=-1$ and $m\alpha=-1$, we get
\begin{eqnarray}
\Div_{\blambda}(\ba^A) + \ba^A\cdot \bD^{-1}\ba^S &=& \frac{1-\beta}{2}\sum_{i\neq j} K^S_{ij} \cos(\theta_i-\theta_j) \label{Kuramoto_last_obst}\\
&+& \frac{1+\beta}{2D}\sum_{i,j,k} K^S_{ij} \left( \frac{1-\beta}{2}K^S_{ik} + K^A_{ik} \right)\sin(\theta_i-\theta_j)\sin(\theta_i-\theta_k)   \ . \notag
\end{eqnarray}
If $\beta=-1$, the last obstruction to $\mke\mkm\mkt$-reversibility reads 
\begin{equation}
\Div_{\blambda}(\ba^A) + \ba^A\cdot \bD^{-1}\ba^S = 2\sum_{i< j} K^S_{ij} \cos(\theta_i-\theta_j) \ .\label{Kuramoto_last_obst_case1} 
\end{equation}

Since the sum on the right-hand side of eq.~\eqref{Kuramoto_last_obst_case1} involves $N(N-1)/2$ functions of only $N$ independent variables $\theta_i$, the proof that the right-hand side of eq.~\eqref{Kuramoto_last_obst_case1} is non-zero for all $\bK$ is not straightforward. Determining whether this term can vanish for some non-trivial choices of $\bK$ is an interesting problem that we will not investigate here. Nevertheless, in the generic case, we expect this term to be non-zero and consequently the $\mke\mkm\mkt$-map with $\gamma=m\alpha=\beta=-1$ not to be a symmetry of dynamics~\eqref{Kuramoto_dynamics}.

If we now consider the case $\beta=1$, we have 
\begin{equation}
\Div_{\blambda}(\ba^A) + \ba^A\cdot \bD^{-1}\ba^S = \frac{1}{D}\sum_{i,j,k} K^S_{ij} K^A_{ik} \sin(\theta_i-\theta_j)\sin(\theta_i-\theta_k)   \ . \label{Kuramoto_last_obst_case2} 
\end{equation}
Similarly to the case $\beta=-1$, we will not investigate the general case here, but still expect the right-hand of eq.~\eqref{Kuramoto_last_obst_case2} to be non-zero generically for $N\geq 3$.

Interestingly, when we only have $N=2$ coupled oscillators, the right-hand side of eeq.~\eqref{Kuramoto_last_obst_case2} can be re-written as 
 $\frac{1}{D}\sum_{ij}K_{ij}^AK_{ij}^S \sin(\theta_i-\theta_j)^2$, which vanishes as it is the contraction of a symmetric and a skew-symmetric matrices.
Hence, for $N=2$, the $\mke\mkm\mkt$-map\footnote{The singular ``map'' is here used rather that its plural version to emphasize that taking $m=-1$ and $\alpha=1$ has the same effect on dynamics~\eqref{Kuramoto_dynamics} than $m=1$ and $\alpha=-1$.} corresponding to $m\alpha=\gamma=-1$ and $\beta=1$ is a symmetry of dynamics~\eqref{Kuramoto_dynamics}. 
In particular, it follows that its stationary measure satisfies 
\begin{equation}
\rmd \ln \pss^{\blambda}= \bD^{-1}\ba^S = \frac{1}{D}\sum_{i,j=1}^2 K^S_{ij}\sin(\theta_i-\theta_j)\rmd \theta^i \ ,
\end{equation}
from which we conclude, in turn, that the stationary probability density is
\begin{equation}
\pss^{\blambda} = \frac{1}{Z}e^{-K_{12}^S\cos(\theta_1-\theta_2)/D} \ ,
\end{equation}
with $Z$ the normalizing constant. Similarly, we also deduce the stationary probability current 
\begin{equation}
\bJss^{\blambda} = \pss^{\blambda}\left[\omega + K_{12}^A\sin(\theta_1-\theta_2)\right]\begin{pmatrix} 1 \\ 1 \end{pmatrix} \ .
\end{equation} 
Using formula~\eqref{manifoldIEPR01}, a tedious but straightforward calculation gives the entropy production rate:
If $K_{12}^S = 0$,
\begin{equation}
\sigma = (1+m\alpha)\frac{\omega^2}{D} + \frac{(1+\gamma)(K_{12}^A)^2}{2D} \ ,
\end{equation}
while, if $K^S_{12}\neq 0$, 
\begin{equation}
\sigma = (1+m\alpha)\frac{\omega^2}{D} + \frac{\mcI_1\left(\frac{K_{12}^S}{D}\right)}{\mcI_0\left(\frac{K_{12}^S}{D}\right)}\left[(1+\gamma)\frac{(K^A_{12})^2}{K^S_{12}} + (1-\beta)K_{12}^S \right] \ .
\label{Kuramoto_IEPR_2oscill}
\end{equation}
As it involves the modified Bessel functions of the first kind $\mcI_n(x)$ of order $n=0,1$, the expression~\eqref{Kuramoto_IEPR_2oscill} of the IEPR displays a slightly more complicated analytical form than that of the previous examples, echoing the more complex multi-body dynamics of the current example.
Besides, note that $\sigma$ is continuous at $K_{12}^S=0$, since $\mcI_1(x)/\mcI_0(x)\to x/2$ as $x\to 0$. 
Finally, let us emphasize that the expression~\eqref{Kuramoto_IEPR_2oscill} of the entropy production of two non-reciprocally coupled oscillators is valid for any $\mke\mkm\mkt$-transformation we have considered in this section, \ie for any $m,\alpha,\beta,\gamma\in\{-1,1\}$, although we have been able to calculate it explicitly thanks to its cancellation for $m\alpha=\gamma=-\beta=-1$.

\subsection{Construction of the typical MT-reversible process on a simply connected manifold and a simple example}
\label{subsec:typicalMTreversibleProcess}

The purpose of this section differs from that of the previous ones~\ref{subsec:chiralAOUP} \&~\ref{subsec:ChiralABP2d}. We are not going to search for an EMT-symmetry of a precise dynamics and then deduce its stationary measure, but rather characterize the structure of the generic MT-reversible dynamics~\eqref{EDSmanifold} on a simply connected manifold $\mcM$. Once this objective has been achieved, we will give an example of such dynamics in $\mbR^2$.

\paragraph{Generic form of MT-reversible processes.}
Let us consider dynamics~\eqref{EDSmanifold} for a given pair $(\ba_{\blambda},\bD)$, the gauge being chosen as $\mkm$-invariant (we have seen at the end of section~\ref{subsubsec:gaugeTheoryIEPR} that this is always possible), together with a given smooth involution $\mkm$ on $\mcM$, and assume that the process is $\mkm\mkt$-reversible.

Upon denoting by $\bPss$ the stationary probability measure of dynamics~\eqref{EDSmanifold}, we show in appendix~\ref{app:MTreversal} that the stationary measure of its $\mkm\mkt$-dual process is $\mkm_*\bPss$. Therefore, since we are considering a case where dynamics~\eqref{EDSmanifold} is $\mkm\mkt$-reversible, its stationary probability satisfies $\mkm_*\bPss=\bPss$.
In the $\mkm$-invariant gauge $\blambda$, this implies that $\mkm^*\pss^{\blambda}=\pss^{\blambda}$.

Furthermore, dynamics~\eqref{EDSmanifold} being $\mkm\mkt$-reversible, it satisfies conditions~\eqref{manifoldEMTrev-D-03}-\eqref{manifoldEMTrev-div-03}. This first means that $\bD$ is $\mkm$-symmetric: $\mkm_*\bD=\bD$. Then, eq.~\eqref{manifoldEMTrev-a-03} reads $\mcA^+_{\blambda}=\ba^S_{\blambda}=\bD\cdot \rmd\ln\pss^{\blambda}$. 
Note that this implies that 
\begin{equation}
\mkm_*\ba^S_{\blambda}=\mkm_*[\bD\cdot\rmd\ln\pss^{\blambda}]=[\mkm_*\bD]\cdot\rmd\ln\mkm^*\pss^{\blambda}=\bD\cdot \rmd \ln \pss^{\blambda}=\ba^S_{\blambda} \ ,
\end{equation}
\ie $\ba^S_{\blambda}$ is $\mkm$-invariant, as it should be.
Moreover, condition~\eqref{TRS-div-cond-03} implies that there exists a skew-symmetric, contravariant, tensor field $\bT$ of order two such that 
\begin{equation}
\pss^{\blambda}\ba^A_{\blambda} = \Div_{\blambda} \bT \ .
\label{generalMT-aA-shape}
\end{equation}
But by definition $\ba^A$ has to be $\mkm$-skew-symmetric, \ie $\mkm_*\ba^A=-\ba^A$. Hence we must have 
$\mkm_*([\pss^{\blambda}]^{-1}\Div_{\blambda}\bT) =- [\pss^{\blambda}]^{-1}\Div_{\blambda}\bT$.
But $\mkm_*([\pss^{\blambda}]^{-1}\Div_{\blambda}\bT)=[\mkm^*\pss^{\blambda}]^{-1} \Div_{\blambda} \mkm_*\bT$ (see appendix~\ref{App:pushforwardDivergSkewTensor}). Since $\pss^{\blambda}$ is $\mkm$-invariant, the skew-symmetry of $\ba^A_{\blambda}$ with respect to the pushforward by $\mkm$ requires that $\Div_{\blambda} \mkm_*\bT = -\Div_{\blambda} \bT$, \ie $\Div_{\blambda} (\mkm_*\bT+\bT)=0$.
According to Hodge-de Rham theory, $\mcM$ being simply connected, this last equation implies the existence of a skew-symmetric contravariant tensor field $\bS$ of order three such that $\mkm_*\bT+\bT = \Div_{\blambda} \bS$. The tensor $\mkm_*\bT+\bT$ being $\mkm$-symmetric, the latter equation implies that $\mkm_*\Div_{\blambda}\bS=\Div_{\blambda}\bS$, \ie $\Div_{\blambda}\mkm_*\bS=\Div_{\blambda}\bS$. But changing $\bT$ by the divergence of a 3-tensor in eq.~\eqref{generalMT-aA-shape} does not affect $\ba^A_{\blambda}$, and hence the dynamics itself. Thus, we can replace $\bT$ by $\tilde{\bT}\equiv\bT-\Div_{\blambda}(\bS/2)$ without loss of generality. And one can easily show that $\mkm_*\tilde{\bT}=-\tilde{\bT}$, \ie the tensor $\bT$ satisfying eq.~\eqref{generalMT-aA-shape} can be chosen $\mkm$-skew-symmetric.
\\

We can now build the most general form of dynamics~\eqref{EDSmanifold} which is $\mkm\mkt$-reversible, for a given $\mkm$, in a gauge $\blambda$ that is $\mkm$-invariant.
Take a symmetric, contravariant, 2-tensor field $\widehat{\bD}$ and set $\bD\equiv (\widehat{\bD} + \mkm_*\widehat{\bD})/2$. It can easily be checked that $\bD$ is again symmetric (in its indices) and, by construction, $\mkm_*\bD=\bD$. Reciprocally, any symmetric and $\mkm$-invariant 2-tensor can be obtained in this way, since the map $\widehat{\bD}\to\bD$ is a linear projection and is hence surjective onto the space of desired tensors.

Then consider a real-valued function $\widehat{\psi}(\bx)$ such that $Z\equiv \int_{\mcM}\exp(-\psi)\rmd\blambda$ is finite, with $\psi\equiv (\widehat{\psi}+\mkm^*\widehat{\psi})/2$, and set $\pss^{\blambda}\equiv \exp(-\psi)/Z$ and $\ba^S\equiv \bD\cdot\rmd\ln\pss^{\blambda}$. Similarly to $\bD$, this procedure gives all the possible $\mkm$-symmetric (positive) densities $\pss^{\blambda}$.

Further, take a skew-symmetric contravariant tensor field $\widehat{\bT}$ of order two, set first $\bT \equiv (\widehat{\bT}-\mkm_*\widehat{\bT})/2$, and then $\ba^A_{\blambda}\equiv [\pss^{\blambda}]^{-1}\Div_{\blambda} \bT$.
Finally, define $\ba_{\blambda}\equiv \ba^S_{\blambda}+\ba^A_{\blambda}$. 
By construction, the process~\eqref{EDSmanifold} with the resulting $(\ba_{\blambda},\bD)$ is $\mkm\mkt$-reversible. Furthermore, thanks to the analysis conducted above, all $\mkm\mkt$-reversible process can be constructed in that way.
This conclusion, that we rapidly drew thanks to the results previously established in this paper, must presumably coincide with recent results obtained in~\cite{barp2021unifying} by different means.

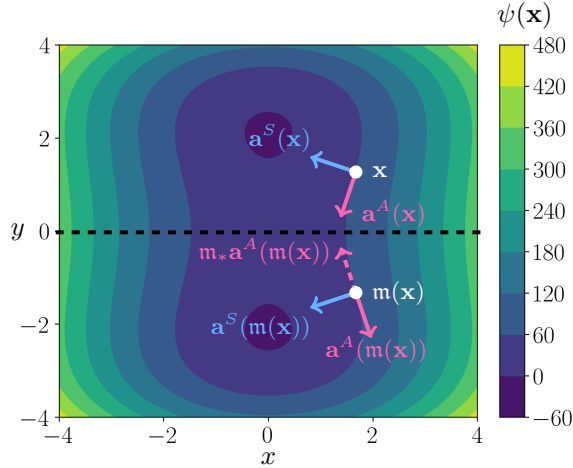
\begin{figure}[h!]
\begin{center}
\begin{tikzpicture}[scale=1]
\draw (0,0) node {\includegraphics[width=0.5\textwidth]{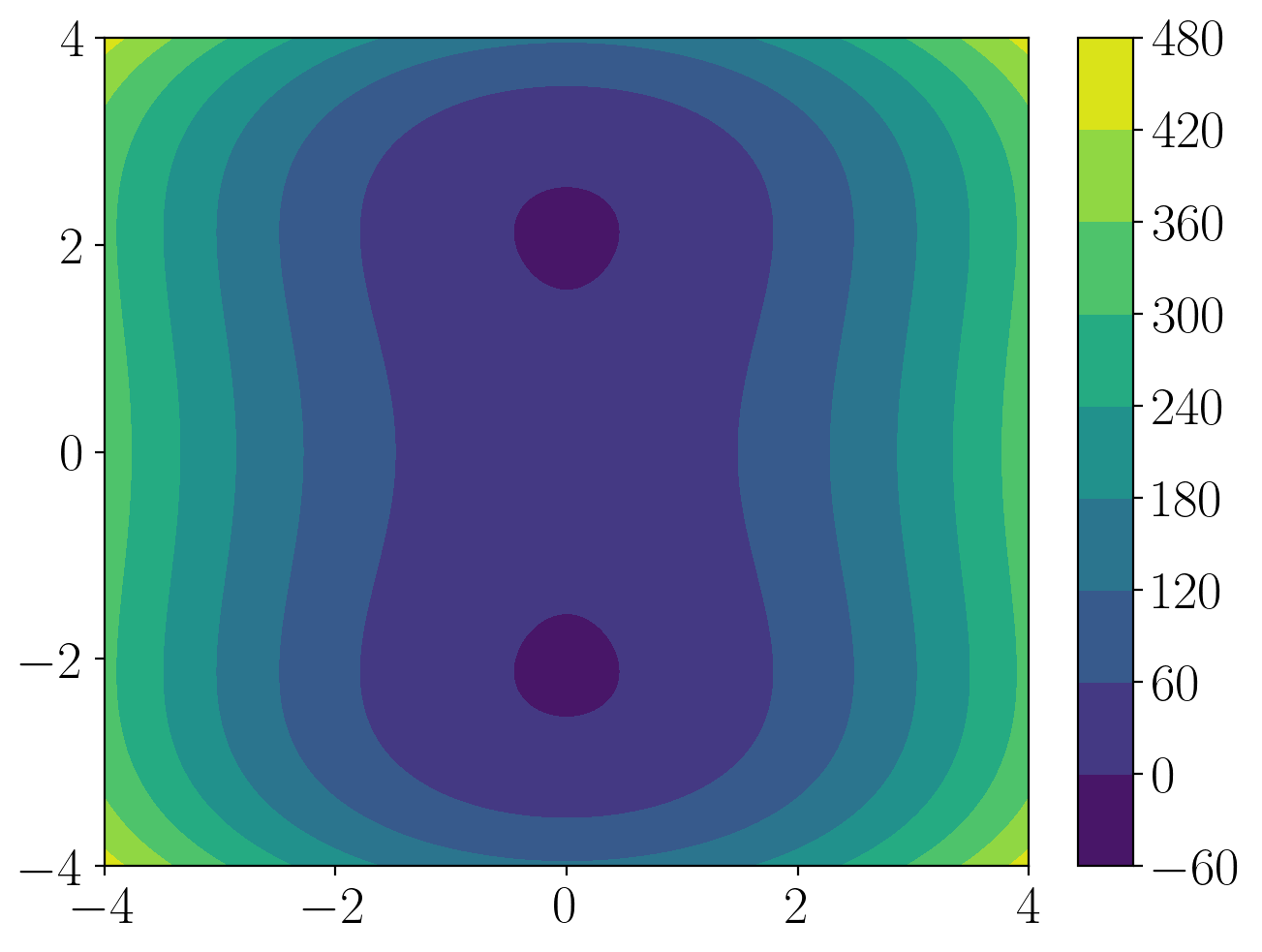}};

	\def\Ox{-0.375}
	\def\Oy{0.125}

\draw (3,3) node[inner sep=0.75pt] [align=left] {$\psi(\bx)$};
\draw (\Ox,\Oy-3) node[inner sep=0.75pt] [align=left] {$x$};
\draw (\Ox-3.3,\Oy) node[inner sep=0.75pt] [align=left] {$y$};

\def\Dx{2.8}
\def\Dy{0}
\draw[dashed,black,line width=1.5pt] (\Ox-\Dx,\Oy-\Dy)--(\Ox+\Dx,\Oy+\Dy);

	\def\x{1.15}
	\def\y{-0.8}

	\def\dx{-0.6}
	\def\dy{-0.2}
	
	\draw[->, bleuClair2, line width=1.5pt] (\Ox+\x,\Oy+\y)--(\Ox+\x+\dx,\Oy+\y+\dy);
	\draw (\Ox+\x+\dx-0.65,\Oy+\y+\dy-0.25) node{\small \textcolor{bleuClair2}{$\ba^S(\mkm(\bx))$}};
	\draw[->, monRose, line width=1.5pt] (\Ox+\x,\Oy+\y)--(\Ox+\x-\dy,\Oy+\y+\dx); 
	\draw (\Ox+\x-\dy+0.07,\Oy+\y+\dx-0.15) node{\small \textcolor{monRose}{$\ba^A(\mkm(\bx))$}};
	\draw[->, monRose, line width=1.5pt, densely dashed] (\Ox+\x,\Oy+\y)--(\Ox+\x+\dy,\Oy+\y-\dx); 
	\draw (\Ox+\x+\dy-1.,\Oy+\y-\dx-0.05) node{\small \textcolor{monRose}{$\mkm_*\ba^A(\mkm(\bx))$}};

	\draw (\Ox+\x,\Oy+\y)node [inner sep=0.75pt][circle,fill,inner sep=1.75pt,color=white][label = right:{\textcolor{white}{\small $\mkm(\bx)$}}]   [align=left] {};



	
	\draw[->, bleuClair2, line width=1.5pt] (\Ox+\x,\Oy-\y)--(\Ox+\x+\dx,\Oy-\y-\dy);
	\draw (\Ox+\x+\dx-0.4,\Oy-\y-\dy+0.25) node{\small \textcolor{bleuClair2}{$\ba^S(\bx)$}};
	\draw[->, monRose, line width=1.5pt] (\Ox+\x,\Oy-\y)--(\Ox+\x+\dy,\Oy-\y+\dx); 
	\draw (\Ox+\x+\dy+0.7,\Oy-\y+\dx+0.05) node{\small \textcolor{monRose}{$\ba^A(\bx)$}};

	\draw (\Ox+\x,\Oy-\y)node [inner sep=0.75pt][circle,fill,inner sep=1.75pt,color=white][label = right:{\textcolor{white}{\small $\bx$}}]   [align=left] {};


\end{tikzpicture}
\caption{\label{fig:ExampleMTreversible} 
Plot of the level sets of the map $\psi(x,y)= -y^2 + 20 x^2 + (y^2 - 4)^2$. We added two points $\bx$ and $\mkm(\bx)$ (the white disks), as well as the value of $\ba^S$ (respectively $\ba^A$) in blue (respectively in pink) at each of these points. We also drew the vector $\mkm_*\ba^A(\mkm(\bx))$ in dashed pink. The dashed, black, line corresponds to the axis of symmetry with respect to which $\mkm$ is a mirror transformation. We see that $\mkm^*\psi=\psi$, \ie the potential $\psi$ is symmetric with respect to this horizontal dashed line. Similarly $\mkm_*\ba^S(\mkm(\bx))=\ba^S(\mkm(\bx))$, \ie the two vectors are the mirror image one of the another with respect to the $x$ axis. On the contrary we see that $\mkm_*\ba^A(\mkm(\bx))=-\ba^A(\mkm(\bx))$.
}
\end{center}
\end{figure}

\paragraph{An illustration in $\mbR^2$.}
Let us consider $\mkm(x,y)=(x,-y)$. We choose the Lebesgue measure $\blambda\equiv\blambda_0$ as the gauge, which is indeed $\mkm$-invariant. Then we take $\bD=D\bI_2$, where $D$ is a constant and $\bI_2$ is the identity $2\times 2$ matrix, so that we clearly have $\mkm_*\bD = \bD$, \ie condition~\eqref{manifoldEMTrev-D-03} is satisfied.
Then we choose any function $\psi(\bx)$ such that $\psi(x,-y)=\psi(x,y)$. For instance, in figure~\ref{fig:ExampleMTreversible}, we took $\psi(\bx)=-y^2 + 20 x^2 + (y^2 - 4)^2$. We further set $\pss^{\blambda}\equiv \exp(-\psi)/Z$ and then  $\ba^S_{\blambda}\equiv D\nabla\ln\pss^{\blambda}$, $\nabla$ denoting the Euclidean gradient.
Since we are in dimension $d=2$, picking a field of skew-symmetric matrices $\bT(\bx)$ comes down to picking a real-valued function $\phi(\bx)$ and then setting
\begin{equation}
\bT(\bx) = \begin{pmatrix}0 & \phi(\bx) \\ -\phi(\bx) & 0 \end{pmatrix} \ .
\end{equation}
Let us choose $\phi=\pss^{\blambda}$. Denoting by $\bJ_\mkm$ the Jacobian matrix of $\mkm$ and $\bJ_\mkm^\top$ its transpose, we then have
\begin{eqnarray*}
\mkm_*\bT(\bx) &\equiv & \bJ_\mkm\bT(\mkm(\bx))\bJ_\mkm^\top \\
&=& \begin{pmatrix} 1 & 0 \\ 0 & -1  \end{pmatrix} \begin{pmatrix}0 & \pss(\mkm(\bx)) \\ -\pss(\mkm(\bx)) & 0 \end{pmatrix}   \begin{pmatrix} 1 & 0 \\ 0 & -1  \end{pmatrix} \\
&=& \begin{pmatrix}0 & -\pss(\mkm(\bx)) \\ \pss(\mkm(\bx)) & 0 \end{pmatrix} \\ 
&=& -\bT(\bx) \ ,
\end{eqnarray*}
where the last equality is due to $\pss(\mkm(\bx))= \pss(\bx)$ since it is the case for $\psi$.
Thus, upon setting
\begin{equation}
\ba^A \equiv \frac{1}{\pss^{\blambda}}\Div_{\blambda}\bT= \frac{1}{\pss^{\blambda}}\begin{pmatrix} -\partial_y\pss^{\blambda} \\ \partial_x\pss^{\blambda}  \end{pmatrix} =  \begin{pmatrix} \partial_y\psi \\ -\partial_x\psi  \end{pmatrix} \ ,
\end{equation}
we have that $\ba^S$ and $\ba^A$ are respectively $\mkm_*$- symmetric and skew-symmetric, and are such that the process with drift $\ba^S+\ba^A$ and diffusion tensor $\bD$ is $\mkm\mkt$-reversible.

We conclude this section by emphasizing that the explicit construction of an $\mkm\mkt$-reversible process \via $\bPss$ is of course not the usual task statistical physicists tackle. Normally, one starts from a given Langevin equation and tries to decide whether it satisfies an $\mkm\mkt$-symmetry for candidate choice(s) of $\mkm$. The results on chiral particles in sections~\ref{subsec:chiralAOUP} and~\ref{subsec:ChiralABP2d} exemplify how the results of this paper can also inform that task.

\newpage
\section{Conclusion}

In this article, we studied the EMT-reversal of a generic Markov process with Gaussian noise on an arbitrary smooth, oriented, connected manifold. For clarity, we first presented our results in $\mbR^d$ before obtaining their generalization to more general manifolds.

We started by showing that the description of such stochastic dynamics in a prescription-free and coordinate-free way requires the choice of a reference measure that plays the role of a gauge or a prior, with all our subsequent results about the EMT-reversals of this dynamics remaining gauge-invariant.

Then we rigorously constructed an extended notion of time-reversal, the EMT-reversal -- that had only been studied in a few specific cases previously -- and uncover the commutativity relation that needs to be satisfied for this time-reversal to be well defined. We further showed that any stochastic dynamics of the type under study actually possesses at least one EMT-symmetry and, if only such a symmetry can be found in an explicit manner, the stationary probability measure and current -- as well as the IEPR associated to other EMT-reversals -- can be obtained analytically.

In practice the construction of such a symmetry involves a search among candidate choices for the mirror operation $\mkm$ and the extension operation $\mke$. 
In screening such candidates it is crucial to have criteria that can be tested using knowledge of the Markov process alone without also requiring us to know its steady state measure. To this end we gave a triad of sufficient and necessary conditions for a stochastic dynamics to be EMT-reversible. These conditions can always be tested analytically, something which is without precedent in the literature, although individual instances may have been found before. We showed that all these results extend to the case of EMT-reversal on a manifold.
They involve: a symmetry of the diffusion tensor that is akin to the Onsager-Casimir symmetry; a constraint of exactness of a differential one-form; and the requirement that the resulting density -- given by the exponential of the potential of this exact one-form -- lies in the kernel of the Fokker-Planck operator.
When the first of these three conditions is satisfied, we show that the entropy production rate can be decomposed as the superposition of 
two independent terms, each one proportional to a quantity measuring the obstruction to fulfillment of one of the two remaining reversibility conditions.

Besides, we gave an interpretation of the irreversible cycle affinity as the vorticity of the symmetric part of the drift for a Riemannian geometry imposed by the diffusion tensor, and studied the relation between the EMT-reversibility of a stochastic process and the deterministic EMT-reversibility of its Stratonovitch average dynamics.

We then proved all the results presented in this article to be gauge invariant, and gave gauge-theoretical interpretations of our reversibility conditions.

Extending our expression of the entropy production in the absence of the assumption over the Onsager-like symmetry of the diffusion tensor could give a more complete correspondence between our reversibility conditions and the various independent sources of entropy production.
Furthermore, finding a better physical interpretation of the scalar field whose cancellation corresponds to the third reversibility condition -- similar to what we did for the irreversible cycle affinity as a vorticity -- remains an open problem whose solution could further improve our understanding of this scalar field and provide physical intuition on its possible manifestations.

We finally illustrated the general, abstract, results of this article on several concrete examples of increasing complexity. In particular, the last of these is that of a system of interacting agents, opening the way for the companion paper of the present article.
Indeed, in a separate paper (part II of this pair), we shall extend these results to infinite dimensional systems and show how it clarifies various properties of out-of-equilibrium stochastic field theories -- which are often a coarse-grained description of multi-body systems --, along lines that parallel those given here for the finite dimensional case. These developments create a new tool set for the classification and study of such theories which are ubiquitous in particular in the description of active matter~\cite{cates2022stochastic,fodor2022irreversibility,marchetti2013hydrodynamics}.

\section*{Acknowledgments}
The authors thank Ronojoy Adhikari and Bal\'asz N\'emeth for their careful comments on the manuscript.
Work funded in part by the European Research Council under the Horizon 2020 Programme, ERC Grant Agreement No. 740269.

\newpage
\appendix

\section{Some properties of the MT-dual process}
\label{app:MTreversal}

In this appendix we obtain the generator of the $\mkm\mkt$-dual process of dynamics~\eqref{EDSmanifold} on a smooth manifold $\mcM$ without boundary. 
In order to lighten the notations, we will rather write the path probability measure $\mkT\mcP$, two-point measure $\mkT \bP_2$, and stationary probability measure $ \mkT\bPss$ of the $\mkm\mkt$-dual process as $\reversed{\mcP}, \reversed{\bP}_2$, and $\reversed{\bP}_{\rm ss}$, respectively, although this notation is reserved for the EMT-dual process in the main text.
Similarly, for any gauge measure $\blambda$, we will denote by $\reversed{p_2}^{\blambda}$, $\reversed{p}_{\rm ss}^{\blambda}$, and $\reversed{\mcW}_{\blambda}$ respectively the two-point probability $\blambda$-density, stationary probability $\blambda$-density and the Fokker-Planck operator in the gauge $\blambda$. When there will be no ambiguity, we will not explicitly write down the subscript or superscript $\blambda$ in the latter quantities. 
\\

We start by choosing a reference measure $\bmu$ that is $\mkm$-invariant.

\subsection{MT-dual two-point and one-point measures}

For any pairs of points $\bx,\by\in\mcM$ and times $\tau \leq t \in \mbT$, the two-point probability measure\footnote{Note that, to lighten notations, we often write $\bP(\bx)$ the value of a measure $\bP$ evaluated on an infinitesimal volume around a point $\bx$, instead of something like $\bP(\bx\pm \Delta \bx)$, which would be less misleading but more cumbersome.} $\reversed{\bP}_2(x,t,y,\tau)$ is obtained by integrating $\bar{\mcP}[(\bx_t)_{t\in\mbT}]=\mcP[(\mkm(\bx_{\mcT-t}))_{t\in\mbT}]$ over all trajectories that pass in the vicinity of $y$ at $\tau$ and $\bx$ at $t$. This gives the relation
\begin{equation}
\reversed{\bP}_2(\bx,t,\by,\tau) = \bP_2 (\mkm(\by),\mcT-\tau,\mkm(\bx),\mcT-t) \ .
\end{equation}
Since the gauge $\bmu$ is $\mkm$-invariant, the same relation holds for the two-point probability $\bmu$-density
\begin{equation}
\reversed{p}_2^{\bmu}(\bx,t,\by,\tau) = p_2^{\bmu} (\mkm(\by),\mcT-\tau,\mkm(\bx),\mcT-t) \ .
\label{eq:proba2ptsDual}
\end{equation}
Similarly, assuming we are in steady state, the $\mkm\mkt$-dual stationary measure $\reversed{\bP}_{\rm ss}$ around a point $\bx\in\mcM$ is obtained by integrating the relation $\bar{\mcP}[(\bx_t)_{t\in\mbT}]=\mcP[(\mkm(\bx_{\mcT-t}))_{t\in\mbT}]$ on all paths that start in the vicinity of $\bx$: $\reversed{\bP}_{\rm ss}(\bx)=\bPss(\mkm(\bx))$, \ie
\begin{equation}
\reversed{\bP}_{\rm ss} = \mkm_*\bPss \ ,
\end{equation}
where $\mkm_*\bPss$ here denotes the ``pushforward measure'' of $\bPss$ by $\mkm$, whose coordinate expression reads
\begin{equation}
[\mkm_*\bPss] = [\mkm_*\Pss] (\bx)\rmd \bx = \Pss(\mkm(\bx)) \left|\mathrm{det} \left[(\partial_j\mkm^i)(\bx)\right] \right|\rmd \bx \ .
\end{equation}
Again thanks to the $\mkm$-invariance of the gauge $\bmu$, we get a similar relation for the stationary probability $\bmu$-density: $\reversed{p}_{\rm ss}^{\bmu}(\bx)=\pss^{\bmu}(\mkm(\bx))$, \ie
\begin{equation}
\reversed{p}_{\rm ss}^{\bmu}=\mkm^*\pss^{\bmu} \ ,
\end{equation}
where $\mkm^*$ here denotes the ``pullback of function'' which is simply defined as $\mkm^*\pss^{\bmu}(\bx)\equiv \pss^{\bmu}(\mkm(\bx))$.

\subsection{MT-dual Fokker-Planck operator in an $\mkm$-invariant gauge}
\label{App:AdjointFormel}
In order to determine, in the $\mkm$-invariant gauge $\bmu$, the explicit expression of the Fokker-Planck operator $\reversed{\mcW}_{\bmu}$ of the $\mkm\mkt$-dual process, let us start by showing that it formally reads:
\begin{equation}
\reversed{\mcW}_{\bmu}=\mkm^* p_{\rm ss}^{\bmu} \mcW^{\dagger} (p_{\rm ss}^{\bmu})^{-1} \mkm^*
\label{eq:OpAdFormel}
\end{equation}
where $\mcW^\dagger$, the $\mathbb{L}^2(\bmu)$-adjoint operator of $\mcW_{\bmu}$, is the generator of the dynamics and is independent of $\bmu$. In order to lighten notations, we are going to forget for a moment the subscript or superscript $\bmu$, for all the objects related to $\bmu$.

In order to prove eq.~\eqref{eq:OpAdFormel}, we use the fact that the so called propagator $p(\bx,t|\by,\tau)$ (the probability $\bmu$-density for the solution $\bx_t$ of dynamics~\eqref{EDSmanifold} to be at $\bx$ at time t, given that it was at $\by$ at time $\tau\leq t$) satisfies the backward Kolmogorov equation: 
\begin{equation}
\partial_{\tau} p(\bx,t|\by,\tau) = -\mcW^{\dagger}_\by \, p(\bx,t|\by,\tau) \ ,
\label{eq:KolmogorovBackward}
\end{equation}
where the subscript of $\mcW^{\dagger}$ now momentarily indicates the variable on which the operator acts,
while the propagator of the $\mkm\mkt$-dual process satisfies the forward Kolmogorov equation (\ie Fokker-Planck equation): 
\begin{equation}
\partial_t \reversed{p}(\bx,t|\by,\tau) = \reversed{\mcW}_\bx \, \reversed{p}(\bx,t|\by,\tau) \ .
\label{eq:KolmogorovForward}
\end{equation}
Combining eq.~\eqref{eq:KolmogorovBackward} with the relation $p_2(\bx,t,\by,\tau)=p(\bx,t|\by,\tau)\pss(\by)$, we get:
\begin{eqnarray*}
\partial_t \reversed{p}(\bx,t|\by,\tau) 
&=& \partial_t \reversed{p}_2(\bx,t,\by,\tau)\reversed{p}_{\rm ss}^{-1}(\by)\\
&=& \reversed{p}_{\rm ss}^{-1}(\by) \partial_t p_2(\mkm(\by),T-\tau,\mkm(\bx),T-t)\\
&=& -\reversed{p}_{\rm ss}^{-1}(\by) \frac{\partial}{\partial (T-t)} p(\mkm(\by),T-\tau|\mkm(\bx),T-t)p_{\rm ss}(\mkm(\bx))\\
&=& \reversed{p}_{\rm ss}^{-1}(\by)p_{\rm ss}(\mkm(\bx)) \mcW^{\dagger}_{\mkm(\bx)} \, p(\mkm(\by),T-\tau|\mkm(\bx),T-t)\\
&=& \reversed{p}_{\rm ss}^{-1}(\by)p_{\rm ss}(\mkm(\bx)) \mcW^{\dagger}_{\mkm(\bx)} \, p_2(\mkm(\by),T-\tau,\mkm(\bx),T-t)p_{\rm ss}^{-1}(\mkm(\bx))\\
&=& \reversed{p}_{\rm ss}^{-1}(\by)p_{\rm ss}(\mkm(\bx)) \mcW^{\dagger}_{\mkm(\bx)}\, p_{\rm ss}^{-1}(\mkm(\bx))\reversed{p}_2(\bx,t,\by,\tau)\\
&=& p_{\rm ss}(\mkm(\bx)) \mcW^{\dagger}_{\mkm(\bx)}\, p_{\rm ss}^{-1}(\mkm(\bx))\reversed{p}(\bx,t|\by,\tau)\\
&=& \mkm^* p_{\rm ss}(\bx) \mcW^{\dagger}_{\bx}\, p_{\rm ss}^{-1}(\bx) \mkm^* \reversed{p}(\bx,t|\by,\tau) \ .
\end{eqnarray*}
Finally, this last equality, together with eq.~\eqref{eq:KolmogorovForward}, leads to the result~\eqref{eq:OpAdFormel} that we wanted to prove.
\\

Let us now show that, in the $\mkm$-invariant gauge $\bmu=\mu(\bx)\rmd x^1\dots\rmd x^d$, the Fokker-Planck operator $\reversed{\mcW}_{\bmu}$ of the MT-reversed dynamics applied to a $\bmu$-density $p^{\bmu}$ reads
\begin{equation}
\reversed{\mcW}_{\bmu}p^{\bmu} = -\Div_{\bmu} \Big[\Big( \mkm_*\left[ 2\bD \cdot \rmd \ln\pss^{\bmu} -\ba_{\bmu} \right] \Big)p^{\bmu} -[\mkm_*\bD]\cdot \rmd p^{\bmu}\Big] \ ,
\label{eq:OpAdExplicit}
\end{equation}
where the definitions $\mkm_*$ (here the pushforward of contravariant tensor fields) and $\mkm^*$ (the pullback of real-valued functions) are given in section~\ref{subsec:MT-dual}. To this end, let us pick a smooth map $f:\mcM\to\mbR$, and compute the effect of $\reversed{\mcW}^\dagger$ on it. Taking the $\mbL^2(\bmu)$-adjoint of expression~\eqref{eq:OpAdFormel} of $\reversed{\mcW}$, and using the fact that $(\mkm^*)^\dagger=\mkm^*$ since $\bmu$ is $\mkm$-invariant, we get
\begin{eqnarray}
\reversed{\mcW}^\dagger f &=& -\mkm^* \pss^{-1}\mu^{-1}\partial_i \mu \left[ a^i \pss \mkm^* f - D^{ij}\partial_j \pss \mkm^* f \right] \\
&=& -\mkm^*\pss^{-1} (\mkm^*f) \mu^{-1}\partial_i \mu \left[a^i \pss -D^{ij}\partial_j \pss \right] \label{uneLigne}\\
& & -\mkm^*\pss^{-1} \left[ a^i\pss-D^{ij}\partial_j \pss \right] \partial_i \mkm^*f \notag\\
 & & + \mkm^*\pss^{-1} \mu^{-1}\partial_i\mu \left[ \pss D^{ij} \partial_j \mkm^* f \right] \notag \ ,
\end{eqnarray}
where we used the chain rule to go from the first to the second equality. Now, since $\pss$ is the stationary $\bmu$-density, the first line of eq.~\eqref{uneLigne} vanishes and, upon using the chain rule again on the last line, we get
\begin{equation}
\reversed{\mcW}^\dagger f = -\mkm^*\left[ a^i-2D^{ij}\partial_j \ln\pss \right] \partial_i \mkm^*f  + \mkm^* \mu^{-1}\partial_i\mu \left[ D^{ij} \partial_j \mkm^* f \right] \ .
\label{app:mcWDaggerFtemp}
\end{equation}
Then, by using the definition of the pullback and pushforward (given in section~\ref{subsec:MT-dual}), the first term on the right-hand side of eq.~\eqref{app:mcWDaggerFtemp} can be re-written as
\begin{eqnarray*}
\left\{-\mkm^*\left[ a^i-2D^{ij}\partial_j \ln\pss \right] \partial_i \mkm^*f\right\}(\bx) &=& \mkm^*\left\{\left[2D^{ij}\partial_j \ln\pss - a^i\right](\bx) \frac{\partial\mkm^k}{\partial x^i}(\bx) \frac{\partial f}{\partial x^k}(\mkm(\bx)) \right\} \quad \\
&=& \frac{\partial \mkm^k}{\partial x^i} (\mkm(\bx))\left[2D^{ij}\partial_j \ln\pss - a^i\right](\mkm(\bx)) \frac{\partial f}{\partial x^k}(\bx) \\
&=& \left( \mkm_*\left[ 2\bD \cdot \rmd \ln\pss -\ba \right]\right) (\bx) \cdot \rmd f (\bx) \ .
\end{eqnarray*}
Then, $\mkm$ being an involution, we have that $\frac{\partial\mkm^i}{\partial x^n}(\mkm(\bx))\frac{\partial\mkm^n}{\partial x^\ell}(\bx)=\delta^i_\ell$, which implies that the term into the squared brackets in the last term on the right-hand side of eq.~\eqref{app:mcWDaggerFtemp} reads
\begin{eqnarray*}
D^{ij}(\bx)\frac{\partial \mkm^k}{\partial x^j}(\bx)\frac{\partial f}{\partial x^k}(\mkm(\bx)) &=& D^{\ell j}(\bx)\frac{\partial\mkm^i}{\partial x^n}(\mkm(\bx))\frac{\partial\mkm^n}{\partial x^\ell}(\bx) \frac{\partial \mkm^k}{\partial x^j}(\bx)\frac{\partial f}{\partial x^k}(\mkm(\bx)) \\ 
&=& \frac{\partial\mkm^i}{\partial x^n}(\mkm(\bx))[\mkm_*\bD]^{nk}(\mkm(\bx))\frac{\partial f}{\partial x^k}(\mkm(\bx)) \\
&=& \left[ \mkm_*\left([\mkm_*\bD]\cdot \rmd f\right)\right]^i(\bx) \ .
\end{eqnarray*}
The last term of eq.~\eqref{app:mcWDaggerFtemp} can then be re-written as
\begin{equation}
\mkm^* \mu^{-1}\partial_i\mu \left[ D^{ij} \partial_j \mkm^* f \right] = \mkm^* \Div_{\bmu} \left[ \mkm_*\left([\mkm_*\bD]\cdot \rmd f\right)\right] \ .
\end{equation}
Further, we prove in appendix~\ref{App:pushforwardDivergSkewTensor} that the pushforward and the divergence operator commutes (on skew-symmetric tensors), \ie $\Div_{\bmu} \left[ \mkm_*\left([\mkm_*\bD]\cdot \rmd f\right)\right]=\mkm_*\Div_{\bmu} \left([\mkm_*\bD]\cdot \rmd f\right)$. But $\Div_{\bmu} \left([\mkm_*\bD]\cdot \rmd f\right)$ being only a scalar, its pushforward by the involution $\mkm$ coincides with its pullback, which means that 
\begin{equation}
\mkm^* \Div_{\bmu} \left[ \mkm_*\left([\mkm_*\bD]\cdot \rmd f\right)\right] = \mkm^* \mkm^*\Div_{\bmu} \left([\mkm_*\bD]\cdot \rmd f\right)= \Div_{\bmu} \left([\mkm_*\bD]\cdot \rmd f\right) \ ,
\end{equation}
where we used the fact that $\mkm$ is involutive.
In turn, the adjoint of $\reversed{\mcW}$ applied to $f$ reads
\begin{equation}
\reversed{\mcW}^\dagger f = \left( \mkm_*\left[ 2\bD \cdot \rmd \ln\pss -\ba \right]\right) \cdot \rmd f  + \Div_{\bmu} \left([\mkm_*\bD]\cdot \rmd f\right) \ .
\end{equation}
Finally, upon taking the adjoint of this last equation by the $\mbL^2(\bmu)$-scalar product (for which $\rmd$ and $\Div_{\bmu}$ are each other's adjoints), we get the Fokker-Planck operator~\eqref{eq:OpAdExplicit} of the MT-reversed process of~\eqref{EDSmanifold} in the $\mkm$-invariant gauge $\bmu$.

\subsection{MT-dual Fokker-Planck operator: arbitrary gauge}
\label{App:AdjointExplicit}

Let us now determine the Fokker-Planck operator $\reversed{\mcW}_{\blambda}$ of the MT-reversed process of dynamics~\eqref{EDSmanifold} in an arbitrary gauge $\blambda$. To this end, we are simply going to start from the invariant gauge $\bmu$ and make a gauge transformation $\bmu\to\blambda$, as explained in section~\ref{subsec:gaugeTheory}.
In section~\ref{subsec:gaugeTheory}, we show that such a gauge transformation changes the probability density and the drift respectively as $p^{\bmu}\to p^{\lambda}=p^{\bmu}\mu/\lambda$ and $\ba_{\bmu}\to\ba_{\blambda}=\ba_{\bmu}+\bD\cdot \rmd \ln \frac{\mu}{\lambda}$. To get the Fokker-Planck equation of the MT-dual in this new gauge, we just multiply that in the initial gauge $\bmu$ by $\mu/\lambda$, the latter being given by eq.~\eqref{eq:OpAdExplicit}. This gives
\begin{eqnarray}
\partial_t p^{\blambda}_t = -\frac{\mu}{\lambda}\mu^{-1}\partial_i \mu \left\{ \left[\mkm_*\left(2\bD\cdot\rmd\ln \pss^{\blambda}\frac{\lambda}{\mu} - \ba_{\blambda} -\bD\cdot \rmd \ln \frac{\lambda}{\mu}\right)\right]^i p_t^{\blambda}\frac{\lambda}{\mu} - [\mkm_*\bD]^{ij}\partial_j p^{\blambda}_t \frac{\lambda}{\mu} \right\}  \ .
\end{eqnarray}
Then factorizing by $\lambda/\mu$ in the curly bracket, using the chain rule and the fundamental property of the logarithm, we get
\begin{eqnarray}
\partial_t p^{\blambda}_t = -\Div_{\blambda} \left\{ \mkm_*\left[2\bD\cdot\rmd\ln \pss^{\blambda} - \ba_{\blambda} +\bD\cdot \rmd \ln \frac{\lambda}{\mu}\right] p_t^{\blambda} - \left( [\mkm_*\bD]\cdot \rmd \ln \frac{\lambda}{\mu}\right) p^{\blambda}_t - [\mkm_*\bD]\cdot\rmd p^{\blambda}_t  \right\}  \ .
\end{eqnarray}
Finally, using the $\mkm$-invariance of $\bmu$, together with the fact that $\mkm_*(\bD\cdot \rmd \ln \frac{\mkm_*\lambda}{\mkm_*\mu})=[\mkm_*\bD]\cdot\rmd\ln \frac{\lambda}{\mu}$, we get
\begin{eqnarray}
\partial_t p^{\blambda}_t = -\Div_{\blambda} \left\{ \mkm_*\left[2\bD\cdot\rmd\ln \pss^{\blambda} - \ba_{\blambda} -\bD\cdot \rmd \ln \frac{\mkm_*\lambda}{\lambda}\right] p_t^{\blambda} - [\mkm_*\bD]\cdot\rmd p^{\blambda}_t  \right\}  \ ,
\end{eqnarray}
In turn, this allows to read out the $\blambda$-drift and diffusion tensor of the MT-dual of dynamics~\eqref{EDSmanifold}, as given in section~\eqref{subsec:manifolds}.

\section{Entropy production rate with respect to an EMT-reversal}
\label{app:IEPR}

In this appendix, we compute the entropy production rate of dynamics~\eqref{EDSmanifold} with respect to a given $\mke\mkm\mkt$-reversal. In particular, we assume that the reversibility condition~\eqref{manifoldEMTrev-D-01} is satisfied, \ie that 
\begin{equation}
\mkm_*\mke(\bD)=\bD \ ,
\label{eq:OnsagerSym}
\end{equation}
and that the tensor field $\bD$ is everywhere invertible. As discussed in the main text, this implies that we can take $\bD^{-1}$ as a Riemannian metric on $\mcM$. Let us denote by $\bnu$ the corresponding Riemannian volume measure.

The (Stratonovitch) Onsager-Machlup action of dynamics~\eqref{EDSmanifold} is then~\cite{cates2022stochastic,cugliandolo2017rules,de2023path,lau2007state}:
\begin{equation}
\mcS[(\bx_t)_{t\in[0,\mcT]}]= \int_0^\mcT\rmd t  \left\{ \frac{1}{4}[\dot{\bx}_t - \ba_{\bnu}]\cdot\bD^{-1}[\dot{\bx}_t - \ba_{\bnu}] + \frac{1}{2}\Div_{\bnu} (\ba_{\bnu}) + f(\bx_t,[\bD]) \right\} \ ,
\end{equation}
where $f(\bx,[\bD])$ is a function of the $D^{ij}$ and their derivatives (up to the second order) at $\bx$ but does not depend on $\dot{\bx}$ nor $\ba$. Similarly, the action of the $\mke\mkm\mkt$-dual process reads
\begin{equation}
\reversed{\mcS}[(\bx_t)_{t\in[0,\mcT]}]= \int_0^\mcT\rmd t  \left\{ \frac{1}{4}[\dot{\bx}_t - \reversed{\ba}_{\bnu}]\cdot\bD^{-1}[\dot{\bx}_t - \reversed{\ba}_{\bnu}] + \frac{1}{2}\Div_{\bnu} (\reversed{\ba}_{\bnu}) + f(\bx_t,[\bD]) \right\} \ ,
\end{equation}
where $\reversed{\ba}_{\bnu}$ is the $\bnu$-drift of the dual process, whose expression is given by eq.~\eqref{manifoldEMTrev-a-01} (or alternatively by eq.~\eqref{manifoldEMTrev-a-02}), and we used that $\reversed{\bD}=\mkm_*\mke(\bD)=\bD$ thanks to eqs.~\eqref{manifoldEMTrev-D-02} \&~\eqref{eq:OnsagerSym}.
Using the symmetric and skew-symmetric parts of $\ba_{\bnu}$ \textit{with respect to the $\mke\mkm\mkt$-reversal}, as defined in section~\ref{subsec:firstEMTcond}, a straightforward computation gives that
\begin{equation}
\reversed{\mcS}-\mcS = \int_0^\mcT \left[ {}^A\ba_{\bnu} \cdot \bD^{-1}\dot{\bx}_t	 - \Div_{\bnu}({}^A\ba_{\bnu}) - {}^A\ba_{\bnu}\cdot\bD^{-1}{}^S\ba_{\bnu} \right] \rmd t \ .
\label{appEq:entropy01}
\end{equation}

Let us now show that eq.~\eqref{appEq:entropy01} is actually gauge-invariant. From the way $\ba_{\bnu}$ and $\reversed{\ba}_{\bnu}$ transform upon a gauge change $\bnu\to\blambda$ (see section~\ref{subsec:gaugeTheory}) and using the hypothesis $\mkm_*\mke(\bD)=\bD$, it can easily be shown that 
\begin{eqnarray}
{}^S\ba_{\bnu} &\to & {}^S\ba_{\blambda}={}^S\ba_{\bnu} + \bD\cdot \rmd \ln \frac{\nu}{\lambda} \ , \\
{}^A\ba_{\bnu} &\to & {}^A\ba_{\blambda}={}^A\ba_{\bnu} \ .
\end{eqnarray}
In particular, this implies that
\begin{equation}
\Div_{\bnu}({}^A\ba_{\bnu}) = \Div_{\bnu}({}^A\ba_{\blambda}) = \nu^{-1}\partial_i \left(\nu \lambda^{-1}\lambda \left[{}^Aa_{\blambda}\right]^i \right) = \Div_{\blambda}({}^A\ba_{\blambda}) + {}^A\ba_{\blambda}\cdot \rmd \ln \frac{\nu}{\lambda} \ ,
\end{equation}
where the last equality follows from the chain rule.
In turn, we can re-write eq.~\eqref{appEq:entropy01} in an arbitrary gauge $\blambda$ as
\begin{equation}
\reversed{\mcS}-\mcS = \int_0^\mcT \left[ {}^A\ba_{\blambda} \cdot \bD^{-1}\dot{\bx}_t - \Div_{\blambda}({}^A\ba_{\blambda}) - {}^A\ba_{\blambda}\cdot \rmd \ln \frac{\nu}{\lambda} - {}^A\ba_{\blambda}\cdot\bD^{-1}\left({}^S\ba_{\blambda}+\bD\cdot \rmd\ln\frac{\lambda}{\nu}\right) \right] \rmd t \ ,
\end{equation}
\ie 
\begin{equation}
\reversed{\mcS}-\mcS = \int_0^\mcT \left[ {}^A\ba_{\blambda} \cdot \bD^{-1}\dot{\bx}_t - \Div_{\blambda}({}^A\ba_{\blambda}) - {}^A\ba_{\blambda}\cdot\bD^{-1} {}^S\ba_{\blambda} \right] \rmd t \ ,
\label{appEq:entropy02}
\end{equation}
which proves the gauge-invariance of the expression~\eqref{appEq:entropy01} of $\reversed{\mcS}-\mcS$.
\\

Because the explicit knowledge of $\reversed{\ba}_{\blambda}$ requires that of the stationary measure, so do that of ${}^{S/A}\ba_{\blambda}$. We thus need to replace the latter by terms that can be obtained analytically from the specification of the Markov process without also knowing its stationary measure.

To do so, we first note that, from the definition of ${}^{S/A}\ba_{\blambda}$ and expression~\eqref{eq:EMTdualabis} of $\reversed{\ba}_{\blambda}$ (which also reads $\reversed{\ba}_{\blambda}=2\bD\cdot \ln \mkm^*(\pss^\mke)^{\blambda} -\mkm_*\mke(\ba_{\blambda})$), it directly follows that
\begin{eqnarray}
{}^S\ba_{\blambda} &=& \ba_{\blambda}^A + \frac{1}{2}\bD\cdot \rmd \ln \frac{\mkm_*\lambda}{\lambda} + \bD\cdot \rmd \ln \mkm^*(\pss^\mke)^{\blambda} \ , \label{EMTdecomp-MEdecomp-S-01}\\
{}^A\ba_{\blambda} &=& \ba_{\blambda}^S - \frac{1}{2}\bD\cdot \rmd \ln \frac{\mkm_*\lambda}{\lambda} - \bD\cdot \rmd \ln \mkm^*(\pss^\mke)^{\blambda} \ .
\label{EMTdecomp-MEdecomp-A-01}
\end{eqnarray}
In eqs.~\eqref{EMTdecomp-MEdecomp-S-01} \&~\eqref{EMTdecomp-MEdecomp-A-01}, $\ba_{\blambda}^S$ and $\ba_{\blambda}^A$ are respectively the symmetric and skew-symmetric parts of $\ba_{\blambda}$ under the map $\mkm_*\mke$, as defined in section~\ref{subsec:firstEMTcond}, and $(\pss^\mke)^{\blambda}$ is the $\blambda$-density of the stationary measure $\bPss^{\mke}$ of dynamics~\eqref{EDSmanifold} under replacement of $\ba_{\blambda}$ and $\bD$ by $\mke(\ba_{\blambda})$ and $\mke(\bD)$, respectively.

Besides, in appendix~\ref{app:MTreversal}, we showed that the MT-reversal of a process whose stationary measure is $\bPss$ has itself a stationary measure given by $\mkm_*\bPss$. Hence, since the $\mke\mkm\mkt$-reversal can be obtained by first applying the map $\mke$ and then the $\mkm\mkt$-reversal, we conclude that the stationary measure of the $\mke\mkm\mkt$-dual, denoted by $\reversed{\bP}_{\rm ss}$, reads $\reversed{\bP}_{\rm ss}=\mkm_*\bPss^\mke$. In turn, the $\blambda$-densities are related through 
\begin{equation}
\reversed{p}_{\rm ss}^{\blambda} = \frac{\mkm_*\lambda}{\lambda}\mkm^*(\pss^\mke)^{\blambda} \ .
\end{equation}
This allows to reformulate eqs.~\eqref{EMTdecomp-MEdecomp-S-01} \&~\eqref{EMTdecomp-MEdecomp-A-01} as
\begin{eqnarray}
{}^S\ba_{\blambda} &=& \ba_{\blambda}^A - \frac{1}{2}\bD\cdot \rmd \ln \frac{\mkm_*\lambda}{\lambda} + \bD\cdot \rmd \ln \reversed{p}_{\rm ss}^{\blambda} \ , \\
{}^A\ba_{\blambda} &=& \ba_{\blambda}^S + \frac{1}{2}\bD\cdot \rmd \ln \frac{\mkm_*\lambda}{\lambda} - \bD\cdot \rmd \ln \reversed{p}_{\rm ss}^{\blambda} \ ,
\end{eqnarray}
or alternatively, using the vector fields $\mcA^{\pm}_{\blambda}$ defined in section~\ref{subsec:manifolds}:
\begin{eqnarray}
{}^S\ba_{\blambda} &=&  \mcA^- + \bD\cdot \rmd \ln \reversed{p}_{\rm ss}^{\blambda} \ , \label{EMTdecomp-MEdecomp-S-02}\\
{}^A\ba_{\blambda} &=&  \mcA^+_{\blambda} -  \bD\cdot \rmd \ln \reversed{p}_{\rm ss}^{\blambda} \ .
\label{EMTdecomp-MEdecomp-A-02}
\end{eqnarray}

Injecting eqs.~\eqref{EMTdecomp-MEdecomp-S-02} \&~\eqref{EMTdecomp-MEdecomp-A-02} into eq.~\eqref{appEq:entropy02} and rearranging the terms gives
\begin{eqnarray}
\reversed{\mcS} -\mcS &=& - \ln \bbpssl (\bx_\mcT) + \ln \bbpssl (\bx_0) +\int_0^\mcT \rmd t \Big\{ \mcA^+_{\blambda}\cdot \bD^{-1} \dot{\bx} - \Div_{\blambda} \mcA^- - \mcA^-\cdot \bD^{-1}\mcA^+_{\blambda} \notag \\
& & + \ \Div_{\blambda}\left[\mcA^- - \mcA^+_{\blambda} + \bD\cdot \rmd \ln \bbpssl \right] + \left[\mcA^- - \mcA^+_{\blambda} + \bD\cdot \rmd \ln \bbpssl \right]\cdot \rmd \ln \bbpssl \Big\} \ .
\label{appEq:entropy03}
\end{eqnarray}
The last ingredient that we need follows from the fact that, by definition, the stationary probability current $\reversed{\bJ}^{\blambda}_{\rm ss}$ of the $\mke\mkm\mkt$-dual is divergence free: $\Div_{\blambda} \reversed{\bJ}^{\blambda}_{\rm ss}=0$. This also reads
\begin{eqnarray}
0= \Div_{\blambda} \left[ \left({}^S\ba -{}^A\ba_{\blambda}\right)\bbpssl - \bD\cdot \rmd \bbpssl \right] \ .
\end{eqnarray}
Factorizing $\bbpssl$ in the squared brackets, using relations~\eqref{EMTdecomp-MEdecomp-S-02} \&~\eqref{EMTdecomp-MEdecomp-A-02}, then using the chain rule and dividing by $\bbpssl$ gives
\begin{equation}
\Div_{\blambda}\left[\mcA^- - \mcA^+_{\blambda} + \bD\cdot \rmd \ln \bbpssl \right] + \left[\mcA^- - \mcA^+_{\blambda} + \bD\cdot \rmd \ln \bbpssl \right]\cdot \rmd \ln \bbpssl =0 \ .
\end{equation}
In turn, the second line in eq.~\eqref{appEq:entropy03} vanishes and the entropy production rate
reads 
\begin{equation}
\sigma \equiv \lim_{\mcT\to\infty} \frac{\reversed{\mcS}-\mcS}{\mcT} = \int_{\mcM} \bJss^{\blambda}\cdot \bD^{-1}\mcA^+_{\blambda} \rmd \blambda -\int_{\mcM} \pss^{\blambda}\left[\Div_{\blambda} \mcA^- + \mcA^-\cdot \bD^{-1}\mcA^+_{\blambda}\right] \rmd\blambda 
\end{equation}
where we used the ergodicity of the dynamics together with relation~\eqref{StratoAverageDyncs_intermed} below.

\section{Stratonovitch average dynamics}
\label{app:StratoAverageDyncs}

In this section, we show that, for any one form $\balpha$ on $\mcM$, its steady-state average against the (Stratonovitch) velocity $\dot{\bx}_t$ of dynamics~\eqref{EDSmanifold} is equal to the spatial integral of this observable against the stationary probability current $\bJss$:
\begin{equation}
\llangle \balpha(\bx_t)\cdot \dot{\bx}_t \rrangle = \int_{\mbR^d} \balpha(\bx)\cdot \bJss^{\blambda}(\bx) \rmd\blambda \ ,
\label{StratoAverageDyncs_intermed}
\end{equation}
where the left-hand side is defined as
\begin{eqnarray}
\llangle \balpha(\bx_t) \cdot \dot{\bx}_t \rrangle \equiv  \lim_{\Delta t\to 0} 
\llangle \alpha_k\left(\bx_t +\frac{\Delta \bx_t}{2}\right) \frac{\Delta x_t^k}{\Delta t} \rrangle \ ,
\end{eqnarray}
$\llangle \cdot \rrangle$ standing for the steady-state average.
But Taylor expanding $\alpha_k(\bx_t +\Delta\bx_t/2)$ around $\bx_t$ and using the Ito prescription in dynamics~\eqref{EDSmanifold} during a time increment $\Delta t$, we get:
\begin{eqnarray}
\llangle \alpha_k\left(\bx_t +\frac{\Delta \bx_t}{2}\right) \Delta x_t^k \rrangle = \llangle \Big[ \alpha_k(\bx_t) + \partial_i\alpha_k(\bx_t) \frac{\Delta x^i_t}{2} \Big] \Big[ A_{(0)}^k(\bx_t)\Delta t + B^{kj}(\bx_t)\Delta\eta^j_t \Big] \rrangle  + \mcO(\Delta t^{3/2}) \ ,
\end{eqnarray}
where $\bA_{(0)}$ is the Ito-total drift of dynamics~\eqref{EDSmanifold} (see the beginning of section~\ref{subsec:generalContext}).
Since the random increment $\Delta \boldeta_t$ between time $t$ and $t+\Delta t$ and $\bx_t$ are independent, and $\langle \Delta x^i_t \Delta\eta^\beta_t \rangle = \langle b^{i}_\gamma(\bx_t)\langle \Delta \eta^\gamma_t \Delta\eta^\beta_t  | \bx_t \rangle\rangle = 2 \Delta t \langle b^i_\beta\rangle $ (the notation $\langle\cdot|\bx_t\rangle$ here stands for the conditional average),  we have
\begin{eqnarray}
\llangle \alpha_k\left(\bx_t +\frac{\Delta \bx_t}{2}\right) \Delta x_t^k \rrangle =  \llangle \alpha_k(\bx_t) A^k(\bx_t) + b^i_\beta b^k_\beta\partial_i\alpha_k(\bx_t) \rrangle  \Delta t + \mcO(\Delta t^{3/2}) \ .
\end{eqnarray}
Finally, we divide each side by $\Delta t$ and take the limit $\Delta t\to 0$.
Writing the average explicitly and integrating by parts gives the required result~\eqref{StratoAverageDyncs_intermed}:
\begin{equation}
\llangle \balpha\left(\bx_t \right) \cdot \dot{\bx}_t \rrangle  = \int \balpha(\bx) \cdot \Big[p_{\rm ss}^{\blambda} \bA_{(0)}(\bx_t) - \Div_{\blambda} ( \bD p^{\blambda}_{\rm ss}) \Big] \lambda(\bx)\rmd \bx = \int \balpha(\bx) \cdot \bJss^{\blambda} (\bx)\rmd \blambda \ .
\end{equation}

\section{Exterior derivative and vorticity operator in curved space}
\label{app:vorticity_operator}

Let $(\mcM,g)$ be a Riemannian manifold, with Levi-Civita connection denoted by $\bnabla$. Let $\bu$ be a vector field over $\mcM$, with dual 1-form $\bu^\flat$.
In this appendix we prove that 
\begin{equation}
g(\bnabla^A_\bv \bu , \bw) = \frac{1}{2}\rmd \bu^\flat(\bv,\bw) \ ,
\label{VorticityOperator_step00}
\end{equation}
where $\bnabla^A\bu$ is the skew-symmetric part (with respect to $g$) of $\bnabla\bu$ seen as an operator from the space of vector fields to itself.
\\

Let $\balpha$ be a one-form and $\bv,\bw$ two vector fields over $\mcM$.
Let us first prove that
\begin{equation}
(\bnabla_\bv \balpha)(\bw) - (\bnabla_\bw \balpha)(\bv) = \rmd \balpha(\bv,\bw) \ .
\label{VorticityOperator_step01}
\end{equation}
By compatibility of the connection with respect tensor product, we  have
\begin{equation}
\bnabla_\bw(\balpha\otimes\bv) = \bnabla_\bw\balpha \otimes \bv + \balpha \otimes \bnabla_\bw \bv \ .
\end{equation}
Then the compatibility of $\bnabla$ with tensor contraction gives
\begin{equation}
\bnabla_\bw(\balpha(\bv)) = (\bnabla_\bw\balpha)(\bv) + \balpha (\bnabla_\bw\bv) \ .
\end{equation}
Since $\balpha(\bv)$ is just a function on $\mcM$, its covariant derivative in the direction of $\bw$ is simply its Lie derivative along $\bw$:  $\bnabla_\bw(\balpha(\bv)) = \bw(\balpha(\bv))$. The above equation thus reads
\begin{equation}
(\bnabla_\bw\balpha)(\bv) = \bw(\balpha(\bv)) -  \balpha (\bnabla_\bw\bv) \ .
\end{equation}
This implies that 
\begin{eqnarray}
(\bnabla_\bv \balpha)(\bw) - (\bnabla_\bw \balpha)(\bv) &=& \bv(\balpha(\bw)) - \bw(\balpha(\bv)) - \left( \balpha(\bnabla_\bv\bw) -  \balpha(\bnabla_\bw\bv) \right) \\
&=& \bv(\balpha(\bw)) - \bw(\balpha(\bv)) -  \balpha(\bnabla_\bv\bw-\bnabla_\bw\bv)
\end{eqnarray}
where the second line follows from the linearity of $\balpha$.
But the Levi-Civita connection is torsion-free, \ie $\bnabla_\bv\bw-\bnabla_\bw\bv = [\bv,\bw]$, where $[\cdot , \cdot ]$ stands for the Lie bracket. Hence we have that
\begin{equation}
(\bnabla_\bv \balpha)(\bw) - (\bnabla_\bw \balpha)(\bv) = \bv(\balpha(\bw)) - \bw(\balpha(\bv)) -  \balpha([\bv,\bw]) \ .
\end{equation}
We then recognize the covariant expression of the exterior derivative of $\balpha$ applied to $\bv,\bw$ on the right-hand side of the latter equation, which thus proves~\eqref{VorticityOperator_step01}.
\\

Now, note that $\bnabla$ commutes with the operator $\flat$. Indeed, by compatibility of $\bnabla$ with the tensor product:
\begin{equation}
\bnabla_\bv(g\otimes \bu) = \bnabla_\bv g \otimes \bu + g \otimes \bnabla_\bv\bu \ ,
\end{equation}
But the Levi-Civita connection preserves the metric tensor, \ie $\bnabla_\bv g = 0$, hence $\bnabla_\bv(g\otimes \bu) = g\otimes \bnabla_\bv \bu $. Hence, the definition of $\flat$ (\ie the contraction of the vector field with the metric), together with the compatibility of $\bnabla$ with tensor contraction, gives
\begin{equation}
\bnabla_\bv(\bu^\flat) = (\bnabla_\bv\bu)^\flat \ .
\label{VorticityOperator_step02}
\end{equation}

Upon denoting by $(\bnabla \bu)^\top$ the adjoint of $\bnabla\bu$ with respect to $g$, and $\bnabla_\bv\bu^\top$ its action on a vector $\bv$, we have
\begin{eqnarray}
g(\bnabla^A_\bv\bu,\bw) &=& \frac{1}{2} g(\bnabla_\bv\bu - \bnabla_\bv\bu^\top,\bw)  \\
&=& \frac{1}{2} \Big[ g(\bnabla_\bv\bu,\bw) - g(\bv,\bnabla_\bw\bu) \Big] \\
&=& \frac{1}{2} \Big[ (\bnabla_\bv\bu)^\flat(\bw) - (\bnabla_\bw\bu)^\flat(\bv) \Big] \ , \label{VorticityOperator_step03}
\end{eqnarray}
where we have just used the definition of $\bnabla\bu^\top$ and that of $\flat$.
Then, using ~\eqref{VorticityOperator_step03} together with~\eqref{VorticityOperator_step02}, we have that
\begin{equation}
g(\bnabla^A_\bv\bu,\bw) = \frac{1}{2} \Big[ (\bnabla_\bv\bu^\flat)(\bw) - (\bnabla_\bw\bu^\flat)(\bv) \Big] \ . 
\label{VorticityOperator_step04}
\end{equation}
Finally, combining~\eqref{VorticityOperator_step04} and~\eqref{VorticityOperator_step01} gives the result we wanted to prove~\eqref{VorticityOperator_step00}.

\section{Pushforward and divergence of skew-symmetric contravariant tensors}
\label{App:pushforwardDivergSkewTensor}
Let $(\mcM,g)$ be a smooth oriented Riemannian manifold without boundary. 
We denote by $\flat$ the isomorphism that lowers all indices with $g$, \ie that maps a contravariant tensors $\ba=a^{i_1\dots i_n}\partial_{i_1}\otimes\dots\otimes\partial_{i_n}$ to the covariant tensor $\ba^\flat \equiv a_{i_1\dots i_n} \rmd x^1\otimes\dots\otimes \rmd x^d$, with $a_{i_1\dots i_n} \equiv a^{j_1\dots j_n}g_{j_1 i_1}\dots g_{j_ni_n}$.
Finally, we denote by $\sharp$ the inverse map of $\flat$, \ie the map that raises all indices.

The metric $g$ induces for all $\bx\in\mcM$ an inner product on the space of covariant tensors of order $n$ at $\bx$:
\begin{equation}
\llangle \balpha,\bbeta \rrangle_{T^n} \equiv\alpha_{i_1\dots i_n}g^{i_1j_1}(\bx)\dots g^{i_nj_n}(\bx) \beta_{j_1\dots j_n} \ .
\end{equation}
(Note that all angular brackets in this appendix denote inner products and must not be confused with averages.)
For skew-symmetric covariant tensors of order $n$, it is also useful to define the rescaled scalar product:
\begin{equation}
\llangle \balpha,\bbeta \rrangle_{\Omega^n} \equiv  \frac{1}{n!}\llangle \balpha,\bbeta \rrangle_{T^n} \ .
\label{localInnerPdtForms}
\end{equation}
Denoting by $\blambda$ the Riemannian volume measure associating to $g$, we can define a (global) scalar product on skew-symmetric covariant tensor fields of order $n$ by integrating their scalar product a each point:
\begin{equation}
\langle\langle \balpha ,\bbeta \rangle\rangle \equiv \int_{\mcM}\llangle \balpha(\bx),\bbeta(\bx) \rrangle_{\Omega^n} \lambda(\bx) \rmd\bx \ .
\label{globalInnerPdtForms}
\end{equation}
The adjoint of the exterior derivative with respect to the scalar product~\eqref{globalInnerPdtForms} is denoted by $\rmd^\dagger$. They both act on differential forms \ie on skew-symmetric covariant tensor fields.
\\

Let us now pick a skew-symmetric contravariant tensor field $\ba$ of order $n$.
We recall that 
\begin{equation}
\rmd^\dagger (\ba^{\flat}) = - (\Div_{\blambda} \ba)^\flat \ ,
\end{equation}
where $[\Div_{\blambda} \ba]^{i_1\dots i_{n-1}} = \lambda^{-1}\partial_j (\lambda a^{ji_1\dots i_{n-1}})$.
Let $\mkm$ be a diffeomorphism on $\mcM$ that preserves the measure $\blambda$, \ie such that $\mkm_*\blambda=\blambda$. The objective of this appendix is to prove that 
\begin{equation}
\Div_{\blambda} (\mkm_*\ba) = \mkm_*\Div_{\blambda} \ba \ ,
\end{equation}
\ie that the pushforward by $\mkm$ commutes with the divergence operator on skew-symmetric contravariant tensor fields.
Let $\balpha$ be a skew-symmetric covariant tensor field of order $n-1$. 
For any contravariant tensor field $\bb$ of order $n-1$ we have that $\balpha(\bb)=\alpha_{i_1\dots i_{n-1}}b^{i_1\dots i_{n-1}}=\langle \balpha,\bb^\flat \rangle_{T^{n-1}}$ at every point $\bx\in\mcM$. Hence, using definition~\eqref{localInnerPdtForms}, we get
\begin{eqnarray}
\int_\mcM \balpha\left( \left[ \rmd^\dagger ( (\mkm_*\ba)^\flat)\right]^\sharp\right) \rmd \blambda = \int_\mcM \llangle  \balpha ,   \rmd^\dagger [ (\mkm_*\ba)^\flat] \rrangle_{T^{n-1}} \rmd \blambda = (n-1)! \int_\mcM \llangle  \balpha ,   \rmd^\dagger [ (\mkm_*\ba)^\flat] \rrangle_{\Omega^{n-1}} \rmd \blambda \ . \
\end{eqnarray}
But the last term corresponds to $(n-1)! \langle\langle \balpha ,   \rmd^\dagger [ (\mkm_*\ba)^\flat] \rangle\rangle$, which is the scalar product for which $\rmd^\dagger$ is the adjoint of $\rmd$. We thus have
\begin{eqnarray*}
\int_\mcM \balpha\left( \left[ \rmd^\dagger ( (\mkm_*\ba)^\flat)\right]^\sharp\right) \rmd \blambda &=&
(n-1)! \int_\mcM \llangle  \rmd\balpha ,   (\mkm_*\ba)^\flat \rrangle_{\Omega^n} \rmd \blambda \\
&=& \frac{(n-1)!}{n!} \int_\mcM \llangle  \rmd\balpha ,   (\mkm_*\ba)^\flat \rrangle_{T^n} \rmd \blambda \\
&=& \frac{1}{n} \int_\mcM   \rmd\balpha \left( \mkm_*\ba\right)  \rmd \blambda \ ,
\end{eqnarray*}
where once again we used definition~\eqref{localInnerPdtForms} together with the relation between $\flat$ and $\langle,\rangle_{T^n}$. The definition of the pullback of differential forms gives
\begin{equation}
\rmd\balpha_\bx (\mkm_*\ba) = [\mkm^*\rmd\balpha]_{\mkm^{-1}(\bx)}(\ba)=[\mkm^*\rmd\balpha]_{\mkm(\bx)}(\ba) \ ,
\label{pullbackPushforward}
\end{equation}
where the fact that $\mkm^{-1}=\mkm$ is due to $\mkm$ being involutive.
Further, we assumed that $\blambda$ is $\mkm$-invariant, which means that 
\begin{equation}
\rmd \blambda(\bx) = \rmd \mkm_*\blambda(\bx) = \lambda(\mkm(\bx)) |\mathrm{det[(\partial_j\mkm^i)(\bx)]}|\rmd \bx \ .
\label{preparChangeVar}
\end{equation}
Relations~\eqref{pullbackPushforward} \&~\eqref{preparChangeVar} then give
\begin{eqnarray}
\int_{\mcM} \rmd \balpha_\bx(\mkm_*\ba) \lambda(\bx)\rmd \bx = \int_\mcM [\mkm^*\rmd\balpha]_{\mkm(\bx)}(\ba) \lambda(\mkm(\bx)) |\mathrm{det[(\partial_j\mkm^i)(\bx)]}|\rmd \bx = \int_{\mcM} [\mkm^*\rmd\balpha]_{\by}(\ba)\lambda(\by)\rmd\by  \ , \
\end{eqnarray}
where the last equality is obtained by a change of variable $\by=\mkm(\bx)$.
In turn we have 
\begin{equation}
\int_\mcM \balpha\left( \left[ \rmd^\dagger ( (\mkm_*\ba)^\flat)\right]^\sharp\right) \rmd \blambda = \frac{1}{n} \int_\mcM   \mkm^*\rmd\balpha \left(\ba\right)  \rmd \blambda  \ .
\end{equation}

Now, we need to use the well-known commutativity between the pullback and the exterior derivative, $\mkm^*\rmd \balpha=\rmd \mkm^*\balpha$, and then do the same transformations in the integral as the ones we have just performed, but in the opposite direction.
The outcome reads
\begin{eqnarray*}
\int_\mcM \balpha\left( \left[ \rmd^\dagger ( (\mkm_*\ba)^\flat)\right]^\sharp\right) \rmd \blambda
&=& \frac{1}{n} \int_\mcM   \rmd\mkm^*\balpha \left(\ba\right)  \rmd \blambda \\
&=& \frac{1}{n} \int_\mcM   \llangle \rmd\mkm^*\balpha , \ba^\flat \rrangle_{T^n}   \rmd \blambda \\
&=& \frac{n!}{n} \int_\mcM   \llangle \rmd\mkm^*\balpha , \ba^\flat \rrangle_{\Omega^n}   \rmd \blambda \\
&=& (n-1)! \int_\mcM   \llangle \mkm^*\balpha , \rmd^\dagger (\ba^\flat) \rrangle_{\Omega^{n-1}}   \rmd \blambda \\
&=&  \int_\mcM   \llangle \mkm^*\balpha , \rmd^\dagger (\ba^\flat) \rrangle_{T^{n-1}}   \rmd \blambda \\
&=&  \int_\mcM   \mkm^*\balpha \left( \left[\rmd^\dagger (\ba^\flat) \right]^\sharp \right)   \rmd \blambda \\
&=&  \int_\mcM   \balpha \left(\mkm_* \left[\rmd^\dagger (\ba^\flat) \right]^\sharp \right)   \rmd \mkm_*\blambda \\
&=&  \int_\mcM   \balpha \left(\mkm_* \left[\rmd^\dagger (\ba^\flat) \right]^\sharp \right)   \rmd \blambda \ .
\end{eqnarray*}
Since this is true for all $\balpha$, we conclude that $\left[ \rmd^\dagger ( (\mkm_*\ba)^\flat)\right]^\sharp = \mkm_* \left[\rmd^\dagger (\ba^\flat) \right]^\sharp$, \ie 
$\Div_{\blambda} (\mkm_*\ba)=\mkm_*	\Div_{\blambda}\ba$.

\bibliographystyle{ieeetr}
\bibliography{../../Biblio}

\end{document}